\documentclass[a4paper,pra,aps,12pt,nofootinbib]{revtex4}

\usepackage[T1]{fontenc}
\usepackage[utf8]{inputenc}
\usepackage{units}
\usepackage{textcomp}
\usepackage{amsmath}    
\usepackage{graphicx}   
\usepackage{verbatim}   
\usepackage{color}      
\usepackage{subfigure}  
\usepackage{hyperref}   
\hypersetup{citebordercolor=.2 .8 .2,urlbordercolor=0 .75 1}
\usepackage{nicefrac}
\usepackage{amsfonts}
\usepackage{amsthm}
\usepackage{amssymb}
\usepackage{booktabs}
\usepackage{braket}
\usepackage{fancyhdr}
\usepackage{moreverb}
\usepackage[brazil]{babel}
\usepackage{mathtools}

\usepackage{dsfont}

\makeatletter
\renewcommand*{\email}[1][E-mail: ]{\begingroup\sanitize@url\@email{#1}}
\makeatother

\begin{document}

	\title{A tutorial on Dirac quantisation by analysing the problem of a ball on an inclined plane as a Hamiltonian system with constraints} 

	\author{M. F. Araujo de Resende}
	\email{resende@if.usp.br}
	\affiliation{Instituto de Física, Universidade de São Paulo, 05508-090 São Paulo SP, Brasil}
	\affiliation{Centro de Ciências Naturais e Humanas, Universidade Federal do ABC, 09210-580 Santo André SP, Brasil}
	
	\author{Thales Machado F.}
	\email{thales.machado@aluno.ufabc.edu.br}
	\affiliation{Centro de Ciências Naturais e Humanas, Universidade Federal do ABC, 09210-580 Santo André SP, Brasil}

	\date{\today }

	\begin{abstract}
		In this paper, we present a detailed review/analysis of the Dirac quantisation of Hamiltonian systems with constraints. To this end, we use, as a guide, the physical example provided by the dynamics of a solid ball rolling, without slipping, down an inclined plane under the action of gravity. After all, however simple this physical system may be, it provides a rich framework for this analysis since, in addition to allowing us to discuss scenarios involving holonomic and non-holonomic constraints, it is also a gauge system. Indeed, due to this latter fact, we have carefully detailed how the transition, from classical to quantum mechanics, must be guided by the Dirac-Bergmann algorithm and by the consequent replacement of Dirac brackets with commutators. As a central result, we demonstrate that the restriction of the Hamiltonian operator of this system with constraints to the physical Hilbert subspace (which is identified with the quantisation of these constraints) reproduces the same Schrödinger equation that can be originally obtained in intrinsic terms, a fact that only reinforces the consistency of the Dirac quantisation method.
	\end{abstract}

	\maketitle

	\section{Introduction}
	
		The entire development, which culminated in the creation of Quantum Mechanics in the first quarter of the 20th century, made it very clear to Paul A. M. Dirac (1902 -- 1984) that the quantum description of a system, which already has a classical analogue, can be expressed via a mathematical correspondence
		\begin{equation}
			F \mapsto \mathcal{Q} \left( F \right) = \hat{F} \label{q-correspondence}
		\end{equation}
		between the functions $ F $, which describe the observables of this classical analogue, and the operators $ \hat{F} $ that act in a Hilbert space \cite{dirac}. Indeed, as the advent of Matrix and Wave Mechanics took place within a Hamiltonian framework \cite{matrix-lewis,heisenberg-3,born-1,born-2,schrodinger-1,schrodinger-2}, upon analysing them, Dirac realised that, for a system whose classical analogue already admits a Hamiltonian formulation, this $ \mathcal{Q} $ must satisfy certain properties, namely \cite{assirati}:
		\begin{enumerate}
			\item[\textbf{I.} \ ] If $ F \left( q , p \right) $ and $ G \left( q , p \right) $ are two distinct functions, then the operators $ \hat{F} = \mathcal{Q} \left( F \right) $ and $ \hat{G} = \mathcal{Q} \left( G \right) $ must be distinct, which characterises $ \mathcal{Q} $ as an injective mapping.
			
			\item[\textbf{II.} \ ] If any two functions $ F $ and $ G $ describing physical observables are linear, then $ \mathcal{Q} $ must also be a linear mapping, since this implies that $ \mathcal{Q} $ satisfies
			\begin{equation*}
				\mathcal{Q} \left( \alpha F + \beta G \right) = \alpha \hspace*{0.04cm} \mathcal{Q} \left( F \right) + \beta \hspace*{0.04cm} \mathcal{Q} \left( G \right)
			\end{equation*}
			for any complex constants $ \alpha $ and $ \beta $.
			
			\item[\textbf{III.} \ ] If $ F $ is a complex function and, therefore, $ F^{\ast } $ is its complex conjugate, then it must hold that
			\begin{equation*}
				\left[ \mathcal{Q} \left( F \right) \right] ^{\dagger } = \mathcal{Q} \left( F^{\ast } \right)
			\end{equation*}
			since this is, for example, what ensures that, when $ F $ is a real function, $ \hat{F} $ is a self-adjoint operator. In other words, when $ F $ is a real function, all the eigenvalues of this operator $ \hat{F} $ are real numbers.
			
			\item[\textbf{IV.} \ ] Given a pair of canonically conjugate (classical) variables
			\begin{equation*}
				\left( q , p \right) = \left( \left( q^ {1} , q^{2} , \ldots , q^{n} \right) , \left( p_{1} , p_{2} , \ldots , p_{n} \right) \right) \ ,
			\end{equation*}
			the quantisation of the (coordinate) functions $ q^{\mu } $ and $ p_{\mu } $ (i.e. the results $ \mathcal{Q} \left( q^{\mu } \right) $ and $ \mathcal{Q} \left( p_{\mu } \right) $) must lead to the respective self-adjoint operators $ \hat{q} ^{\mu } $ and $ \hat{p} _{\mu } $ which satisfy the commutation relations
			\begin{equation}
				\left[ \hat{q} ^{\mu } , \hat{q} ^{\nu } \right] = \left[ \hat{p} _{\mu } , \hat{p} _{\nu } \right] = 0 \quad \textnormal{and} \quad \left[ \hat {q} ^{\mu } , \hat{p} _{\nu } \right] = i \hbar \hspace*{0.04cm} \delta ^{\mu } _{\nu } \hat{I} \ . \label{comutadores}
			\end{equation}
			Here, $ \hbar $ is Planck’s constant and $ \mu , \nu = 1 , 2 , \dots , n $, where $ n $ is a non-zero natural number. Incidentally, observe that, since the presence of the operator $ \hat{I} $ in these commutation relations stems from the fact that, for any two operators $ \hat{F} $ and $ \hat{G} $, the result of
			\begin{equation*}
				\bigl[ \hat{F} , \hat{G} \bigr] = \hat{F} \hat{G} - \hat{G} \hat{F}
			\end{equation*}
			must also be identified as an operator (which is not necessarily self-adjoint), it is necessary that the quantisation of the constant function $ 1 $ results in the identity operator $ \hat{I} = \mathcal{Q} \left( 1 \right) $.
			
			\item[\textbf{V.} \ ] The Hilbert space $ \mathfrak{H} _{\mathrm{ph}} $, on which all operators $ \hat{F} = \mathcal{Q} \left( F \right) $ act, must be irreducible with respect to the operators $ \hat{q} ^{\mu } $ and $ \hat{p} _{\mu } $. This means that any operator $ \hat{O} $, such that
			\begin{equation*}
				\bigl[ \hat{O} , \hat{q} ^{\mu } \bigr] = \bigl[ \hat{O} , \hat{p} _{\mu } \bigr] = 0
			\end{equation*}
			for all values $ \mu = 1 , 2 , \dots , n $, must be a multiple of the identity operator $ \hat{I} $.
			
			\item[\textbf{VI.} \ ] The result of the quantisation performed by $ \mathcal{Q} $ must have a classical limit. Roughly speaking, this means stating, for instance, that for any two operators $ \hat{F} = \mathcal{Q} \left( F \right) $ and $ \hat{G} = \mathcal{Q} \left( G \right) $, the relation
			\begin{equation}
				\lim _{\hbar \rightarrow 0} \left( \frac{1}{i \hbar } \bigl[ \hat{F} , \hat{G} \bigr] \right) = \left\{ F , G \right\} \label{pre-dirac}
			\end{equation}
			must hold because, in general, the product of such operators depends on Planck’s constant $ \hbar $.
		\end{enumerate}
		
		Note that, although we have written expression (\ref{pre-dirac}) using Poisson brackets \cite{ref-poisson}
		\begin{equation}
			\left\{ F , G \right\} = \sum ^{n} _{\mu = 1} \left( \frac{\partial F}{\partial q^{\mu }} \frac{\partial G}{\partial p_{\mu }} - \frac{\partial G}{\partial q^{\mu }} \frac{\partial F}{\partial p_{\mu }} \right)  \ , \label{poisson}
		\end{equation}
		where $ F \left( q , p \right) $ and $ G \left( q , p \right) $ are arbitrary functions\footnote{Although we are stating that these two functions may be arbitrary, note that, here, we are implicitly assuming that they must be, at the very least, of class $ C^{1} $: i.e., we are assuming that, at the very least, all their first-order partial derivatives exist and are continuous in the domains where they are defined \cite{spivak}.} that describe two physical observables classically, this expression (\ref{pre-dirac}) is only valid when $ \left( q , p \right) $ are intrinsic parameters: i.e., this expression (\ref{pre-dirac}) is only valid when the parameters contained in $ \left( q , p \right) $ are not related to one another. After all, when, for example, a set of relations
		\begin{equation}
			\Phi \left( q , p \right) = \left( \Phi _{\left( 1 \right) } \left( q , p \right) , \Phi _{\left( 2 \right) } \left( q , p \right) , \ldots , \Phi _{\left( m \right) } \left( q , p \right) \right) \approx \bigl( \underbrace{0 , 0 , \ldots , 0} _{m \ \text{times}} \bigr) \label{general-constraints}
		\end{equation}
		appears in the description of a physical system, where $ m $ is a natural number necessarily less than $ n $, Dirac demonstrated that the right-hand side of the expression (\ref{pre-dirac}) must be written using parentheses \cite{dirac}
		\begin{equation}
			\left\{ F , G \right\} _{D} = \left\{ F , G \right\} - \sum ^{m} _{j, \ell = 1} \left\{ F , \Phi _{\left( j \right) } \right\} \Theta ^{j \ell } \left\{ \Phi _{\left( \ell \right) } , G \right\} \ , \label{dirac-parenteses}
		\end{equation}
		which incorporates this term, where $ j ,\ell = 1 , 2 , \ldots , m $. And here, as all the terms
		\begin{equation}
			\Theta _{j \ell } = \left\{ \Phi _{\left( j \right) } , \Phi _{\left( \ell \right) } \right\} \label{terms}
		\end{equation}
		are to be interpreted as the elements of a matrix $ \Theta $ of order $ m $, all terms $ \Theta ^{j \ell } $ appearing in (\ref{dirac-parenteses}) must be interpreted as the elements of the inverse matrix $ \Theta ^{-1} $. Of course, much can be said about this $ \Theta ^{-1} $ since, as is well noted in the literature, this set of constraints (\ref{general-constraints}) does not always lead to a $ \Theta $ that is, in fact, invertible \cite{gitman}. However, what must be made very clear to you, the reader, at this early stage, is that the way we have written property \textbf{VI} should, in fact, be interpreted as a special case of a statement that is somewhat more general. In this case, as the special case of a statement that replaces the expression (\ref{pre-dirac}) with
		\begin{equation}
			\lim _{\hbar \rightarrow 0} \left( \frac{1}{i \hbar } \bigl[ \hat{F} , \hat{G} \bigr] \right) = \left\{ F , G \right\} _{D} \ , \label{dirac-condition}
		\end{equation}
		since the Dirac brackets (\ref{dirac-parenteses}), in fact, reduce to Poisson brackets when the matrix $ \Theta $ cannot be defined because no set of constraints is available for this purpose. Incidentally, note that what ensures the fulfilment of (\ref{dirac-condition}) is the fact that this quantisation process, in which a set of constraints $ \Phi \left( q , p \right) = 0 $ is present, is based on the correspondence \cite{dirac,gitman}
		\begin{equation}
			\bigl[ \hat{F} , \hat{G} \bigr] = i \hbar \hspace*{0.04cm} \mathcal{Q} \left( \left\{ F , G \right\} _{D} \right) \ , \label{correspondence-dirac}
		\end{equation}
		it can be shown that, in this case, the commutation relations (\ref{comutadores}) are not necessarily valid.
		
		It is also worth noting here that, due to the development of other quantum theories throughout the 20th century, various other quantisation processes have also emerged, driven by a range of motivations and requirements \cite{ali}. But, however interesting these other quantisation processes may be, they must all lead to results that are, in some way, compatible with those identified by Dirac, whose properties \textbf{IV} and \textbf{VI} are justified in the light of the correspondence (\ref{correspondence-dirac}). After all, given that both Werner Heisenberg (1901 -- 1976) and Erwin Schrödinger (1887 -- 1961) used (intentionally or not) the classical Hamiltonian formulation as the basis for the conception of Matrix and Wave Mechanics respectively \cite{heisenberg-3,born-2,schrodinger-1,schrodinger-2}, Dirac eventually realised that both Mechanics were, in fact, mere outcomes of the quantisation process outlined above. In other words, even though other quantisation processes have also come to light, they all need to be, in some way, compatible with the one identified by Dirac, since it was he who defined what we recognise today as Quantum Mechanics \cite{dirac-paper-1,dirac-paper-2,dirac-paper-3}.
		
		Of course, there are several questions that remain open regarding the compatibility just mentioned in the previous paragraph \cite{carosso}. One such question arises, for instance, from the fact that, since some classical systems are described by functions containing products $ q^{\mu } p_{\mu } $, such products prevent us from establishing this correspondence $ \mathcal{Q} $ in a unique way. After all, since properties \textbf{III} and \textbf{IV}, taken together, require that the operator, which must correspond to such a product, must be self-adjoint, we are unable to establish this $ \mathcal{Q} $ unambiguously simply because there is no unique way to define such an operator. In fact, by according to the Groenewold–van Hove theorem, for example, it is impossible to construct this correspondence $ \mathcal{Q} $ when classical systems are described by functions in which such products correspond to polynomials of degree greater than or equal to $ 3 $ \cite{groenewold}. However, whilst it is certainly very interesting to explore such open questions in order, for instance, better understand other nuances involved in this Dirac quantisation process, there are also several very simple systems, already well known to the general public, which, in addition to allowing us to have a good discussion of their description in terms of a constrained Hamiltonian system, also allow us to carry out a univocal quantisation that is quite consistent with reality. One such system is, for example, that consisting of a massive three-dimensional ball moving
		\begin{itemize}
			\item[(i)] under the action of a gravitational field $ \vec{g} $, and
			\item[(ii)] rolling, without any slippage, on a two-dimensional inclined plane.
		\end{itemize}
		And since this physical system is one whose classical description is already well established in the literature \cite{thornton,tipler,walker,sears}, it is precisely this system that we shall use to present this discussion throughout this manuscript.
		
		Indeed, in order to ensure that this discussion is conducted successfully, we shall use the next Section to present a fairly detailed review of the classical description of this physical system in Lagrangian terms. And as the next Section will be subdivided into ‘two parts’, with one providing a description of this system in intrinsic terms whilst the other explores a formalism involving a set of constraints, the next two Sections will focus on analogous descriptions, but in Hamiltonian terms. In the case of Section \ref{h-description}, for instance, as it shows us that the classical Hamiltonian description of this system in intrinsic terms contains no products between canonically conjugate variables, its Subsection \ref{intrinsic-q-description} will be devoted to presenting the univocal quantisation of this system. As for Section \ref{constraints-h-description}, which is the longest of all, it provides a very detailed analysis of this physical system in constrained Hamiltonian terms: after all, as it is this Section that reveals to us that this physical system is a gauge model, such detail is necessary since this situation leads us to one of the cases where the matrix $ \Theta $ is not invertible. In light of the description given in Section \ref{constraints-h-description}, the unambiguous quantisation of this Hamiltonian system with constraints is presented in Section \ref{q-constraints-description} and, given the discussion this prompts, we demonstrate that both quantisations describe the same physics.
		
	\section{\label{lag-description}Classical Lagrangian description}
	
\begin{figure}[!t]
			\centering
			\includegraphics[viewport=250 0 0 140,scale=1.2]{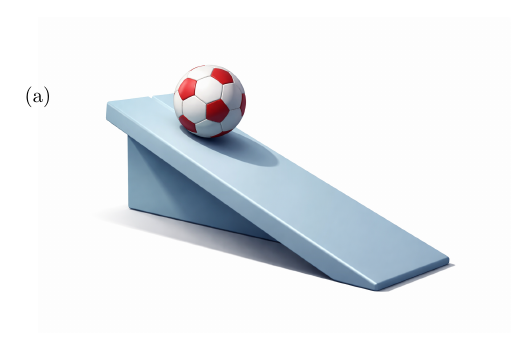} \\
			\includegraphics[viewport=250 20 0 132,scale=1.2]{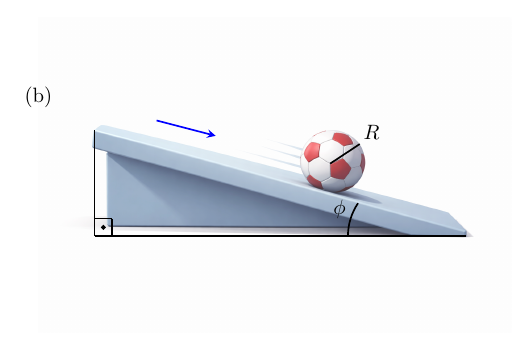}
			\caption{In the figure above (a), we see an image, merely illustrative, of a football rolling down an inclined plane. Although the system analysed in this manuscript is that of a massive ball, which is not necessarily identical to a football, the figure below (b) provides some details that clearly characterise this system, such as the fact that this ball has a non-zero radius $ R $ and that the plane is inclined at an angle $ \phi $ relative to the ground. Note that, in (b), the image of the ball rolling down the inclined plane has been deliberately rendered in duller colours to highlight the mathematical information it contains. One piece of this information is the direction of the projection of the Earth’s gravitational field onto this inclined plane, which is necessarily parallel to the hypotenuse of the right-angled triangle visible in this same figure.}
			\label{plano-inclinado}
		\end{figure}

		Indeed, in order to begin analysing the situation of this three-dimensional massive ball, which moves whilst satisfying conditions (i) and (ii) mentioned in the Introduction, it is quite useful to consider the situation illustrated in Figure \ref{plano-inclinado}.
		After all, it is precisely with the help of this Figure \ref{plano-inclinado} that we are able to identify certain pieces of information which, although quite simple, are of great value in describing this motion, such as, for example,
		\begin{itemize}
			\item the fact that the ball has a non-zero radius $ R $ and
			\item the existence of an angle $ \phi $, also non-zero, which allows us to gauge how inclined this plane is relative to the ground.
		\end{itemize}
		
\begin{figure}[!t]
			\centering
			\includegraphics[viewport=280 20 0 140,scale=1.2]{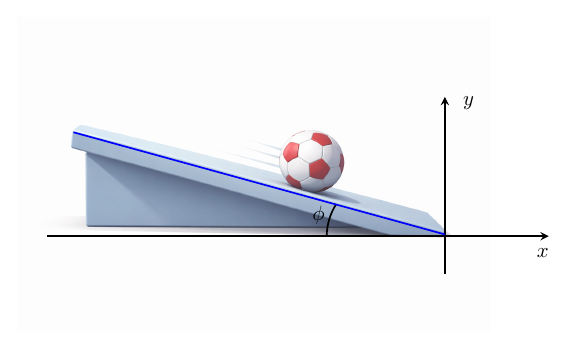}
			\caption{Here, we see the same content as shown in Figure \ref{plano-inclinado}(b), but now superimposed on a Cartesian coordinate system. Note that the projection of the inclined plane, which lies in the second quadrant of the $ xy $-plane, corresponds to the graph defined by (\ref{declive}).}
			\label{reta}
		\end{figure}

		Note that, although this massive ball is three-dimensional, the fact that its motion is restricted to a two-dimensional inclined plane seems to imply that the translation of this ball along that plane can be perfectly described using only two spatial variables \cite{kreyszig}: i.e., these two variables are the ones that, in some way, are capable of describing this plan intrinsically. However, as the projection of the gravitational force onto this inclined plane corresponds to a vector field whose direction is the same as that already highlighted (in blue) in Figure \ref{plano-inclinado}(b), this ultimately reveals that, in fact, the translation of the ball along this plane can be described using a single spatial variable provided that $ \phi $ is constant. In order to understand why this is so, one need only note a very important fact that Figure \ref{reta}
		already illustrates. What fact? That it is possible to choose a coordinate system in which this plane is such that
		\begin{equation}
			y \left( x \right) = - \left( \tan \phi \right) x \ . \label{declive}
		\end{equation}
		And why is this fact so important? Because, when this massive ball moves an infinitesimal distance $ ds $ along this inclined plane, in the same direction as that determined by the projection of the gravitational field, it is precisely the non-zero value of $ \phi $ that allows us to deal, for instance, with the Pythagorean relation
		\begin{equation}
			\left( ds \right) ^{2} = \left( dx \right) ^{2} + \left( dy \right) ^{2} \ , \label{pitagoras}
		\end{equation}
		where $ dx $ and $ dy $ are the (infinitesimal) projections of $ ds $ that are, respectively, parallel and perpendicular to the ground. However, since (\ref{declive}) also makes it quite clear that these (infinitesimal) projections $ dx $ and $ dy $ are related by
		\begin{equation*}
			dy = \frac{dy}{dx} \ dx = - \left( \tan \phi \right) dx \ ,
		\end{equation*}
		this is what allows us to conclude that this Pythagorean relation (\ref{pitagoras}) can be reduced to
		\begin{equation}
			ds = \left( \sec \phi \right) dx \label{reduction}
		\end{equation}
		due to the identity
		\begin{equation}
			\sec ^{2} \phi = 1 + \tan ^{2} \phi \ . \label{trigonometry}
		\end{equation}
		In other words, even though this two-dimensional inclined plane can be described, intrinsically, by only two spatial variables, this result (\ref{reduction}) already makes it quite clear that any distance $ s $, travelled by the massive ball along this inclined plane, can be parameterised using only the spatial variable $ x $. In this context, as the expression\footnote{Note that, here, we are using the convention $ \dot{F} = dF / dt $, which will prove very useful later on, where $ F $ denotes any physically meaningful function whose total derivative can be evaluated with respect to the parameter $ t $ (time).}
		\begin{equation}
			K_{\parallel } \left( \dot{s} \right) = \frac{m \dot{s} ^{2}}{2} \ \Leftrightarrow \ K_{\parallel } \left( \dot{x} , \dot{y} \right) = \left[ \frac{m \dot{x} ^{2}}{2} + \frac{m \dot {y} ^{2}}{2} - \lambda ^{\left( 1 \right) } f_{1} \left( x , y \right) \right] _{f_{\left( 1 \right) } = 0} \label{constrained-k}
		\end{equation}
		actually describes the kinetic energy of the translation of this ball, where
		\begin{equation}
			f_{\left( 1 \right) } \left( x , y \right) = \left( \tan \phi \right) x + y \label{f1-constraint}
		\end{equation}
		and $ \lambda ^{\left( 1 \right) } $ is a multiplier that implements the constraint
		\begin{equation*}
			f_{\left( 1 \right) } \left( x , y \right) \approx 0
		\end{equation*}
		that follows from (\ref{declive}), it is not difficult to see that the expression for this same kinetic energy in intrinsic terms can be written as
		\begin{equation}
			K_{\parallel } \left( \dot{x} \right) = \frac{m \left( \sec ^{2} \phi \right) \dot{x} ^{2}}{2} \ . \label{k-paralela}
		\end{equation}
		
\begin{figure}[!t]
			\centering
			\includegraphics[viewport=380 60 0 170,scale=1.3]{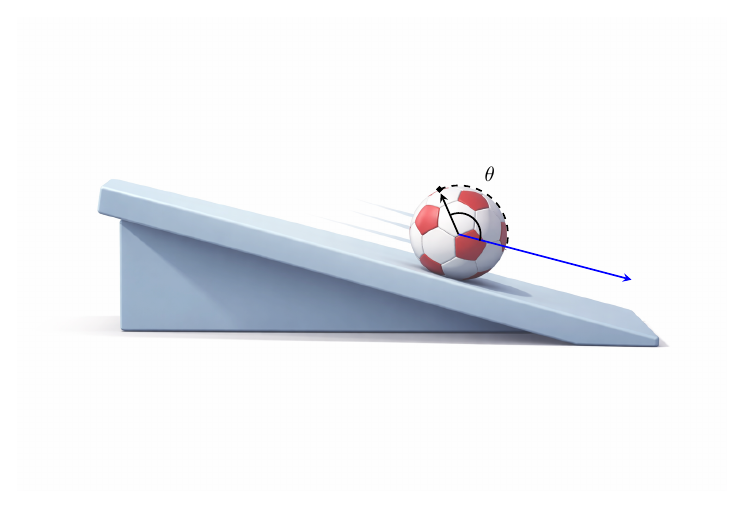}
			\caption{Here, once again, we see the same content as in Figure \ref{plano-inclinado}(b), only now slightly enlarged to highlight the presence of the parameter $ \theta $, which helps us to assess how this ball (which is a rigid body) rotates about its own axis: more specifically, how any of the fixed points on this ball (such as the one highlighted in black) rotates about the axis that passes through the centre of the ball (an axis which, although not shown here, must be perpendicular to the same $ xy $ plane highlighted in Figure \ref{reta}) as it rolls down the inclined plane. And, as this figure makes clear, one way (though not the only one) to do this is by defining $ \theta $ as the angle between a direction defined by the projection of the Earth’s gravitational field (highlighted in blue in accordance with what has already been shown in Figures \ref{plano-inclinado}(b) and \ref{reta}), and another direction defined by connecting the centre of this ball to any of its fixed points.}
			\label{rolando}
		\end{figure}

		Of course, in light of what we have just said, a more attentive reader might, for instance, point out that (\ref{k-paralela}) is the expression of the centre of mass energy of this massive ball. After all, as we have already stated in the Introduction, this massive ball moves by rolling, without any sliding, along this inclined plane. As a consequence, it is quite reasonable that this more attentive reader will realise that there is another form of kinetic energy that must also be taken into account here. In other words, since the variable $ x $ is being used to describe the translational motion (i.e. the movement of the centre of mass) of this sphere along the inclined plane, it is quite natural for this more attentive reader to observe that (\ref{k-paralela}) is not yet accounting for the energy due to the fact that this sphere also rotates about an axis which, in addition to being perpendicular to the $ xy $ plane, intersects its centre, as illustrated in Figure \ref{rolando}.
		And, as this more attentive reader is quite right to draw our attention to this fact, it is worth recalling that the expression for this other form of kinetic energy, which is also associated with the motion of this massive ball on the plane, is given by
		\begin{equation}
			K_{\circlearrowright } \bigl( \dot{\theta } \bigr) = \frac{I \dot{\theta } ^{2}}{2} \ , \label{k-rotating}
		\end{equation}
		where $ \dot{\theta } $ is the angular velocity (frequency) characterising this rolling motion \cite{savelyev-1}. Certainly, many other considerations can be raised by regarding the movement of this ball on the inclined plane, since, due to the nature of the contact between the surface of the ball and that of the plane, a variety of things can happen. However, when we consider the various reasons that lead us to assume that, from a certain point in time, this rolling occurs without any slippage, and that this absence of slippage can be characterised by the expression \cite{poisson-book}
		\begin{equation}
			ds = R d \theta \label{ds} \ ,
		\end{equation}
		it is not difficult to see that, in this case, the expression (\ref{reduction}) reduces to
		\begin{equation}
			d \theta = \frac{\sec \phi }{R} \ dx \ \Leftrightarrow \ \dot{\theta } = \frac{\sec \phi }{R} \ \dot{x} \ . \label{rotation}
		\end{equation}
		Thus, since the momentum of inertia of a hollow ball ($ a = 0 $) or a solid ball ($ a = 2 $), rotating about one of its infinite axes of symmetry, is defined as \cite{symon}
		\begin{equation*}
			I = \frac{2mR^{2}}{a + 3} \ ,
		\end{equation*}
		it is not difficult to conclude that, whether this sphere is actually hollow or solid, the expression (\ref{rotation}) can be expanded as
		\begin{equation}
			K_{\circlearrowright } \bigl( \dot{x} , \dot{\theta } \bigr) = \left[ \frac{m \bigl( R \dot{\theta } \bigr) ^{2}}{a + 3} - \lambda ^{\left( 2 \right) } f_{\left( 2 \right) } \bigl( \dot{x} , \dot{\theta } \bigr) \right] _{f_{\left( 2 \right) } = 0} \ , \label{constrained-rotation-k}
		\end{equation}
		where
		\begin{equation}
			f_{\left( 2 \right) } \bigl( \dot{x} , \dot{\theta } \bigr) = \left( \sec \phi \right) \dot{x} - R \dot{\theta } \label{f2-constraint}
		\end{equation}
		and $ \lambda ^{\left( 2 \right) } $ is a multiplier that implements the constraint
		\begin{equation*}
			f_{\left( 2 \right) } \bigl( \dot{x} , \dot{\theta } \bigr) \approx 0
		\end{equation*}
		which follows from expression (\ref{rotation}). In other words, it is not difficult to conclude that the kinetic energy, which is associated solely with rolling without slipping (i.e., with rolling that satisfies condition (\ref{ds})) of this ball down the plane, can be expressed as
		\begin{equation}
			K_{\circlearrowright } \left( \dot{x} \right) = \frac{m \left( \sec ^{2} \phi \right) \dot{x} ^{2}}{a + 3} \label{k-rotating-x}
		\end{equation}
		in intrinsic terms.
		
		\subsection{Further points regarding the intrinsic description}
		
			Note that, if we assume that the minimum value of the potential energy associated with this ball is attained at the point $ \left( x , y \right) = \left( 0 , 0 \right) $, it is not difficult to infer that one of the functions that best models this potential energy is given by
			\begin{equation}
				U \left( x , y \right) = \left[ mgy - \lambda ^{\left( 1 \right) } f_{\left( 1 \right) } \left( x , y \right) \right] _{f_{\left( 1 \right) } = 0} \ \Leftrightarrow \ U \left( x \right) = - mg \left( \tan \phi \right) x \label{potential}
			\end{equation}
			due to relation (\ref{declive}). Here, $ g = \left\vert \vec{g} \right\vert $. And this is a result which, when analysed together with the others (\ref{k-paralela}) and (\ref{k-rotating-x}), leads us to the conclusion that the only spatial variable that is, in fact, relevant to the description of this problem is $ x $ since, for example, the total kinetic energy of this ball, whether hollow or solid, can be expressed as
			\begin{equation}
				K \left( \dot{x} \right) = K_{\parallel } \left( \dot{x} \right) +  K_{\circlearrowright } \left( \dot{x} \right) = \frac{m \left( \sec ^{2} \phi \right) \dot {x} ^{2}}{2} \left( \frac{a + 5}{a + 3} \right) \ . \label{k-total}
			\end{equation}
			In this context, since the expressions (\ref{potential}) and (\ref{k-total}), taken together, allow us to define a Lagrangian function via \cite{landau-mechanics}
			\begin{equation}
				L \left( x , \dot{x} \right) = K \left( \dot{x} \right) - U \left( x \right) \ , \label{intrisic-lagrangean}
			\end{equation}
			it is not difficult to see that, in accordance with the Euler-Lagrange formalism, the equation of motion for this ball can be obtained via
			\begin{equation*}
				\frac{d}{dt} \left( \frac{dL}{d \dot{x}} \right) - \frac{dL}{dx} = 0 \ .
			\end{equation*}
			That is, in accordance with all the results we have obtained above, the equation of motion for this ball is given by
			\begin{equation}
				\ddot{x} - g \left( \frac{a + 3}{a + 5} \right) \frac{\sin \left( 2 \phi \right) }{2} = 0 \ . \label{mov-eq}
			\end{equation}
			
		\subsection{\label{sec:vak}On the description in terms of a constrained system}
		
			However, as much as everything we did in the previous Subsection is correct, it is worth noting that one of the things that the expressions (\ref{constrained-k}), (\ref{constrained-rotation-k}) and (\ref{potential}) are doing is pointing to the possibility of describing this system (which consists of a massive ball moving whilst satisfying conditions (i) and (ii)), by exploiting a Lagrangian formulation with constraints. After all, note that, for these expressions (\ref{constrained-k}), (\ref{constrained-rotation-k}) and (\ref{potential}) could be well defined, it was necessary to identify two expressions (\ref{f1-constraint}) and (\ref{f2-constraint}) which, with the aid of some multipliers, impose the constraints
			\begin{subequations}
				\begin{align*}
					f_{\left( 1 \right) } \bigl( x , y \bigr) & \approx 0 \quad \textnormal{and} \\
					f_{\left( 2 \right) } \bigl( \dot{x} , \dot{\theta } \bigr) & \approx 0
				\end{align*}
			\end{subequations}
			that are, respectively, capable of characterising the motion of this ball moving in accordance with conditions (i) and (ii). In other words, one of the things that these three expressions (\ref{constrained-k}), (\ref{constrained-rotation-k}) and (\ref{potential}) demonstrate is that it is perfectly possible to construct an alternative Lagrangian function that, despite having a greater number of variables, allows us to obtain the same equations of motion as (\ref{intrisic-lagrangean}).
			
			Speaking of which, as regards deriving this alternative Lagrangian function, all the results we have obtained so far suggest that it seems possible to express it, for instance, as
			\begin{eqnarray}
				\lefteqn{L_{\mathrm{vak}} \left( \bigl( x , y , \theta , \lambda ^{\left( 1 \right) } , \lambda ^{\left( 2 \right) } \bigr) , \bigl( \dot{x} , \dot{y} , \dot{\theta } , \dot{\lambda } ^{\left( 1 \right) } , \dot{\lambda } ^{\left( 2 \right) } \bigr) \right) } \hspace*{2.0cm} \notag \\
				& = & \frac{m \dot{x} ^{2}}{2} + \frac{m \dot {y} ^{2}}{2} + \frac{m \bigl( R \dot{\theta } \bigr) ^{2}}{a + 3} - mgy - \lambda ^{\left( 1 \right) } f_{\left( 1 \right) } \left( x , y \right) - \lambda ^{\left( 2 \right) } f_{\left( 2 \right) } \bigl( \dot{x} , \dot{\theta } \bigr) \ , \label{constrained-lagrangean-vak}
			\end{eqnarray}
			where not only $ \lambda ^{\left( 1 \right) } $ and $ \lambda ^{\left( 2 \right) } $, now interpreted as Lagrange multipliers, but also their respective generalised velocities $ \dot{\lambda } ^{\left( 1 \right) } $ and $ \dot{\lambda } ^{\left( 2 \right) } $ must be raised to the status of variables in this new formulation. However, although (\ref{constrained-lagrangean-vak}) does indeed seem like an excellent candidate for describing this physical system because
			\begin{equation*}
				L \left( x , \dot{x} \right) = \left. L_{\mathrm{vak}} \left( \bigl( x , y , \theta , \lambda ^{\left( 1 \right) } , \lambda ^{\left( 2 \right) } \bigr) , \bigl( \dot{x} , \dot{y} , \dot{\theta } , \dot{\lambda } ^{\left( 1 \right) } , \dot{\lambda } ^{\left( 2 \right) } \bigr) \right) \right\vert _{\vec{f} = \vec{0}} \ ,
			\end{equation*}
			where $ \vec{f} = \left( f_{\left( 1 \right) } , f_{\left( 2 \right) } \right) $, this expression raises an important point that needs to be addressed here. After all, the presence of the term
			\begin{equation*}
				\lambda ^{\left( 2 \right) } f_{\left( 2 \right) } \bigl( \dot{x} , \dot{\theta } \bigr)
			\end{equation*}
			in (\ref{constrained-lagrangean-vak}) demands that we deal with Vakonomic Mechanics \cite{kozlov-1,kozlov-2,kozlov-3,cardin}: i.e., with an alternative model of Analytical Mechanics that, in addition to exploring a principle of least action\footnote{Here, it is worth noting that although this variational principle has become popularly known as the principle of least action for historical and philosophical reasons, it merely requires that the action associated with any mechanical system be stationary in order to obtain the equations of motion.} which differs slightly from the original to arrive at the equations of motion, it also requires that, in the transition from Lagrangian to Hamiltonian formalism, we deal with affine Poisson brackets in order to establish an equivalence with those of Dirac \cite{leon-1,jimenez}. In this fashion, as the focus of our manuscript is to serve as a sort of tutorial on the quantisation method originally identified by Dirac, we shall not pursue any analysis involving Vakonomic Mechanics, however fruitful the discussions to which it might lead us may be \cite{benito,borisov,michal,vak-lemos,huang-1,huang-2}.
			
			\subsubsection{A more appropriate implementation of the non-holonomic constraint}
			
				Incidentally, since we have just referred to the principle of least action, it is important to note that, because (\ref{f2-constraint}) is defined by a function that depends on the generalised velocities $ \dot{x} $ and $ \dot{\theta } $ (rather than the corresponding coordinates $ x $ and $ \theta $), the theory of Lagrangian systems with constraints suggests that a good way for us to define this alternative Lagrangian function $ L_{\mathrm{alt}} $ (without having to deal with a principle of least action different from the original) one is by considering the Euler-Lagrange equations \cite{lemos-book}               
				\begin{equation}
					\frac{d}{dt} \left( \frac{\partial L_{\mathrm{alt}}}{\partial \dot{q} ^{\mu }} \right) - \frac{\partial L_{\mathrm{alt}}}{\partial q^{\mu }} = \lambda ^{\left( 2 \right) } \frac{\partial f_{\left( 2 \right) }}{\partial \dot{q} ^{\mu }} \label{alt-el}
				\end{equation}
				modified with respect to the Lagrange-D’Alembert principle, by taking into account that $ q = \bigl( x , y , \theta , \lambda ^{\left( 1 \right) } , \lambda ^{\left( 2 \right) } \bigr) $ since
				\begin{eqnarray}
					\lefteqn{L_{\mathrm{alt}} \left( \bigl( x , y , \theta , \lambda ^{\left( 1 \right) } \bigr) , \bigl( \dot{x} , \dot{y} , \dot{\theta } , \dot{\lambda } ^{\left( 1 \right) } \bigr) \right) } \hspace*{3.0cm} \notag \\
					& = & \frac{m \dot{x} ^{2}}{2} + \frac{m \dot{y} ^{2}}{2} + \frac{m \bigl( R \dot{\theta } \bigr) ^{2}}{a + 3} - mgy - \lambda ^{\left( 1 \right) } f_{\left( 1 \right) } \left( x , y \right) \bigr) \ . \label{constrained-lagrangean-non-holo}
				\end{eqnarray}
				After all, since the application of (\ref{alt-el}) leads us to
				\begin{subequations} \label{nh-lag}
					\begin{align}
						\frac{d}{dt} \left( \frac{\partial L_{\mathrm{alt}}}{\partial \dot{x}} \right) - \frac{\partial L_{\mathrm{alt}}}{\partial x} = \lambda ^{\left( 2 \right) } \frac{\partial f_{\left( 2 \right) }}{\partial \dot{x}} \ & \Rightarrow \ \ddot{x} + \frac{\lambda ^{\left( 1 \right) } \left( \tan \phi \right) }{m} + \frac{\lambda ^{\left( 2 \right) } \left( \sec \phi \right) }{m} = 0 \ , \label{nh-lag-1} \\
						\frac{d}{dt} \left( \frac{\partial L_{\mathrm{alt}}}{\partial \dot{y}} \right) - \frac{\partial L_{\mathrm{alt}}}{\partial y} = 0 \ & \Rightarrow \ \ddot{y} - \left( g - \frac{\lambda ^{\left( 1 \right) }} {m} \right) = 0 \ , \label{nh-lag-2} \\
						\frac{d}{dt} \left( \frac{\partial L_{\mathrm{alt}}}{\partial \dot{\theta }} \right) - \frac{\partial L_{\mathrm{alt}}}{\partial \theta } = \lambda ^{\left( 2 \right) } \frac{\partial f_{\left( 2 \right) }}{\partial \dot{\theta }} \ & \Rightarrow \ \ddot{\theta } = - \frac{\lambda ^{\left( 2 \right) } \left( a + 3 \right) }{2mR} \ , \label{nh-lag-3} \\
						\frac{d}{dt} \left( \frac{\partial L_{\mathrm{alt}}}{\partial \dot{\lambda } ^{1}} \right) - \frac{\partial L_{\mathrm{alt}}}{\partial \lambda ^{\left( 1 \right) }} = 0 \ & \Rightarrow \ f_{\left( 1 \right) } \left( x , y \right) = 0 \ \ \textnormal{and} \label{nh-lag-4} \\
						\frac{d}{dt} \left( \frac{\partial L_{\mathrm{alt}}}{\partial \dot{\lambda } ^{2}} \right) - \frac{\partial L_{\mathrm{alt}}}{\partial \lambda ^{\left( 2 \right) }} = 0 \ & \Rightarrow \ f_{\left( 2 \right) } \bigl( \dot{x} , \dot{\theta } \bigr) = 0 \ , \label{nh-lag-5}
					\end{align}
				\end{subequations}
				it is quite remarkable that, in addition to (\ref{nh-lag-4}) and (\ref{nh-lag-5}) turn out to be the same as the constraints (\ref{f1-constraint}) and (\ref{f2-constraint}) respectively, it is these constraints that, when substituted into the subsequent manipulations of (\ref{nh-lag-2}) and (\ref{nh-lag-3}), help us to realise that
				\begin{subequations}
					\begin{align*}
						\chi ^{\left( 1 \right) } & = m \left( g + \ddot{y} \right) = m \left[ g - \left( \tan \phi \right) \ddot{x} \right] \quad \textnormal{and} \\
						\chi ^{\left( 2 \right) } & = - \frac{2mR}{a + 3} \ \ddot{\theta } = - \frac{2m \left( \sec \phi \right) }{a + 3} \ \ddot{x} \ .
					\end{align*}
				\end{subequations}
				That is, we are faced with two results which, when substituted into (\ref{h-lag-1}), yield
				\begin{equation}
					\left[ 1 + \left( \tan ^{2} \phi \right) \right] \ddot{x} + \frac{2 \left( \sec ^{2} \phi \right) }{a + 3} \ \ddot{x} - g \left( \tan \phi \right) = 0 \ . \label{soma-trigo}
				\end{equation}
				And, since the trigonometric relation (\ref{trigonometry}) allows us to rewrite this last result as
				\begin{equation}
					\left( \frac{a + 5}{a + 3} \right)\left( \sec ^{2} \phi \right) \ddot{x} - g \left( \tan \phi \right) = 0 \ , \label{mov-equation-l2}
				\end{equation}
				it is precisely this that makes it clear that, in fact, $ L $ and $ L_{\mathrm{alt}} $ are describing the same physical reality, since the further expansion of (\ref{mov-equation-l2}) leads us, exactly, to the same equation (\ref{mov-eq}).
			
			\subsubsection{A holonomic solution}
			
				Although all the Lagrangian formulations we have presented above help us to describe the same physical reality, it is quite striking that the constraints (\ref{f1-constraint}) and (\ref{f2-constraint}) are of a different nature. After all, note that, whilst (\ref{f1-constraint}) dictates where our physical system can be, the expression (\ref{f2-constraint}) dictates how our physical system can move. However, even though relation (\ref{rotation}) has led us to identify this constraint (\ref{f2-constraint}), one cannot fail to recognise that this same relation (\ref{rotation}) also points to the existence of an isomorphism between the parameters $ x $ and $ \theta $, which can be perfectly defined as
				\begin{equation}
					\theta \left( x \right) = \frac{\left( \sec \phi \right) x}{R} \ \Leftrightarrow \  x \left( \theta \right) = \frac{R \theta }{\sec \phi } \ . \label{isomorphism}
				\end{equation}
				And why is it interesting to recognise this? Because what this isomorphism is showing us is that, in fact, it also defines a constraint that this physical system must satisfy. Spoken in more specific terms, it is not at all difficult to see that this isomorphism is, in fact, defining a constraint
				\begin{equation*}
					\left( \sec \phi \right) x - R \theta \approx 0 \ ,
				\end{equation*}
				which seems far more interesting to consider, since it is precisely from this that the constraint (\ref{f2-constraint}) arises.
				
				Of course, the expression (\ref{isomorphism}) that we have chosen for this isomorphism might well be questioned by a more attentive reader. After all, since (\ref{isomorphism}) arises from an integration, which must be performed over the equality on the left-hand side of (\ref{rotation}), such integration may lead to non-zero constants that it seems we are ignoring here. However, even if such an integration were to lead us to a more general isomorphism, defined by
				\begin{equation*}
					\theta \left( x \right) - \theta _{0} = \frac{\sec \phi }{R} \left( x - x_{0} \right) \ \Leftrightarrow \ x \left( \theta \right) - x_{0} = \frac{R}{\sec \phi } \left( \theta - \theta _{0} \right) \ ,
				\end{equation*}
				where $ \theta _{0} $ and $ x_{0} $ are these non-zero constants, observe that, since there exists a variable transformation $ \left( x , \theta \right) \rightarrow \left( x^{\prime } , \theta ^{\prime } \right) $ such that
				\begin{equation}
					\left( \sec \phi \right) x_{0} - R \theta _{0} = 0 \ , \label{initial-condition}
				\end{equation}
				these are precisely the variables we are considering here. In other words, we are considering that $ \left( x , \theta \right) $ are the variables that allow us to model the initial rolling condition, without slippage, by using the relation (\ref{initial-condition}) because this does not detract, in any way from, the description of our physical problem. In this way, as the definition of a new function
				\begin{equation*}
					F_{\left( 2 \right) } \left( x , \theta \right) = \left( \sec \phi \right) x - R \theta \ ,
				\end{equation*}
				and the consequent use of a new Lagrangian
				\begin{eqnarray}
					\lefteqn{L_{\mathrm{hol}} \left( \bigl( x , y , \theta , \chi ^{\left( 1 \right) } , \chi ^{\left( 2 \right) } \bigr) , \bigl( \dot{x} , \dot{y} , \dot{\theta } , \dot{\chi } ^{\left( 1 \right) } , \dot{\chi } ^{\left( 2 \right) } \bigr) \right) } \hspace*{2.0cm} \notag \\
					& = & \frac{m \dot{x} ^{2}}{2} + \frac{m \dot {y} ^{2}}{2} + \frac{m \bigl( R \dot{\theta } \bigr) ^{2}}{a + 3} - mgy - \chi ^{\left( 1 \right) } f_{\left( 1 \right) } \left( x , y \right) - \chi ^{\left( 2 \right) } F_{\left( 2 \right) } \bigl( x , \theta \bigr) \ , \label{constrained-lagrangean-holo}
				\end{eqnarray}
				leads us to the set of Euler-Lagrange equations
				\begin{subequations} \label{h-lag}
					\begin{align}
						\frac{d}{dt} \left( \frac{\partial L_{\mathrm{hol}}}{\partial \dot{x}} \right) - \frac{\partial L_{\mathrm{hol}}}{\partial x} = 0 \ & \Rightarrow \ \ddot{x} + \frac{\chi ^{\left( 1 \right) } \left( \tan \phi \right) }{m} + \frac{\chi ^{\left( 2 \right) } \left( \sec \phi \right) }{m} = 0 \ , \label{h-lag-1} \\
						\frac{d}{dt} \left( \frac{\partial L_{\mathrm{hol}}}{\partial \dot{y}} \right) - \frac{\partial L_{\mathrm{hol}}}{\partial y} = 0 \ & \Rightarrow \ \ddot{y} + \left( g + \frac{\chi ^{\left( 1 \right) }} {m} \right) = 0 \ , \label{h-lag-2} \\
						\frac{d}{dt} \left( \frac{\partial L_{\mathrm{hol}}}{\partial \dot{\theta }} \right) - \frac{\partial L_{\mathrm{hol}}}{\partial \theta } = 0 \ & \Rightarrow \ \ddot{\theta } - \frac{\chi ^{\left( 2 \right) } \left( a + 3 \right) }{2mR} = 0 \ , \label{h-lag-3} \\
						\frac{d}{dt} \left( \frac{\partial L_{\mathrm{hol}}}{\partial \dot{\chi } ^{\left( 1 \right) }} \right) - \frac{\partial L_{\mathrm{hol}}}{\partial \chi ^{\left( 1 \right) }} = 0 \ & \Rightarrow \ f_{\left( 1 \right) } \left( x , y \right) = 0 \ \ \textnormal{and} \label{h-lag-4} \\
						\frac{d}{dt} \left( \frac{\partial L_{\mathrm{hol}}}{\partial \dot{\chi } ^{\left( 2 \right) }} \right) - \frac{\partial L_{\mathrm{hol}}}{\partial \chi ^{\left( 2 \right) }} = 0 \ & \Rightarrow \ F_{\left( 2 \right) } \bigl( x , \theta \bigr) = 0 \ , \label{h-lag-5}
					\end{align}
				\end{subequations}
				it is precisely these equations that make it crystal clear that $ L $ and $ L_{\mathrm{hol}} $ do, in fact, describe the same physical reality\footnote{Just to prevent you, the reader, from eventually confusing, later on, the results that have already followed from the Lagrangian $ L_{\mathrm{alt}} $ with those that will follow from $ L_{\mathrm{hol}} $, we shall denote the Lagrange multipliers, which implement the constraints
				\begin{equation}
					f_{\left( 1 \right) } \left( x , y \right) = 0 \quad \textnormal{and} \quad F_{\left( 2 \right) } \left( x , \theta \right) = 0 \label{new-constraints}
				\end{equation}
				to $ L_{\mathrm{hol}} $, as $ \chi ^{\left( 1 \right) } $ and $ \chi ^{\left( 2 \right) } $ respectively.}. After all, since these equations contain (\ref{h-lag-4}) and (\ref{h-lag-5}) as the necessary constraints for this physical description to be successfully realised, note that replacing them into equations (\ref{h-lag-2}) and (\ref{h-lag-3}) leads us to the respective expressions
				\begin{subequations}
					\begin{align*}
						\chi ^{\left( 1 \right) } & = - m \left( g + \ddot{y} \right) = - m \left[ g - \left( \tan \phi \right) \ddot{x} \right] \quad \textnormal{and} \\
						\chi ^{\left( 2 \right) } & = \frac{2mR}{a + 3} \ \ddot{\theta } = \frac{2m \left( \sec \phi \right) }{a + 3} \ \ddot{x}
					\end{align*}
				\end{subequations}
				that, when substituted into (\ref{h-lag-1}), lead us to the same result (\ref{soma-trigo}). And, as we have already stated that the expansion of (\ref{soma-trigo}) leads us to equation (\ref{mov-eq}), it is precisely this that allows us to state that the Lagrangian functions $ L $ and $ L_{\mathrm{hol}} $ describe the same physical reality.
				
	\section{\label{h-description}Hamiltonian description in intrinsic terms}
	
		Because the Dirac quantisation process requires that the quantum formulation of a system, which has a classical analogue, must be derived using the classical Hamiltonian description of that analogue as a starting point \cite{dirac}, what we need to do from here on is to find the Hamiltonian function $ H $ that classically describes the physical system we are analysing: i.e., the one defined by a three-dimensional massive ball moving in accordance with conditions (i) and (ii). And, by according to the literature, this can be done provided we remember that the energy of a conservative system can be expressed, in Lagrangian terms, as \cite{gitman,mf-constraints}
		\begin{equation}
			\mathcal{E} = \frac{\partial L}{\partial \dot{q} ^{\mu }} \ \dot{q} ^{\mu } - L \ . \label{energy-lagrangian}
		\end{equation}
		After all, since the physical system we are analysing here is conservative, it is precisely this expression for the energy that allows us to construct such a Hamiltonian function via \cite{lag-ham}
		\begin{equation}
			H = \left. \mathcal{E} \right\vert _{\dot{q} = v \left( q , p \right) } \ , \label{generic-hamilton}
		\end{equation}
		by recognising that
		\begin{equation}
			p_{\mu } := \frac{\partial L}{\partial \dot{q} ^{\mu}} \label{p-generic-definition}
		\end{equation}
		defines the momentum $ p = \left( p_{1} , p_{2} , \ldots , p_{n} \right) $ that is canonically conjugate to the variable $ q = \left( q^{1} , q^{2} , \ldots , q^{n} \right) $.
		
		\subsection{\label{intrinsic-h-description}On the classical description}
		
			Observe that, since (\ref{generic-hamilton}) was defined by restricting the expression (\ref{energy-lagrangian}) to
			\begin{equation*}
				\dot{q} = v \left( q , p \right) \ ,
			\end{equation*}
			the definition (\ref{p-generic-definition}) must be used to express the velocities $ \dot{q} = \left( \dot{q} ^{1} , \dot{q} ^{2} , \ldots , \dot{q} ^{n} \right) $ in terms of the variables $ q $ and $ p $. And since this Section begins by exploring this need, in presenting the classical Hamiltonian description of the system we are analysing, we shall start this presentation by showing how it is possible to express these velocities using the Lagrangian (\ref{intrisic-lagrangean}) that describes this system in intrinsic terms. By the way, deriving this expression is very straightforward: just note that, since using expression (\ref{p-generic-definition}) leads us to
			\begin{equation*}
				P = \frac{\partial K}{\partial \dot{x}} = m \left( \sec ^{2} \phi \right) \left( \frac{a + 5}{a + 3} \right) \dot{x} \ ,
			\end{equation*}
			algebraic manipulation of this result shows us that
			\begin{equation}
				\dot{x} = v \left( x , P \right) = \frac{P}{m \left( \sec ^{2} \phi \right) } \left( \frac{a + 3}{a + 5} \right) \ . \label{intrinsic-v}
			\end{equation}
			As a consequence, since the replacement of the latter result in (\ref{intrisic-lagrangean}) yields
			\begin{equation*}
				\left. L \left( x , \dot{x} \right) \right\vert _{\dot{x} = v \left( x , P \right) } = \frac{\left( P \right) ^{2}}{2m \left( \sec ^{2} \phi \right) } \left( \frac{a + 3}{a + 5} \right) + mg \left( \tan \phi \right) x \ ,
			\end{equation*}
			 it is immediate to conclude that the Hamiltonian function, which classically describes the system under consideration in intrinsic terms, can be defined as \cite{gitman,mf-constraints}
			\begin{equation}
				H \left( x , P \right) = \frac{\left( P \right) ^{2}}{2m \left( \sec ^{2} \phi \right) } \left( \frac{a + 3}{a + 5} \right) - mg \left( \tan \phi \right) x \ . \label{hamil-intrinsic}
			\end{equation}
			
			\subsubsection{Equations of motion and some applications of Poisson brackets}
			
				Incidentally, since the Introduction was used, amongst other things, to briefly mention the central role that Poisson brackets play in the canonical quantisation process, it is very important to note that these same Poisson brackets allow us to express the equations of motion, which are associated with the classical description of any physical observable $ F \left( q , p \right) $, as
				\begin{equation}
					\dot{F} = \frac{\partial F}{\partial t} + \left\{ F , H \right\} \ . \label{c-time-evolution}
				\end{equation}
				After all, when we realise that both $ x $ and $ P $ are, by definition, interpretable as two physical observables of the system described by (\ref{hamil-intrinsic}), applying this ``recipe'' (\ref{c-time-evolution}) leads us not only to
				\begin{equation*}
					\dot{x} = \left\{ x , H \right\} = \frac{P}{m \left( \sec ^{2} \phi \right) } \left( \frac{a + 3}{a + 5} \right) \ ,
				\end{equation*}
				which is precisely the same result as that already obtained in (\ref{intrinsic-v}), but also to
				\begin{equation}
					\dot{P} = mg \left( \tan \phi \right) \ , \label{p-point}
				\end{equation}
				which is nothing more than the expression for the force, in full accordance with what is stated in Newton’s second law.
				
				Furthermore, note that, as the differentiation of both sides of (\ref{intrinsic-v}) with respect to $ t $ gives us
				\begin{equation*}
					\ddot{x} = \frac{\dot{P}}{m \left( \sec ^{2} \phi \right) } \left( \frac{a + 3}{a + 5} \right) \ ,
				\end{equation*}
				it is precisely this result (\ref{p-point}) that leads us to an expression
				\begin{equation}
					\ddot{x} - g \left( \frac{a + 3}{a + 5} \right) \frac{\sin \left( 2 \phi \right) }{2} = 0 \label{h-intrinsic-mov-equation}
				\end{equation}
				that merely proves that the Hamiltonian function (\ref{hamil-intrinsic}) does, in fact, describe the same physics as the Lagrangian (\ref{intrisic-lagrangean}). However, it is also important to note here that the use of Poisson brackets also shows us that
				\begin{equation*}
					\left\{ x , x \right\} = \left\{ P , P \right\} = 0 \quad \textnormal{and} \quad \left\{ x , P \right\} = 1 \ .
				\end{equation*}
				And why is it important to note this? Because, as the use of these Poisson brackets shows us that these variables $ x $ and $ P $ are canonically conjugate, it is precisely these that must lead us to the self-adjoint operators $ \hat{x} = \mathcal{Q} \left( x \right) $ and $ \hat{P} = \mathcal{Q} \left( P \right) $ required by the canonical quantisation process. After all, remember that, in a scenario where the physical system is described in intrinsic terms, the Dirac brackets reduce to Poisson brackets and this is what causes the relation (\ref{correspondence-dirac}), which underpins the commutation relations (\ref{comutadores}), be reduced to
				\begin{equation*}
					\bigl[ \hat{F} , \hat{G} \bigr] = i \hbar \hspace*{0.04cm} \mathcal{Q} \left( \left\{ F , G \right\} \right) \ .
				\end{equation*}
		
		\subsection{\label{intrinsic-q-description}On the quantum description}
		
			Although the way in which we drafted Section \ref{lag-description} may have led to an expectation that the present Subsection would continue the classical Hamiltonian description of our system, now in terms of a system with constraints, we shall go against this expectation and defer this presentation to Section \ref{constraints-h-description}. And why are we doing this? Because, as we have just mentioned in the previous paragraph about the need to establish the self-adjoint operators $ \hat{x} = \mathcal{Q} \left( x \right) $ and $ \hat{P} = \mathcal{Q} \left( P \right) $, it is not difficult to conclude that, if we are actually able to define such operators, the Hamiltonian operator $ \hat{H} = \mathcal{Q} \left( H \right) $ will also be self-adjoint. After all, if, one the one hand, property \textbf{II} guarantees that we can obtain this Hamiltonian operator via
			\begin{equation}
				\hat{H} = H \bigl( \mathcal{Q} \bigl( \hat{x} \bigr) , \mathcal{Q} \bigl( \hat{P} \bigr) \bigr) = \frac{\hat {P} ^{2}}{2m \left( \sec ^{2} \phi \right) } \left( \frac{a + 3}{a + 5} \right) - mg \left( \tan \phi \right) \hat{x} \ , \label{h-operator}
			\end{equation}
			Linear Algebra, on the other hand, ensures that this sum of the two self-adjoint operators, which appear on the right-hand side of (\ref{h-operator}), will also be a self-adjoint operator \cite{hoff}. In other words, and as valid as it may be to analyse the consequences of quantising a Hamiltonian system with constraints, the Hamiltonian function (\ref{hamil-intrinsic}) that we have obtained already allows us to define a univocal quantum theory for our system, provided that we are actually able to recognise $ \hat{x} = \mathcal{Q} \left( x \right) $ and $ \hat{P} = \mathcal{Q} \left( P \right) $ as two self-adjoint operators. In these terms, and by noting that the postulates of Quantum Mechanics require that the commutation relations (\ref{comutadores}) also be obtained as a consequence of the action of the operators
			\begin{equation*}
				\bigl[ \hat{x} , \hat{x} \bigr] \ , \ \ \bigl[ \hat{P} , \hat{P} \bigr] \quad \textnormal{and} \quad \bigl[ \hat{x} , \hat{P} \bigr]
			\end{equation*}
			on the wave function that, for example, solves the Schrödinger’s wave equation
			\begin{equation}
				\hat{H} \psi \left( x , t \right) = i \hbar \hspace*{0.04cm} \frac{\partial \psi }{\partial t} \label{schr-equation}
			\end{equation}
			in the position representation \cite{mcweeny,liboff}, it is not difficult to see that such self-adjoint operators can, in fact, be defined as
			\begin{equation}
				\hat{x} = x \quad \textnormal{and} \quad \hat{P} = - i \hbar \hspace*{0.04cm} \frac{\partial }{\partial x} \ . \label{representation}
			\end{equation}
			
			\subsubsection{Some considerations regarding the Hamiltonian operator and the corresponding Schrödinger equation}
			
				Given that the coefficients, which multiply the operators $ \hat{P} ^{2} $ and $ \hat{x} $ in (\ref{h-operator}), are mere real constants, it is by no means unreasonable to rewrite this Hamiltonian operator in a more concise form as
				\begin{equation}
					\hat{H} = \frac{\hat{P} ^{2}}{2M} - \mathfrak{f} \hspace*{0.04cm} \hat{x} \label{h-enxuto}
				\end{equation}
				provided that the equalities
				\begin{equation}
					M = m \left( \sec ^{2} \phi \right) \left( \frac{a + 5}{a + 3} \right) \quad \textnormal{and} \quad \mathfrak{f} = mg \left( \tan \phi \right) \label{mf-definitions}
				\end{equation}
				hold. And from this result (\ref{h-enxuto}), it is also not difficult to see that the Schrödinger equation (\ref{schr-equation}) takes the form
				\begin{equation}
					- \left( \frac{\hbar ^{2}}{2M} \frac{\partial ^{2}}{\partial x^{2}} + \mathfrak{f} \hspace*{0.04cm} x \right) \psi \left( x , t \right) = i \hbar \hspace*{0.04cm} \frac{\partial \psi }{\partial t} \label{schr-equation2}
				\end{equation}
				as a consequence of replacing the expressions (\ref{representation}) into (\ref{h-enxuto}). However, since the left-hand side (right-hand side) of (\ref{schr-equation2}) depends only on the spatial (temporal) parameter, it is quite convenient to consider that its solution is such that \cite{strauss}
				\begin{equation}
					\psi \left( x , t \right) = \Psi \left( x \right) \Lambda \left( t \right) \ . \label{separated-psi}
				\end{equation}
				After all, when we replace this expression (\ref{separated-psi}) into equation (\ref{schr-equation2}), this allows us to obtain two equations
				\begin{subequations} \label{desmontadas}
					\begin{align}
						- \left( \frac{\hbar ^{2}}{2M} \frac{d^{2}}{dx^{2}} + \mathfrak{f} \hspace*{0.04cm} x \right) \Psi \left( x \right) & = E \hspace*{0.04cm} \Psi \left( x \right) \quad \textnormal{and} \label{desmontadas-1} \\
						i \hbar \hspace*{0.04cm} \frac{d \Lambda }{dt} & = E \hspace*{0.04cm} \Lambda \left( t \right) \label{desmontadas-2}
					\end{align}
				\end{subequations}
				which are independent of one another despite the same constant $ E $ appearing in both.
				
				In fact, in the case of equation (\ref{desmontadas-2}), it is not difficult to see that its general solution takes the form
				\begin{equation}
					\Lambda \left( t \right) = \Lambda _{0} \hspace*{0.04cm} e^{-i E \left( t - t_{0} \right) / \hbar } \ , \label{lambda-function}
				\end{equation}
				where $ \Lambda _{0} $ is the constant that determines the value of this function at the instant $ t_{0} $. And since the postulates of Quantum Mechanics require that the expression
				\begin{equation*}
					\left\vert \psi \left( x , t \right) \right\vert ^{2} = \psi ^{\ast} \left( x , t \right) \cdot \psi \left( x , t \right)
				\end{equation*}
				be interpreted as the probability density (in this case, the probability per unit volume) of finding a particle, at time $ t $, in a very specific region \cite{mcweeny,liboff}, this result (\ref{lambda-function}) makes it quite clear that we are dealing with a physical system that is in a stationary state: i.e., as (\ref{lambda-function}) allows us to recognise that
				\begin{equation}
					\left\vert \psi \left( x , t \right) \right\vert ^{2} = \left\vert \Lambda _{0} \right\vert ^{2} \left\vert \Psi \left( x \right) \right\vert ^{2} \ , \label{stationary-prob}
				\end{equation}
				it becomes quite clear that this probability density does not change as time passes, a fact that is quite consistent with the fact that the potential
				\begin{equation}
					V \left( x \right) = - \mathfrak{f} \hspace*{0.04cm} x \ , \label{potential-fx}
				\end{equation}
				to which our massive ball is subjected, also does not depend on $ t $.
				
				In the case of equation (\ref{desmontadas-1}), it is worth noting that, as a change of variables
				\begin{equation}
					u = - \left( \frac{2M}{\hbar ^{2} \mathfrak{f} \hspace*{0.04cm} ^{2}} \right) ^{1/3} \left( \mathfrak{f} \hspace*{0.04cm} x + E \right) \label{coordinate-transf}
				\end{equation}
				allows us to observe that
				\begin{equation*}
					\frac{d}{dx} = \frac{du}{dx} \frac{d}{du} = - \left( \frac{2M \mathfrak{f}}{\hbar ^{2}} \right) ^{1/3} \frac{d}{du} \ \Rightarrow \ \frac{d^{2}}{dx^{2}} = \left( \frac{2M \mathfrak{f}}{\hbar ^{2}} \right) ^{2/3} \frac{d^{2}}{du^{2}} \ ,
				\end{equation*}
				it turns out that this equation (\ref{desmontadas-1}) is equivalent to
				\begin{equation*}
					- \left[ \frac{\hbar ^{2}}{2M} \left( \frac{2M \mathfrak{f}}{\hbar ^{2}} \right) ^{2/3} \frac{d^{2}}{du^{2}} - \left( \frac{2M}{\hbar ^{2} \mathfrak{f} \hspace*{0.04cm} ^{2}} \right) ^{-1/3} u \right] \Upsilon \left( u \right) = - \left( \frac{2M}{\hbar ^{2} \mathfrak{f} \hspace*{0.04cm} ^{2}} \right) ^{-1/3} \left( \frac{d^{2}}{du^{2}} - u \right) \Upsilon \left( u \right) = 0
				\end{equation*}
				provided that
				\begin{equation}
					\Psi \left( x \right) = \left. \Upsilon \left( u \right) \right\vert _{u = u \left( x \right) } \ . \label{psiu}
				\end{equation}
				In other words, this result allows us to conclude that we are dealing with an Airy equation \cite{airy-paper}
				\begin{equation*}
					\left( \frac{d^{2}}{du^{2}} - u \right) \Upsilon \left( u \right) = 0 \ ,
				\end{equation*}
				whose general solution is
				\begin{equation}
					\Upsilon \left( u \right) = C \hspace*{0.04cm} \mathrm{Ai} \left( u \right) + D \hspace*{0.04cm} \mathrm{Bi} \left( u \right) \ , \label{general-u-solution}
				\end{equation}
				where $ C $ and $ D $ are two constants, and
				\begin{subequations} \label{airy-funct}
					\begin{align}
						\mathrm{Ai} \left( u \right) & = \frac{1}{\pi } \int ^{\infty } _{0} \cos \left( \frac{\sigma ^{3}}{3} + u \sigma \right) d \sigma \quad \textnormal{and} \label{airy-funct-1} \\
						\mathrm{Bi} \left( u \right) & = \frac{1}{\pi } \int ^{\infty } _{0} \left[ \exp \left( - \frac{\sigma ^{3}}{3} + u \sigma \right) + \sin \left( \frac{\sigma ^{3}}{3} + u \sigma \right) \right] d \sigma \ . \label{airy-funct-2}
					\end{align}
				\end{subequations}
				
		\subsection{Finding some stationary solutions}
		
\begin{figure}[t!]
				\centering
				\includegraphics[viewport=290 10 0 190,scale=1.1]{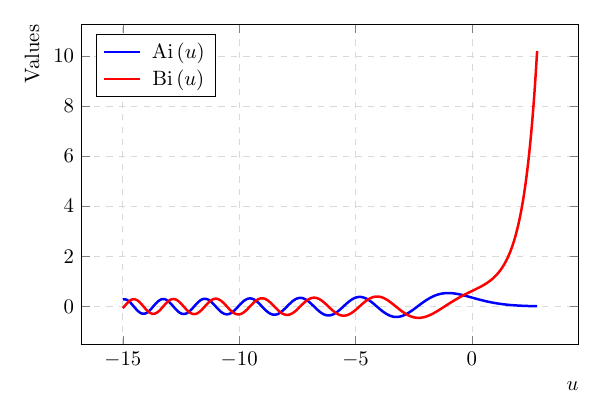}
				\caption{Graphs of the Airy functions $ \mathrm{Ai} \left( u \right) $ (in blue) and $ \mathrm{Bi} \left( u \right) $ (in red), whose definitions were given in (\ref{airy-funct-1}) and (\ref{airy-funct-2}) respectively. Note that, although these graphs may appear to emphasise the real interval $ - 15 < u \lesssim 3 $, they are presented here to illustrate that these two functions are not square-integrable on the real interval $ - \infty < u < \infty $. After all, apart from both exhibiting non-convergent oscillatory behaviour within the interval $ - \infty < u \lesssim - 1.8955 $, only the function converges asymptotically (in this case, to zero) on the interval $ u \gtrsim - 1.8955 $ \cite{nist}.}
				\label{fig:airy}
			\end{figure}

			Clearly, if we wish to continue working with the same canonically conjugate variables that led us to equation (\ref{schr-equation2}), we must take relation (\ref{psiu}) into account. After all, it is this relation that guarantees that, in order to obtain the general solution of the equation (\ref{desmontadas-1}), we need only substitute (\ref{coordinate-transf}) into the solution (\ref{general-u-solution}): i.e., it is the relation (\ref{psiu}) that guarantees that the general solution of (\ref{desmontadas-1}) is, a priori, given by
			\begin{equation}
				\Psi \left( x \right) = C \hspace*{0.04cm} \mathrm{Ai} \left( - \left( \frac{2M}{\hbar ^{2} \mathfrak{f} \hspace*{0.04cm} ^{2}} \right) ^{1/3} \left( \mathfrak{f} \hspace*{0.04cm} x + E \right) \right) + D \hspace*{0.04cm} \mathrm{Bi} \left( - \left( \frac{2M}{\hbar ^{2} \mathfrak{f} \hspace*{0.04cm} ^{2}} \right) ^{1/3} \left( \mathfrak{f} \hspace*{0.04cm} x + E \right) \right) \ . \label{general-airy-solution}
			\end{equation}
			However, since interpreting (\ref{stationary-prob}) as a probability density also requires that
			\begin{equation}
				\int ^{\infty } _{- \infty } \left\vert \psi \left( x , t \right) \right\vert ^{2} dx = \left\vert \Lambda _{0} \right\vert ^{2} \int ^{\infty } _{- \infty } \left\vert \Psi \left( x \right) \right\vert ^{2} dx = 1 \label{integral-vality}
			\end{equation}
			is a valid result, it becomes quite clear that we are faced with a new ``problem'' that needs to be resolved because, as Figure \ref{fig:airy}
			clearly illustrates, the Airy functions $ \mathrm{Ai} \left( u \right) $ and $ \mathrm{Bi} \left( u \right) $ are not square-integrable within the interval under consideration \cite{vallee}. And, from a physical point of view, this new ``problem'' was actually to be expected, since, as
			\begin{equation*}
				\lim _{x \rightarrow - \infty } V \left( x \right) \ \rightarrow \ \infty \quad \textnormal{and} \quad \lim _{x \rightarrow \infty } V \left( x \right) \ \rightarrow \ - \infty \ ,
			\end{equation*}
			this leads us to a continuous and unbounded spectrum for $ E $, which implies, for instance, the existence of unbound states with arbitrarily negative energies.
			
			\subsubsection{\label{boucing}Imposing an infinite potential barrier}
			
				One of the simplest ways to ensure that the result (\ref{integral-vality}) holds is, for example, by imposing the boundary condition
				\begin{equation}
					\Psi \left( x \right) = 0 \ , \ \ \text{when} \ \ x \geqslant 0 \ , \label{infinite-boundary}
				\end{equation}
				which can easily be imposed by introducing an infinite potential barrier at the point $ x = 0 $: i.e., by considering that the expression (\ref{potential-fx}) remains valid, but only within the interval $ - \infty < x < 0 $ because introducing such a barrier implies that
				\begin{equation}
					V \left( x \right) \rightarrow \infty \ , \label{infty-potential}
				\end{equation}
				when $ x \geqslant 0 $. Note that, from a classical point of view, introducing this infinite potential barrier is roughly equivalent to introducing a wall, at the point $ x = 0 $, to prevent this ball from continuing to move, for all eternity, precisely in the region $ x \geqslant 0 $ that leads us to such unbound states with arbitrarily negative energies.
				
\begin{figure}[!t]
					\centering
					\includegraphics[viewport=280 10 0 190,scale=1.2]{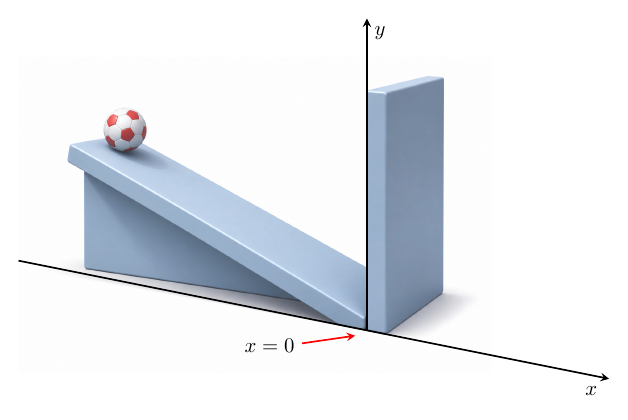}
					\caption{Illustration of how the problem we are analysing can be ``slightly'' reformulated so that we do not have to deal with states associated with arbitrarily negative energies. After all, as this illustration was designed to show us the situation of a football that, in addition to moving in accordance with points (i) and (ii) of the Introduction, is unable to reach the region where $ x \geqslant 0 $ due to the presence of a wall, this system must be regarded as the classical analogue of a quantum system whose potential is defined by: (\ref{potential-fx}), when $ x < 0 $; and (\ref{infty-potential}), otherwise.}
					\label{barreira}
				\end{figure}

				In fact, a good illustration of this new situation is provided by Figure \ref{barreira},
				since in it one can observe the presence of a wall that, although it does not appear to be very large or/and very thick, must be interpreted as having all the properties necessary to prevent the ball from passing through it. However, one question this illustration fails to answer is: why did we introduce this wall, and consequently the associated infinite potential barrier, precisely at this point $ x = 0 $ and not elsewhere? After all, when we recall that Figure \ref{fig:airy} was constructed, amongst other things, to illustrate that the Airy function $ \mathrm{Bi} \left( u \right) $ is strictly increasing when $ u \gtrsim - 1.8955 $ \cite{nist}, it becomes clear that the introduction of such a potential barrier could have been made at another point due, for instance, to relation (\ref{coordinate-transf}). But, regardless of where exactly is the best point to introduce this barrier of infinite potential, one very important observation we need to make here is that this strictly increasing behaviour of $ \mathrm{Bi} \left( u \right) $ already requires us to set $ D = 0 $ in the supposed solution (\ref{general-airy-solution}). Thus, by assuming that the infinite potential barrier has indeed been introduced in
				\begin{equation}
					x = 0 \ \Leftrightarrow \ u_{0} = u \left( 0 \right) = - \left( \frac{2M}{\hbar ^{2} \mathfrak{f} \hspace*{0.04cm} ^{2}} \right) ^{1/3} E \ , \label{negative-u0}
				\end{equation}
				it is not difficult to see that the result (\ref{integral-vality}) ultimately reduces to
				\begin{equation}
					\left\vert \Lambda _{0} \right\vert ^{2} \left\vert C \right\vert ^{2} \int ^{0} _{- \infty } \left[ \mathrm{Ai} \left( - \left( \frac{2M}{\hbar ^{2} \mathfrak{f} \hspace*{0.04cm} ^{2}} \right) ^{1/3} \left( \mathfrak{f} \hspace*{0.04cm} x + E \right) \right) \right] ^{2} dx = 1 \ . \label{prob-reduction}
				\end{equation}
				And why is it interesting to observe this? Because, given that it is precisely the relation (\ref{coordinate-transf}) that allows us to rewrite (\ref{prob-reduction}) as
				\begin{equation*}
					\left\vert \Lambda _{0} \right\vert ^{2} \left\vert C \right\vert ^{2} \left( \frac{\hbar ^{2}}{2M \mathfrak{f}} \right) ^{1/3} \int ^{\infty } _{u_{0}} \left[ \mathrm{Ai} \left( u \right) \right] ^{2} du = 1 \ ,
				\end{equation*}
				it is the result
				\begin{equation}
					\int ^{\infty } _{u_{0}} \left[ \mathrm{Ai} \left( u \right) \right] ^{2} du = - u_{0} \left[ \mathrm{Ai} \left( u_{0} \right) \right] ^{2} + \left\{ \frac{d}{du} \left[ \mathrm{Ai} \left( u \right) \right] \right\} ^{2} _{u = u_{0}} \label{integration-result}
				\end{equation}
				that allows us to recognise that
				\begin{equation}
					\left\vert C \right\vert ^{2} = - \left( \frac{E \left\vert \Lambda _{0} \right\vert ^{2}}{u_{0} \mathfrak{f}} \right) ^{-1} \left( u_{0} \left[ \mathrm{Ai} \left( u_{0} \right) \right] ^{2} - \left\{ \frac{d}{du} \left[ \mathrm{Ai} \left( u \right) \right] \right\} ^{2} _{u = u_{0}} \right) ^{-1} \ . \label{c1-condition-x0}
				\end{equation}
				In other words, the solution of (\ref{desmontadas-1}) that, a priori, best describes our physical system, provided that a potential barrier of infinite height is present at the point $ x = 0 $, is
				\begin{equation}
					\Psi \left( x \right) = C \hspace*{0.04cm} \mathrm{Ai} \left( - \left( \frac{2M}{\hbar ^{2} \mathfrak{f} \hspace*{0.04cm} ^{2}} \right) ^{1/3} \left( \mathfrak{f} \hspace*{0.04cm} x + E \right) \right) \ , \label{psi-function-infinity}
				\end{equation}
				where $ C $ is any constant satisfying condition (\ref{c1-condition-x0}).
				
\begin{figure}[t!]
					\centering
					\includegraphics[viewport=250 20 0 200,scale=1.0]{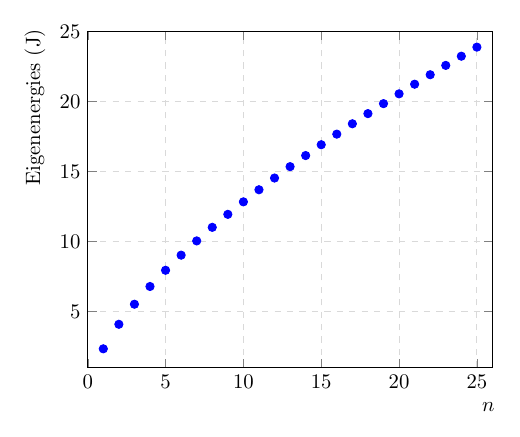}
					\caption{Graph showing the values of the eigenenergies, which characterise our quantum system when an infinite potential barrier is imposed at the point $ x = 0 $, as a function of $ n $. As this graph has been constructed for illustrative purposes only, we are assuming here that $ \left[ \hbar ^{2} / \left( 2M \mathfrak{f} \right) \right] ^{1/3} = 1 $ since, according to (\ref{mf-definitions}), the values of $ M $ may vary for different balls. Note that, as the values of these eigenenergies follow from expression (\ref{eigenenergies-p-infinity}), the case illustrated by this graph corresponds to that in which these eigenenergies are identified with the modulus of the roots of $ \mathrm{Ai} \left( x \right) $.}
					\label{fig:barreira-3}
				\end{figure}

				Observe that, given everything we have just presented, it seems that we have avoided, perhaps deliberately, explaining why we chose to place this barrier of infinite potential at $ x = 0 $ and not at some other point, right? And, now, due to this ``a priori'' assumption we mentioned in the last paragraph, it seems as though we have increased even further the number of justifications we still need to provide you, the reader, in order to lend greater consistency to everything we are saying here. And, in order for us to begin presenting all these justifications that seem to be missing, it is very important that you, the reader, note that dealing with a situation where, for example, $ \Psi \left( 0 \right) = 0 $ means dealing with a situation where
				\begin{itemize}
					\item the argument of the Airy function appearing in (\ref{psi-function-infinity}), when evaluated at the point $ x = 0 $, is equal to the same value $ u_{0} $ that has already been presented in (\ref{negative-u0}) and
					\item the value taken, at $ x = 0 $, by this same Airy function in (\ref{psi-function-infinity}), must also be equal to zero due to the non-zero nature of $ C $.
				\end{itemize}
				And why is it so important that you note this? Because it is precisely when we combine the information contained in these two items that it becomes clear that
				\begin{equation*}
					u_{0} = - \left( \frac{2M}{\hbar ^{2} \mathfrak{f} \hspace*{0.04cm} ^{2}} \right) ^{1/3}
				\end{equation*}
				must be recognised as one of the roots of the Airy function appearing in (\ref{psi-function-infinity}). Thus, since all the eigenvalues of this boundary value problem (\ref{infinite-boundary}) must be proportional to the countable roots ($ a_{n} $) of the Airy function given in (\ref{psi-function-infinity}) \cite{vallee}, this allows us to conclude that such eigenvalues (in this case, the eigenenergies of our physical system) are given by
				\begin{equation}
					E_{n} = - \left( \frac{\hbar ^{2} \mathfrak{f} \hspace*{0.04cm} ^{2}}{2M} \right) ^{1/3} a_{n} \ . \label{eigenenergies-p-infinity}
				\end{equation}
				That is, as $ n = 1 , 2 , \ldots , \infty $, it is the countability of the roots of this Airy function that ensures the discretisation of the energy levels in this case. Indeed, a good illustration of this result can be seen, for instance, in Figure \ref{fig:barreira-3}.
				
\begin{figure}[!t]
					\centering
					\includegraphics[viewport=270 0 -65 190,scale=0.8]{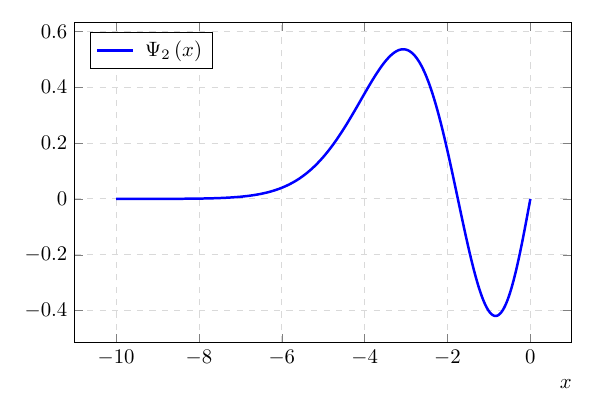}
					\includegraphics[viewport=235 0 15 198,scale=0.8]{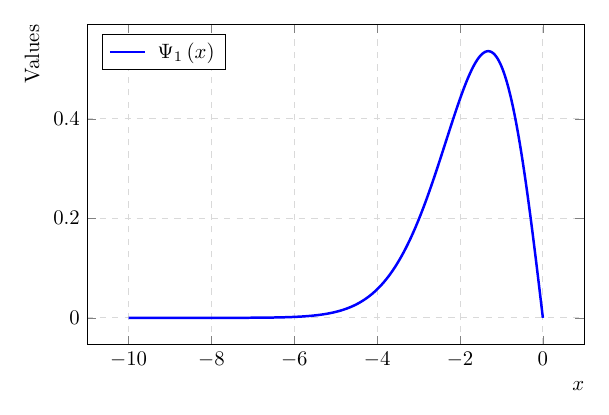} \\
					\includegraphics[viewport=285 0 -45 198,scale=0.8]{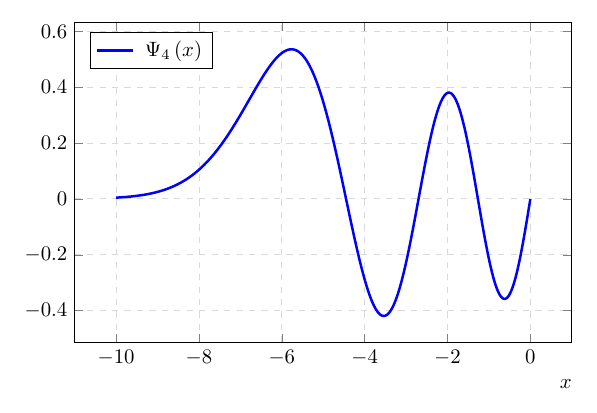}
					\includegraphics[viewport=258 0 8 190,scale=0.8]{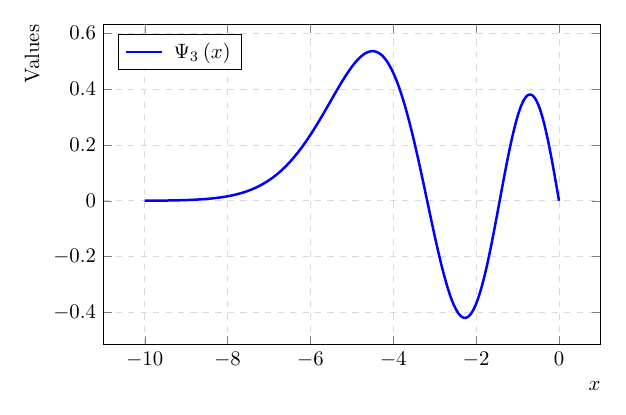}
					\caption{Graphs illustrating the general behaviour of the first four eigenstates of the time-independent Schrödinger equation (\ref{desmontadas-1}) describing the situation under analysis, where an infinite potential barrier has been inserted at the point $ x = 0 $. As these graphs are presented for illustrative purposes only, they have been constructed on the assumption that $ C_{n} = 1 $ in the first four expressions (i.e., for $ n = 1 , 2 , 3 , 4 $) given in (\ref{n-eigenstates}), since it is these expressions that define each of these four eigenstates $ \Psi _{n} \left( x \right) $. Note that, if we also assume that $ \left[ \hbar ^{2} / \left( 2M \mathfrak{f} \right) \right] ^{1/3} = 1 $, the eigenenergy of the $ n $-th eigenstate plotted here can be identified in the graph shown in Figure \ref{fig:barreira-3}. After all, and in accordance with expression (\ref{eigenenergies-p-infinity}), the eigenenergies of these four eigenstates are the four lowest eigenenergies shown in Figure \ref{fig:barreira-3}.}
					\label{fig:barreira-2}
				\end{figure}

\begin{figure}[!t]
					\centering
					\includegraphics[viewport=270 0 -65 190,scale=0.8]{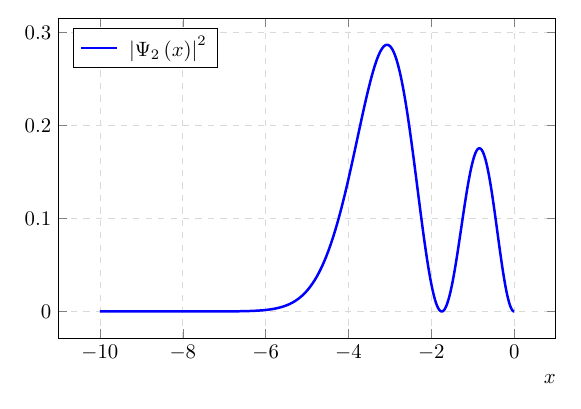}
					\includegraphics[viewport=235 0 15 198,scale=0.8]{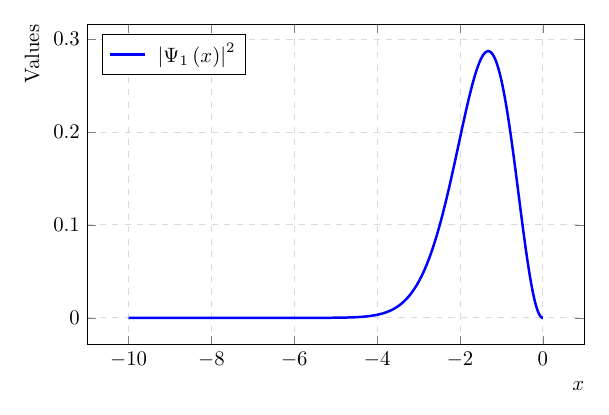} \\
					\includegraphics[viewport=285 0 -45 198,scale=0.8]{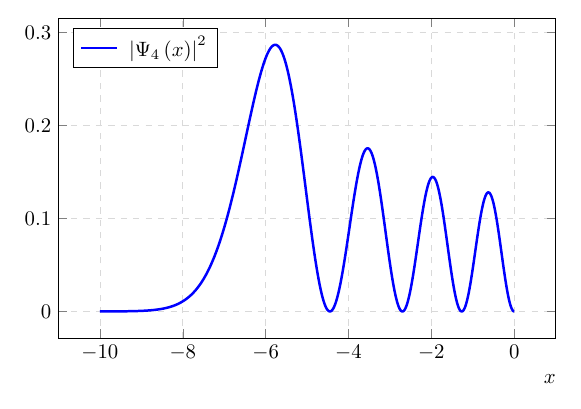}
					\includegraphics[viewport=258 0 8 190,scale=0.8]{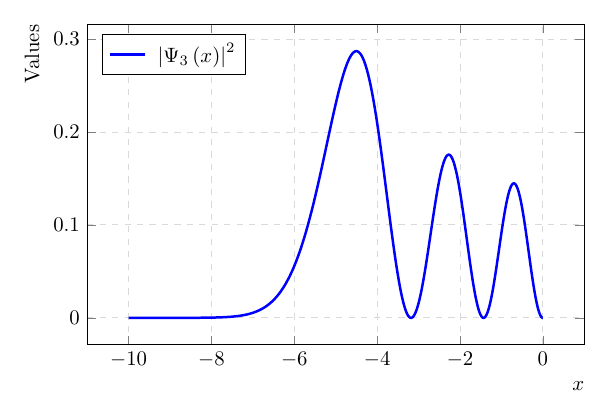}
					\caption{General behaviour of the four probability densities that can be obtained using (\ref{particular-density}), and associated with the four eigenstates shown in Figure \ref{squared-barreira}. Although these four new graphs were constructed, for example, without the slightest concern for ensuring that each of them satisfied condition (\ref{integral-vality}), they all make it quite clear that their maximum points correspond to the points where it is most likely to find the ball rolling down the inclined plane.}
					\label{squared-barreira}
				\end{figure}

				Another very important point we must also note here is that, as the result (\ref{eigenenergies-p-infinity}) makes it quite clear that $ E $ takes on infinite values, it is precisely the presence of this $ E $ in the expression (\ref{psi-function-infinity}) that allows us to explain why we that, ``a priori'', the function (\ref{psi-function-infinity}) was the best solution to (\ref{desmontadas-1}). After all, as the replacement of these infinite eigenvalues $ E_{n} $ in (\ref{psi-function-infinity}) leads us to the infinite functions
				\begin{equation}
					\Psi _{n} \left( x \right) = C_{n} \hspace*{0.04cm} \mathrm{Ai} \left( - \left( \frac{2M}{\hbar ^{2} \mathfrak{f} \hspace*{0.04cm} ^{2}} \right) ^{1/3} \left( \mathfrak{f} \hspace*{0.04cm} x + E_{n} \right) \right) \ , \label{n-eigenstates}
				\end{equation}
				all these infinite functions are solutions of (\ref{desmontadas-1}). In other words, what this replacement tells us is that the function $ \Psi _{n} \left( x \right) $ is the $ n $-th stationary eigenstate of the equation (\ref{desmontadas-1}) whose associated eigenenergy is $ E_{n} $. In fact, a good illustration, showing the first four stationary states (\ref{n-eigenstates}), is given in Figure \ref{fig:barreira-2}.
				And, by assuming that all these infinite eigenstates are normalised (i.e. that all of them satisfy a relation analogous to (\ref{prob-reduction})) given the requirement that each of them defines
				\begin{equation}
					\left\vert \Psi _{n} \left( x \right) \right\vert ^{2} = \Psi ^{\ast } _{n} \left( x \right) \cdot \Psi _{n} \left( x \right) \label{particular-density}
				\end{equation}
				as the probability density of finding this ball, in some very specific region, in that state $ \Psi _{n} \left( x \right) $ which has energy $ E_{n} $, it is not difficult to conclude that all the constants $ C_{n} $, which appear in (\ref{n-eigenstates}), must be such that
				\begin{equation}
					\left\vert C_{n} \right\vert ^{2} = - \left( \frac{E_{n} \left\vert \Lambda _{0} \right\vert ^{2}}{a_{n} \mathfrak{f}} \right) ^{-1} \left( - \left\{ \frac{d}{du} \left[ \mathrm{Ai} \left( u \right) \right] \right\} ^{2} _{u = a_{n}} \right) ^{-1} \ . \label{cn-condition-x0}
				\end{equation}
				A good illustration, showing the general behaviour of the probability density associated with the same four eigenstates plotted in \ref{fig:barreira-2}, can be found in Figure \ref{squared-barreira}.
				
			\subsubsection{\label{symm-pot}Creating symmetrical potential}
			
				Another way of dealing with this same inclined plane, by again ``slightly'' modifying the situation so that we can continue to consider stationary solutions, is to make it symmetrical. And this can be done, for instance, by reflecting this inclined plane about the point $ x = 0 $, as clearly illustrated in Figure \ref{cunha}:
				\begin{figure}[!t]
					\centering
					\includegraphics[viewport=280 40 0 125,scale=1.4]{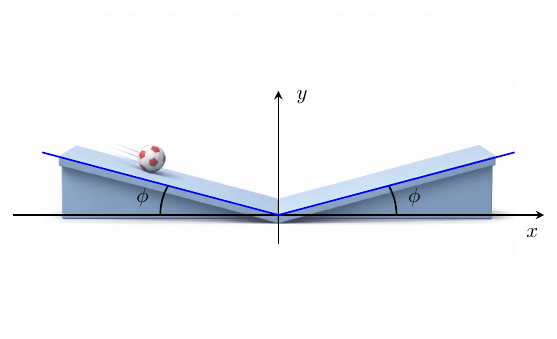}
					\caption{Another illustration of how the problem we are analysing here can, once again, be reformulated so that we do not have to deal with bound states that possess arbitrarily negative energies. Note that, unlike in the illustration in Figure \ref{barreira}, the football now reaches the region where $ x \geqslant 0 $, since the original inclined plane has simply been ``mirrored'' with respect to the plane $ x = z = 0 $. Although we have not mentioned the $ z $-axis at all in this manuscript, it should be understood as the axis that is perpendicular to the $ xy $-plane and intersects this plane at its origin.}
					\label{cunha}
				\end{figure}
				i.e., this can be done by considering that the massive ball moves on a plane defined by
				\begin{equation*}
					y \left( x \right) = \begin{cases}
						\ - \left( \tan \phi \right) x \ , \ \ \text{when} \ \ x < 0 \ , \ \ \textnormal{and} \\
						\hspace*{0.54cm} \left( \tan \phi \right) x \ , \ \ \text{otherwise} \ .
					\end{cases}
				\end{equation*}
				Of course, if we do this, we will be dealing with a situation that, in practice, requires us to replace the original Hamiltonian function (\ref{hamil-intrinsic}) with another
				\begin{equation}
					H \left( x , P \right) = \frac{P^{2}}{2m \left( \sec ^{2} \phi \right) } \left( \frac{a + 3}{a + 5} \right) + V \left( x \right) \ , \label{hamil-intrinsic-mirror}
				\end{equation}
				whose potential is given by
				\begin{equation}
					V \left( x \right) = mg \left( \tan \phi \right) \left\vert x \right\vert \ . \label{v-cunha}
				\end{equation}
				However, although it may seem that we are complicating the formulation of our problem somewhat, one thing that is already quite clear, from this new Hamiltonian function (\ref{hamil-intrinsic-mirror}), is that, since
				\begin{equation*}
					V \left( 0 \right) = 0 \quad \textnormal{and} \quad \lim _{x \rightarrow \pm \infty } V \left( x \right) \ \rightarrow \ \infty \ ,
				\end{equation*}
				it leads us to a situation where the energy spectrum is quantised.
				
				Another rather notable point here is that, although this inverted wedge potential (\ref{v-cunha}) differs from the previous one (\ref{potential}), the Schrödinger equation corresponding to this new formulation is given by
				\begin{equation}
					- \left( \frac{\hbar ^{2}}{2M} \frac{\partial ^{2}}{\partial x^{2}} - \mathfrak{f} \hspace*{0.04cm} \left\vert x \right\vert \right) \psi \left( x , t \right) = i \hbar \hspace*{0.04cm} \frac{\partial \psi }{\partial t} \ , \label{new-schrodinger}
				\end{equation}
				where $ M $ and $ \mathfrak{f} $ remain defined by the same expressions in (\ref{mf-definitions}). In other words, once again we are faced with a scenario where it is possible to find the solution of its Schrödinger equation by employing the method of separation of variables. And since the only change we have made to the original problem concerns the expression of a potential that, like the previous one, depends solely on the position variable, it is not difficult to conclude that the solution of this new Schrödinger equation (\ref{new-schrodinger}) is given by
				\begin{equation}
					\psi \left( x , t \right) = \Psi \left( x \right) e^{-i E \left( t - t_{0} \right) / \hbar } \ , \label{separated-psi-cunha}
				\end{equation}
				provided that $ \Psi \left( x \right) $ is a solution of the time-independent Schrödinger equation
				\begin{equation}
					- \left( \frac{\hbar ^{2}}{2M} \frac{d^{2}}{dx^{2}} - \mathfrak{f} \hspace*{0.04cm} \left\vert x \right\vert \right) \Psi \left( x \right) = E \Psi \left( x \right) \ . \label{schrodinger-module}
				\end{equation}
				
				Since we have just written down the expression for this time-independent Schrödinger equation, it is quite pertinent to note that, if we wish to explore a scenario in which this massive ball can move across the entire extent of this infinite plane (which, now, takes the form of an inverted wedge), it is not incorrect to assume that the general solution to this new Schrödinger equation (\ref{new-schrodinger}) is, a priori, given by the continuous function
				\begin{equation}
					\Psi \left( x \right) = \begin{cases}
						\ \Psi _{-} \left( x \right) \ , \ \ \text{when} \ \ x < 0 \ , \ \ \textnormal{and} \\
						\ \Psi _{+} \left( x \right) \ , \ \ \text{otherwise} \ ,
					\end{cases} \label{solution-pm}
				\end{equation}
				where
				\begin{subequations}
					\begin{align*}
						\Psi _{-} \left( x \right) & = C^{\prime } \hspace*{0.04cm} \mathrm{Ai} \left( - \left( \frac{2M}{\hbar ^{2} \mathfrak{f} \hspace*{0.04cm} ^{2}} \right) ^{1/3} \left( \mathfrak{f} \hspace*{0.04cm} x + E \right) \right) \ \ \textnormal{and} \\
						\Psi _{+} \left( x \right) & = D^{\prime } \hspace*{0.04cm} \mathrm{Ai} \left( \left( \frac{2M}{\hbar ^{2} \mathfrak{f} \hspace*{0.04cm} ^{2}} \right) ^{1/3} \left( \mathfrak{f} \hspace*{0.04cm} x - E \right) \right) \ . 
					\end{align*}
				\end{subequations}
				Note that, because
				\begin{equation*}
					\left. \mathrm{Ai} \left( - \left( \frac{2M}{\hbar ^{2} \mathfrak{f} \hspace*{0.04cm} ^{2}} \right) ^{1/3} \left( \mathfrak{f} \hspace*{0.04cm} x + E \right) \right) \right\vert _{x=0} = \left. \mathrm{Ai} \left( \left( \frac{2M}{\hbar ^{2} \mathfrak{f} \hspace*{0.04cm} ^{2}} \right) ^{1/3} \left( \mathfrak{f} \hspace*{0.04cm} x - E \right) \right) \right\vert _{x=0} = \mathrm{Ai} \left( u_{0} \right) \ ,
				\end{equation*}
				where $ u_{0} $ takes the same value as stated in (\ref{negative-u0}), the continuity of $ \Psi \left( x \right) $ is guaranteed provided that the constants $ C^{\prime } $ and $ D^{\prime } $ are, for example, such that $ \left\vert C^{\prime } \right\vert = \pm \left\vert D^{\prime } \right\vert $. And since this observation, when combined with the need to interpret $ \left\vert \Psi \left( x \right) \right\vert ^{2} $ as a probability density, also leads us to
				\begin{equation}
					\left\vert C^{\prime } \right\vert ^{2} \left[ \int ^{0} _{- \infty } \left\vert \Psi _{-} \left( x \right) \right\vert ^{2} dx + \int ^{\infty } _{0} \left\vert \Psi _{+} \left( x \right) \right\vert ^{2} dx \right] = 1 \ , \label{condition-cunha-pm}
				\end{equation}
				it is not difficult to see, with the help of the change of variables
				\begin{equation*}
					u = \begin{cases}
						\ - \left[ 2M / \left( \hbar ^{2} \mathfrak{f} \hspace*{0.04cm} ^{2} \right) \right] ^{1/3} \left( \mathfrak{f} \hspace*{0.04cm} x + E \right) \ , \ \ \text{when} \ \ x < 0 \ , \ \ \textnormal{and} \\
						\hspace*{0.54cm} \left[ 2M / \left( \hbar ^{2} \mathfrak{f} \hspace*{0.04cm} ^{2} \right) \right] ^{1/3} \left( \mathfrak{f} \hspace*{0.04cm} x - E \right) \ , \ \ \text{otherwise} \ ,
					\end{cases}
				\end{equation*}
				that this condition (\ref{condition-cunha-pm}) is equivalent to
				\begin{equation*}
					\left\vert C^{\prime } \right\vert ^{2} \left( \frac{\hbar ^{2}}{2M \mathfrak{f}} \right) ^{1/3} \int ^{\infty } _{u_{0}} \left[ \mathrm{Ai} \left( u \right) \right] ^{2} du = \frac{1}{2} \ .
				\end{equation*}
				In other words, once again we are faced with a situation where result (\ref{integration-result}) can be applied, which leads us to
				\begin{equation}
					\left\vert C^{\prime } \right\vert ^{2} = - \left( \frac{2E}{u_{0} \mathfrak{f}} \right) ^{-1} \left( - \left\{ \frac{d}{du} \left[ \mathrm{Ai} \left( u \right) \right] \right\} ^{2} _{u = u_{0}} \right) ^{-1} \ .\label{d1-condition-x0}
				\end{equation}
				
				Based on what we have just outlined in the previous paragraph, it is clear that the general solution of equation (\ref{schrodinger-module}) is given by the function (\ref{solution-pm}), where $ C^{\prime } $ is any constant that satisfies condition (\ref{d1-condition-x0}). However, as we have also stated that the constants $ C^{\prime } $ and $ D^{\prime } $ must be such that $ \left\vert C^{\prime } \right\vert = \pm \left\vert D^{\prime } \right\vert $, it is also impossible not to recognise that this condition implies that these solutions of (\ref{schrodinger-module}) can be even or odd. And, in the case of odd solutions, such as the condition
				\begin{equation*}
					\Psi _{-} \left( x \right) = - \Psi _{+} \left( x \right)
				\end{equation*}
				requires that they be zero at $ x = 0 $, this ultimately leads us, once again, to a situation in which the eigenenergies take the values
				\begin{equation}
					E_{2n-1} = - \left( \frac{\hbar ^{2} \mathfrak{f} \hspace*{0.04cm} ^{2}}{2M} \right) ^{1/3} a_{n} \label{eigenenergies-impar}
				\end{equation}
				identical to those already given in (\ref{eigenenergies-p-infinity}). In other words, once again, we are faced with a result that shows that the time-independent Schrödinger equation (now given by (\ref{schrodinger-module})) has infinitely many solutions
				\begin{equation}
					\Psi _{2n-1} \left( x \right) = C^{\prime } _{2n} \hspace*{0.04cm} \left[ \mathrm{Ai} \left( - \left( \frac{2M}{\hbar ^{2} \mathfrak{f} \hspace*{0.04cm} ^{2}} \right) ^{1/3} \left( \mathfrak{f} \hspace*{0.04cm} x + E_{2n-1} \right) \right) - \mathrm{Ai} \left( - \left( \frac{2M}{\hbar ^{2} \mathfrak{f} \hspace*{0.04cm} ^{2}} \right) ^{1/3} \left( \mathfrak{f} \hspace*{0.04cm} x - E_{2n-1} \right) \right) \right] \ , \label{n-eigenstates-impar}
				\end{equation}
				which can be interpreted as the stationary eigenstates whose associated eigenenergies are given by (\ref{eigenenergies-p-infinity}). By the way, note once again that, given the requirement that each of these eigenstates must also satisfy
				\begin{equation}
					\int ^{\infty } _{- \infty } \left\vert \Psi _{2n-1} \left( x \right) \right\vert ^{2} dx = 1 \ , \label{n-condition-prob}
				\end{equation}
				all these constants $ C^{\prime } _{2n-1} $ in (\ref{n-eigenstates-impar}) must be such that
				\begin{equation}
					\left\vert C^{\prime } _{2n-1} \right\vert ^{2} = - \left( \frac{E_{2n-1} \left\vert \Lambda _{0} \right\vert ^{2}}{a_{n} \mathfrak{f}} \right) ^{-1} \left( - \left\{ \frac{d}{du} \left[ \mathrm{Ai} \left( u \right) \right] \right\} ^{2} _{u = a_{n}} \right) ^{-1} \ . \label{cn-prime-condition-x0}
				\end{equation}
				
				In the case of the even solutions to this same equation (\ref{schrodinger-module}), although the condition
				\begin{equation*}
					\Psi _{-} \left( x \right) = \Psi _{+} \left( x \right)
				\end{equation*}
				does not require them to vanish at $ x = 0 $, this same condition ultimately requires not only that the derivatives of these solutions be continuous at $ x = 0 $, but also that these same derivatives vanish at that point. Thus, since
				\begin{itemize}
					\item the argument of the derivatives of the Airy functions appearing in (\ref{solution-pm}), when evaluated at the point $ x = 0 $, has the same value $ u_{0} $ as that already presented in (\ref{negative-u0}) and
					\item the values taken, at $ x = 0 $, by the derivatives of these same Airy functions in (\ref{solution-pm}), must also be equal to zero due to the non-zero nature of $ C^{\prime } $,
				\end{itemize}
				it becomes clear that
				\begin{equation*}
					u_{0} = - \left( \frac{2M}{\hbar ^{2} \mathfrak{f} \hspace*{0.04cm} ^{2}} \right) ^{1/3}
				\end{equation*}
				must now be recognised as one of the roots of the derivatives of those Airy functions appearing in (\ref{solution-pm}) \cite{vallee} and that, therefore, the eigenenergies of our physical system, in this case where $ \Psi \left( x \right) $ is an even function, are given by
				\begin{equation}
					E_{2n} = - \left( \frac{\hbar ^{2} \mathfrak{f} \hspace*{0.04cm} ^{2}}{2M} \right) ^{1/3} a^{\prime} _{n} \ , \label{eigenenergies-par}
				\end{equation}
				where $ a^{\prime } _{n} $ denotes all the countable roots of the derivatives of these Airy functions. Therefore, once again we are dealing with eigenenergies that show us that all the eigenstates, which are associated with them via equation (\ref{schrodinger-module}), are expressed by
				\begin{equation}
					\Psi _{2n} \left( x \right) = C^{\prime } _{2n} \hspace*{0.04cm} \left[ \mathrm{Ai} \left( - \left( \frac{2M}{\hbar ^{2} \mathfrak{f} \hspace*{0.04cm} ^{2}} \right) ^{1/3} \left( \mathfrak{f} \hspace*{0.04cm} x + E_{2n} \right) \right) + \mathrm{Ai} \left( - \left( \frac{2M}{\hbar ^{2} \mathfrak{f} \hspace*{0.04cm} ^{2}} \right) ^{1/3} \left( \mathfrak{f} \hspace*{0.04cm} x - E_{2n} \right) \right) \right] \ , \label{n-eigenstates-par}
				\end{equation}
				where, due to the requirement that each of them must also satisfy (\ref{n-condition-prob}), the constants $ C^{\prime } _{2n} $ assume values
				\begin{equation*}
					\left\vert C^{\prime } _{2n} \right\vert ^{2} = - \left( \frac{E_{2n} \left\vert \Lambda _{0} \right\vert ^{2}}{a_{n} \mathfrak{f}} \right) ^{-1} \left( - \left\{ \frac{d}{du} \left[ \mathrm{Ai} \left( u \right) \right] \right\} ^{2} _{u = a_{n}} \right) ^{-1}
				\end{equation*}
				analogous to those previously stated.
				
	\section{\label{constraints-h-description}Classical Hamiltonian description in terms of constraints}
	
		Even though we may continue to present new stationary examples, and even though we have already put forward some sound arguments, at the start of Subsection \ref{intrinsic-q-description}, which justify why the description in intrinsic terms is already sufficient to construct a quantum description of our three-dimensional massive ball, perhaps a reader who prefers a slightly more complete and consistent text might already have been missing the presentation that, for example, justifies the name of the present Section. After all, given that we use Section \ref{lag-description} to present the classical Lagrangian formulation of the system under analysis in both intrinsic and non-intrinsic terms, why not do the same from the Hamiltonian point of view? By the way, what difference does it make if we quantise the Hamiltonian formulation of this system in the presence of constraints?
		
		And so that we may begin to present this to you, the reader, to show what happens differently in the presence of constraints, it is worth turning our attention to the Lagrangian function (\ref{constrained-lagrangean-holo}). Why? Because, amongst the many reasons we might give to justify the use of (\ref{constrained-lagrangean-holo}), it is not difficult to see that it allows us to define the components of the momentum $ p = \left( P_{x} , P_{y} , P_{\theta } , \Pi _{\left( 1 \right) } , \Pi _{\left( 2 \right) } \right) $, which is conjugate to the variable $ q = \left( x , y , \theta , \chi ^{\left( 1 \right) } , \chi ^{\left( 2 \right) } \right) $, as
		\begin{subequations} \label{pxytheta-pi34}
			\begin{align}
				& P_{x} = \frac{\partial L_{\mathrm{hol}}}{\partial \dot{x}} = m \dot{x} \ , \ \ P_{y} = \frac{\partial L_{\mathrm{hol}}}{\partial \dot{y}} = m \dot{y} \ , \ \ P_{\theta } = \frac{\partial L_{\mathrm{hol}}}{\partial \dot{\theta }} = \frac{2mR^{2}}{a + 3} \ \dot{\theta } \ , \label{pxytheta} \\
				& \Pi _{\left( 1 \right) } = \frac{\partial L_{\mathrm{hol}}}{\partial \dot{\chi } ^{\left( 1 \right) }} = 0 \quad \textnormal{and} \quad \Pi _{\left( 2 \right) } = \frac{\partial L_{\mathrm{hol}}}{\partial \dot{\chi } ^{\left( 2 \right) }} = 0 \ . \label{pi13}
			\end{align}
		\end{subequations}
		Moreover, note that, given the expressions in (\ref{pxytheta}), it is also not at all difficult to recognise, for instance, that
		\begin{equation*}
			\dot{x} = v_{x} \left( q , p \right) = \frac{P_{x}}{m} \ , \ \ \dot{y} = v_{y} \left( q , p \right) = \frac{P_{y}}{m} \quad \textnormal{and} \quad \dot{\theta } = v_{\theta } \left( q , p \right) = \frac{a + 3}{2mR^{2}} \ P_{\theta } \ .
		\end{equation*}
		However, as the expressions in (\ref{pi13}) show us that the momenta $ \Pi _{\left( 1 \right) } $ and $ \Pi _{\left( 2 \right) } $, which are conjugate to $ \chi ^{\left( 1 \right) } $ and $ \chi ^{\left( 2 \right) } $ respectively, are identically zero and, therefore, these momenta cannot be used to express the velocities $ \dot{\chi } ^{\left( 1 \right) } $ and $ \dot{\chi } ^{\left( 2 \right) } $ in terms of the variables $ q $ and $ p $ \cite{lima}, it becomes clear that the expressions in (\ref{pi13}) are, in fact, pointing to the existence of two (primary) constraints. Videlicet, to the existence of two constraints which, under the identification of two functions\footnote{As strange as this choice we are making (to enumerate the functions (\ref{Phi-function}) and, consequently, the constraints (\ref{Phi34}) to which they lead) may seem, later on we will explain why we are doing this.}
		\begin{equation}
			\Phi _{\left( 3 \right) } \left( \chi ^{\left( 1 \right) } , \Pi _{\left( 1 \right) } \right) = \Pi _{\left( 1 \right) } \quad \textnormal{and} \quad \Phi _{\left( 4 \right) } \left( \chi ^{\left( 2 \right) } , \Pi _{\left( 2 \right) } \right) = \Pi _{\left( 2 \right) } \ , \label{Phi-function}
		\end{equation}
		can be defined as
		\begin{equation}
			\Phi _{\left( 3 \right) } \left( \chi ^{\left( 1 \right) } , \Pi _{\left( 1 \right) } \right) \approx 0 \quad \textnormal{and} \quad \Phi _{\left( 4 \right) } \left( \chi ^{\left( 2 \right) } , \Pi _{\left( 2 \right) } \right) \approx  0  \ . \label{Phi34}
		\end{equation}
		
		Note that, if these constraints (\ref{Phi34}) did not exist, we would be dealing with a situation where relations
		\begin{equation*}
			\dot{\chi } ^{\left( 1 \right) } = \upsilon ^{\left( 1 \right) } \left( q , p \right) \quad \textnormal{and} \quad \dot{\chi } ^{\left( 2 \right) } = \upsilon ^{\left( 2 \right) } \left( q , p \right)
		\end{equation*}
		would hold: i.e., we would be dealing with a situation that would allow us to express the velocities $ \dot{\chi } ^{\left( 1 \right) } $ and $ \dot{\chi } ^{\left( 2 \right) } $ as functions of $ q $ and $ p $. However, as this is not the case, the identification of these new constraints (\ref{Phi34}) makes it clear that they must be incorporated, in some way, into the formulation of our problem. And since our aim, in calculating the momentum (\ref{pxytheta-pi34}), is to obtain a Hamiltonian function capable of describing the same physics as (\ref{constrained-lagrangean-holo}), the best way to express this Hamiltonian is by implementing these constraints via new Lagrange multipliers. That is, given that all the results we have obtained since the beginning of this Section allow us to observe that
		\begin{equation*}
			\left. L_{\mathrm{hol}} \left( q , \dot{q} \right) \right\vert _{\dot{x} = v \left( q , p \right) } = \frac{P_{x} ^{2}}{2m} + \frac{P_{y} ^{2}}{2m} + \frac{P_{\theta } ^{2}}{2mR^{2}} - mgy - \chi ^{\left( 1 \right) } f_{\left( 1 \right) } \left( x , y \right) - \chi ^{\left( 2 \right) } F_{\left( 2 \right) } \bigl( x , \theta \bigr) \ ,
		\end{equation*}
		it is not difficult to conclude that the Hamiltonian function, which can be obtained using this result, can be expressed as
		\begin{equation}
			H_{\mathrm{nd}} \left( q , p \right) = H_{\mathrm{st}} \left( q , p \right) + \chi ^{\left( 3 \right) } \Phi _{\left( 3 \right) } \left( \chi ^{\left( 1 \right) } , \Pi _{\left( 1 \right) } \right) + \chi ^{\left( 4 \right) } \Phi _{\left( 4 \right) } \left( \chi ^{\left( 2 \right) } , \Pi _{\left( 2 \right) } \right) \ , \label{h-st}
		\end{equation}
		where
		\begin{eqnarray*}
			H_{\mathrm{st}} \left( q , p \right) & = & \left. \frac{\partial L}{\partial \dot{q} ^{\mu }} \ \dot{q} ^{\mu } - L \ \right\vert _{\dot{x} = v \left( q , p \right) } \\
			& = & \frac{P_{x} ^{2}}{2m} + \frac{P_{y} ^{2}}{2m} + \frac{\left( a + 3 \right) P_{\theta } ^{2}}{4mR^{2}} + mgy + \chi ^{\left( 1 \right) } f_{\left( 1 \right) } \left( x , y \right) + \chi ^{\left( 2 \right) } F_{\left( 2 \right) } \bigl( x , \theta \bigr) \ .
		\end{eqnarray*}
		Here, $ \chi ^{\left( 3 \right) } $ and $ \chi ^{\left( 4 \right) } $ should be interpreted as the new Lagrange multipliers that, as we have already mentioned above, are required to incorporate these new constraints (\ref{Phi34}) into the Hamiltonian (\ref{h-st}).
		
		\subsection{A first application of Poisson brackets}
		
			If we draw on everything that has already been presented in Subsection \ref{intrinsic-h-description} to proceed with what still needs to be presented here, it becomes quite natural to use the results we have obtained so far to, for example, derive the equations of motion. And, in accordance with what we have already stated in Subsection \ref{intrinsic-h-description}, this can be done using the expression (\ref{c-time-evolution}). However, before we begin this search for the equations of motion that follow from this Hamiltonian formulation with constraints, we must also note something very important. After all, since the momenta $ \Pi _{\left( 1 \right) } $ and $ \Pi _{\left( 2 \right) } $ are identically zero at all times, the constraints (\ref{Phi34}) they define must also be. In plain English, because the two expressions contained in (\ref{Phi34}) hold at all times, the two expressions
			\begin{equation*}
				\dot{\Phi } _{\left( 3 \right) } \left( \chi ^{\left( 1 \right) } , \Pi _{\left( 1 \right) } \right) \approx 0 \quad \textnormal{and} \quad \dot{\Phi } _{\left( 4 \right) } \left( \chi ^{\left( 2 \right) } , \Pi _{\left( 2 \right) } \right) \approx  0
			\end{equation*}
			must also be. In this context, since these two relations contained in (\ref{Phi34}) can also be interpreted as two physical observables of this Hamiltonian system, it is crucial that we use (\ref{c-time-evolution}) to assess what follows from\footnote{Note that, since the functions contained in (\ref{Phi-function}) do not depend explicitly on $ t $, their partial derivatives with respect to this parameter are identically zero. In other words, although the right-hand side of (\ref{c-time-evolution}) contains a partial derivative with respect to $ t $, it is the lack of explicit dependence of the functions in (\ref{Phi-function}) on that same parameter that justifies the absence of these partial derivatives on the right-hand side of (\ref{Phi34-dot}).}
			\begin{subequations} \label{Phi34-dot}
				\begin{align}
					\dot{\Phi } _{\left( 3 \right) } & = \left\{ \Phi _{\left( 3 \right) } , H_{\mathrm{nd}} \right\} \approx 0 \label{Phi3-dot} \\
					\dot{\Phi } _{\left( 4 \right) } & = \left\{ \Phi _{\left( 4 \right) } , H_{\mathrm{nd}} \right\} \approx 0 \label{Phi4-dot}
				\end{align}
			\end{subequations}
			as a first step, and only then derive the equations of motion\footnote{Further on, we shall present a much stronger argument to explain why it is better to evaluate (\ref{Phi34-dot}) first.}. And the result that follows from this first evaluation is rather ``curious'' because, as
			\begin{equation*}
				\left\{ \Phi _{\left( 3 \right) } , H_{\mathrm{nd}} \right\} = - f_{\left( 1 \right) } \left( x , y \right) \quad \textnormal{and} \quad \left\{ \Phi _{\left( 4 \right) } , H_{\mathrm{nd}} \right\} = - F_{\left( 2 \right) } \left( x , \theta \right) \ ,
			\end{equation*}
			by substituting these last two results in (\ref{Phi3-dot}) and (\ref{Phi4-dot}) respectively, this leads us to the same expressions (\ref{new-constraints}): i.e., this leads us to the same expressions that were chosen as the two constraints that needed to be imposed on the Lagrangian function (\ref{constrained-lagrangean-holo}) so that it could describe the same physics as (\ref{intrisic-lagrangean}). And this is a very welcome result because it sounds like a kind of ``breath of fresh air'', which seems to point to the fact that, although the Hamiltonian function (\ref{h-st}) was obtained in a different way from that which led to (\ref{hamil-intrinsic}), it does seem to describe the same physics as the Lagrangian (\ref{constrained-lagrangean-holo}).
			
			However, given that to obtain (\ref{h-st}) we had to implement the new constraints (\ref{Phi34}) using the new Lagrange multipliers $ \chi ^{\left( 3 \right) } $ and $ \chi ^{\left( 4 \right) } $, it is precisely here that we can explain why we have enumerated these multipliers as such. After all, since the two expressions in (\ref{new-constraints}), by depending solely on the parameters $ \left( x , y , \theta \right) $ in some way, define two constraints that can be interpreted as independent of those in (\ref{Phi34}), $ \chi ^{\left( 1 \right) } $ and $ \chi ^{\left( 2 \right) } $, which have already been interpreted as Lagrange multipliers in (\ref{constrained-lagrangean-holo}), can also be interpreted as Lagrange multipliers here. That is, as the derivation of (\ref{Phi34-dot}) makes it quite clear that the expressions in (\ref{new-constraints}) are also constraints from the Hamiltonian point of view, when we make the identification
			\begin{equation}
				\Phi _{\left( 1 \right) } \left( x , y \right) = f_{\left( 1 \right) } \left( x , y \right) \quad \textnormal{and} \quad \Phi _{\left( 2 \right) } \left( x , \theta \right) = F_{\left( 2 \right) } \left( x , \theta \right) \label{reexpress}
			\end{equation}
			to further emphasise the fact that expressions
			\begin{equation}
				\Phi _{\left( 1 \right) } \left( x , y \right) \approx 0 \quad \textnormal{and} \quad \Phi _{\left( 2 \right) } \left( x , \theta \right) \approx 0 \label{new-phi-constraints}
			\end{equation}
			also define the set of constraints $ \Phi \left( q , p \right) $, it is precisely this that allows us to rewrite (\ref{constrained-lagrangean-holo}) as
			\begin{eqnarray*}
				\lefteqn{H_{\mathrm{nd}} \left( q , p \right) = H \left( \left( x , y , \theta \right) , \left( P_{x} , P_{y} , P_{\theta } \right) \right) } \hspace*{3.0cm} \\
				& + & \chi ^{\left( 1 \right) } \Phi _{\left( 1 \right) } \left( x , y \right) + \chi ^{\left( 2 \right) } \Phi _{\left( 2 \right) } \left( x , \theta \right) \\
				& + & \chi ^{\left( 3 \right) } \Phi _{\left( 3 \right) } \left( \chi ^{\left( 1 \right) } , \Pi _{\left( 1 \right) } \right) + \chi ^{\left( 4 \right) } \Phi _{\left( 4 \right) } \left( \chi ^{\left( 2 \right) } , \Pi _{\left( 2 \right) } \right)
			\end{eqnarray*}
			under the assumption that
			\begin{equation*}
				H \left( \left( x , y , \theta \right) , \left( P_{x} , P_{y} , P_{\theta } \right) \right) = \frac{P_{x} ^{2}}{2m} + \frac{P_{y} ^{2}}{2m} + \frac{\left( a + 3 \right) P_{\theta } ^{2}}{4mR^{2}} + mgy  \ .
			\end{equation*}
			
		\subsection{Discovering new constraints}
		
			Note that, since the expressions in (\ref{new-phi-constraints}) can also be interpreted as two Hamiltonian constraints, it makes perfect sense to also evaluate the following\footnote{Note that, once again, we are faced with a situation where the partial derivatives with respect to $ t $ do not appear on the right-hand side of (\ref{Phi12-dot}) because the functions being differentiated do not depend explicitly on this parameter. Indeed, whenever something similar occurs in the following results, such partial derivatives will be ignored since, in fact, they are identically zero.}
			\begin{subequations} \label{Phi12-dot}
				\begin{align}
					\dot{\Phi } _{\left( 1 \right) } & = \left\{ \Phi _{\left( 1 \right) } , H_{\mathrm{nd}} \right\} \approx 0 \label{Phi1-dot} \\
					\dot{\Phi } _{\left( 2 \right) } & = \left\{ \Phi _{\left( 2 \right) } , H_{\mathrm{nd}} \right\} \approx 0 \ . \label{Phi2-dot}
				\end{align}
			\end{subequations}
			And as the expansions
			\begin{subequations}
				\begin{align*}
					\left\{ \Phi _{\left( 1 \right) } , H_{\mathrm{nd}} \right\} & = \frac{\left( \tan \phi \right) P_{x}}{m} \ + \ \frac{P_{y}}{m} \quad \textnormal{and} \\
					\left\{ \Phi _{\left( 2 \right) } , H_{\mathrm{nd}} \right\} & = \frac{\left( \sec \phi \right) P_{x}}{m} \ - \ \frac{\left( a + 3 \right) P_{\theta }}{2mR} \ ,
				\end{align*}
			\end{subequations}
			when substituted into (\ref{Phi1-dot}) and (\ref{Phi2-dot}) respectively, allow us to recognise, for instance, that
			\begin{subequations} \label{new-appoint-56}
				\begin{align}
					\dot{\Phi } _{\left( 1 \right) } & \approx 0 \ \Leftrightarrow \ \left( \tan \phi \right) P_{x} + P_{y} \approx 0 \quad \textnormal{and} \label{new-appoint-5} \\
					\dot{\Phi } _{\left( 2 \right) } & \approx 0 \ \Leftrightarrow \ \left( \sec \phi \right) P_{x} - \frac{\left( a + 3 \right) P_{\theta }}{2R} \approx 0 \label{new-appoint-6} \ ,
				\end{align}
			\end{subequations}
			it becomes quite clear that, since functions
			\begin{equation}
				\Phi _{\left( 5 \right) } \left( P_{x} , P_{y} \right) = \left( \tan \phi \right) P_{x} + P_{y} \quad \textnormal{and} \quad \Phi _{\left( 6 \right) } \left( P_{x} , P_{\theta } \right) = \left( \sec \phi \right) P_{x} - \frac{\left( a + 3 \right) P_{\theta }}{2R} \label{Phi56-functions}
			\end{equation}
			are expressed in terms of variables that do not appear in (\ref{Phi-function}) and (\ref{reexpress}), these results (\ref{new-appoint-5}) and (\ref{new-appoint-6}) are, in fact, pointing to the existence of two new constraints that are completely independent of those already given in (\ref{Phi34}) and (\ref{new-phi-constraints}). Because of this, since the discovery of the constraints (\ref{Phi34}) already required their incorporation into the Hamiltonian via new Lagrange multipliers, the discovery of these new constraints
			\begin{equation}
				\Phi _{\left( 5 \right) } \left( P_{x} , P_{y} \right) \approx 0 \quad \textnormal{and} \quad \Phi _{\left( 6 \right) } \left( P_{x} , P_{\theta } \right) \approx 0 \label{Phi56}
			\end{equation}
			also requires the same. Thus, since the Hamiltonian function we are dealing with is (\ref{h-st}), it is not difficult to conclude that this incorporation leads us to a new Hamiltonian
			\begin{equation}
				H_{\mathrm{total}} \left( q , p \right) = H_{\mathrm{nd}} \left( q , p \right) + \chi ^{\left( 5 \right) } \Phi _{\left( 5 \right) } \left( P_{x} , P_{y} \right) + \chi ^{\left( 6 \right) } \Phi _{\left( 6 \right) } \left( P_{x} , P_{\theta } \right) \ , \label{h-nh-classic}
			\end{equation}
			where $ \chi ^{\left( 5 \right) } $ and $ \chi ^{\left( 6 \right) } $ can be interpreted as the new Lagrange multipliers. That is,
			\begin{eqnarray*}
				\lefteqn{H_{\mathrm{total}} \left( q , p \right) = H \left( \left( x , y , \theta \right) , \left( P_{x} , P_{y} , P_{\theta } \right) \right) } \hspace*{3.0cm} \\
				& + & \chi ^{\left( 1 \right) } \Phi _{\left( 1 \right) } \left( x , y \right) + \chi ^{\left( 2 \right) } \Phi _{\left( 2 \right) } \left( x , \theta \right) \\
				& + & \chi ^{\left( 3 \right) } \Phi _{\left( 3 \right) } \left( \chi ^{\left( 1 \right) } , \Pi _{\left( 1 \right) } \right) + \chi ^{\left( 4 \right) } \Phi _{\left( 4 \right) } \left( \chi ^{\left( 2 \right) } , \Pi _{\left( 2 \right) } \right) \\
				& + & \chi ^{\left( 5 \right) } \Phi _{\left( 5 \right) } \left( P_{x} , P_{y} \right) + \chi ^{\left( 6 \right) } \Phi _{\left( 6 \right) } \left( P_{x} , P_{\theta } \right) \ .
			\end{eqnarray*}
			
		\subsection{On the Dirac-Bergmann algorithm}
		
			Broadly speaking, everything we have been doing since (\ref{Phi34-dot}) is nothing more than the application of the Dirac-Bergmann algorithm: i.e., what we have been doing since (\ref{Phi34-dot}) is nothing more than the application of a systematic procedure, whose name is justified by the fact that it was developed (independently) by Dirac and P. Bergmann (1915 -- 2002), for the identification of all the constraints that define a classical Hamiltonian system \cite{bergmann-1,bergmann-2,dirac-bergmann}. After all, it is precisely this identification that guarantees, amongst other things, the reliability and consistency of the expressions for the equations of motion of a system with constraints.
			
			Incidentally, in order to understand why this identification ensures such reliability and consistency, it is worth noting that, from a geometric point of view, the domain (phase space) of any Hamiltonian function can be recognised as a cotangent bundle $ T^{\ast } \mathcal{M} _{n} $: i.e., since it is valid to consider that the coordinates $ q = \left( q^ {1} , q^{2} , \ldots , q^{n} \right) $ parametrise a manifold $ \mathcal{M} _{n} $ which we may, for convenience, assume to be infinitely differentiable, such a domain can be interpreted in terms of the disjoint union of all spaces that are cotangent to $ \mathcal{M} _ {n} $ at each of its points, since (\ref{p-generic-definition}) is, in fact, defining a transformation that exchanges the velocities $ \dot{q} = \left( \dot{q} ^{1} , \dot {q} ^{2} , \ldots , \dot{q} ^{n} \right) $ (i.e., the coordinates parameterising a space that is tangent to $ \mathcal{M} _{n} $ at each of its points) with their duals $ p = \left( p_{1} , p_{2} , \ldots , p_{n} \right) $ \cite{manfredo-gr}. However, when these same parameters $ q $ and $ p $ also feature in a set of constraints such as (\ref{general-constraints}), it becomes quite clear that the physics of the Hamiltonian system in question can, in fact, be described by a smaller number of coordinates. After all, since this same geometric perspective also allows us to state that each constraint
			\begin{equation*}
				\Phi _{\left( \ell \right) } \left( q , p \right) \approx 0
			\end{equation*}
			defines a surface in $ T^{\ast } \mathcal{M} _{n} $, the existence of this set of constraints (\ref{general-constraints}) allows us to recognise that the physical parameters (i.e., the parameters that describe, solely, the physical system intrinsically) of this Hamiltonian description belong to a cotangent bundle $ T^{\ast } \mathcal{M} _{n-m}$ which is of lower dimension than $ T^{\ast } \mathcal{M} _{n} $: scilicet, the physical parameters of such a system belong to a cotangent bundle $ T^{\ast } \mathcal{M} _{n-m}$ which is contained within $ T^{\ast } \mathcal{M} _{n} $ \cite{manfredo-gr}.
			
			Observe that what we have just said in the previous paragraph can be illustrated by all the results we have already obtained throughout this Section. And in order to understand why this is so, it suffices to see that, although we initially considered the system under analysis to be parameterised by the pair $ \left( q , p \right) = \left( \left( x , y , \theta , \chi ^{\left( 1 \right) } , \chi ^{\left( 2 \right) } \right) , \left( P_{x} , P_{y} , P_{\theta } , \Pi _{\left( 1 \right) } , \Pi _{\left( 2 \right) } \right) \right) $, it is crystal clear that the expressions of the six constraints appearing in (\ref{Phi34}), (\ref{new-phi-constraints}) and (\ref{Phi56}) do indeed take the form of surface equations. By the way, note that when we place the functions
            \begin{equation}
                \Phi _{\left( 1 \right) } \left( x , y \right) = \left( \tan \phi \right) x + y \quad \textnormal{and} \quad \Phi _{\left( 5 \right) } \left( P_{x} , P_{y} \right) = \left( \tan \phi \right) P_{x} + P_{y} \label{similar-1}
            \end{equation}
            side by side, it also becomes quite clear that they are very similar, a similarity that can also be identified between the functions
			\begin{equation}
				\Phi _{\left( 2 \right) } \left( x , \theta \right) = \left( \sec \phi \right) x - R \theta \quad \textnormal{and} \quad \Phi _{\left( 6 \right) } \left( P_{x} , P_{\theta } \right) = \left( \sec \phi \right) P_{x} - \frac{\left( a + 3 \right) P_{\theta }}{2R} \ . \label{similar-2}
			\end{equation}
			Of course, a more attentive reader might well point out that all these similarities are perfectly justifiable given that, in spite of having arrived at the functions $ \Phi _{\left( 5 \right) } \left( P_{x} , P_{y} \right) $ and $ \Phi _{\left( 6 \right) } \left( P_{x} , P_{\theta } \right) $ by evaluating some Poisson brackets, these brackets were only evaluated so that we could assess what emerged from the consistency equations mentioned in (\ref{Phi12-dot}). In other words, what this more attentive reader can say is that all these similarities stem from the simple fact that the expressions $ \Phi _{\left( 5 \right) } \left( P_{x} , P_{y} \right) $ and $ \Phi _{\left( 6 \right) } \left( P_{x} , P_{\theta } \right) $ were identified by evaluating, respectively, the total derivatives of $ \Phi _{\left( 1 \right) } \left( x , y \right) $ and $ \Phi _{\left( 2 \right) } \left( x , \theta \right) $ with respect to time. And, without a shadow of a doubt, a reader who recognises all this is right. However, since the calculation of $ \Phi _{\left( 5 \right) } \left( P_{x} , P_{y} \right) $ and $ \Phi _{\left( 6 \right) } \left( P_{x} , P_{\theta } \right) $, which involves calculating these time derivatives, subsequently requires that the velocities $ \dot{x} $ and $ \dot{\theta } $ be replaced by their respective duals $ P_{x} $ and $ P_{\theta } $, it is precisely this that further reinforces what we have already said regarding the fact that, for instance, the physical parameters of a Hamiltonian system with constraints actually belong to a cotangent bundle that is dimensionally smaller than that originally parameterised by $ \left( q , p \right) $. After all, since $ P_{x} $ is dual to $ \dot{x} $, the striking similarity in (\ref{similar-1}) can be perfectly justified by recognising that the surfaces, which are defined by the constraints
			\begin{equation}
				\Phi _{\left( 1 \right) } \left( x , y \right) \approx 0 \quad \textnormal{and} \quad \Phi _{\left( 5 \right) } \left( P_{x} , P_{y} \right) \approx 0 \ , \label{sub-f-1}
			\end{equation}
			together define a cotangent subbundle of $ T^{\ast } \mathcal{M} _{n} $ \cite{manfredo-gr}.
			
			Although we concluded the previous paragraph by focusing solely on the geometric interpretation of this similarity between the constraints highlighted in (\ref{sub-f-1}), everything we have just said also applies to the constraints derived from the expressions appearing in (\ref{similar-2}): i.e., since $ P_{\theta } $ is dual to $ \dot{\theta } $, the similarity identified in (\ref{similar-1}) can also be justified by the fact that the surfaces, which are defined by the relations
			\begin{equation}
				\Phi _{\left( 2 \right) } \left( x , \theta \right) \approx 0 \quad \textnormal{and} \quad \Phi _{\left( 6 \right) } \left( P_{x} , P_{\theta } \right) \approx 0 \ , \label{sub-f-2}
			\end{equation}
			together form another cotangent subbundle of $ T^{\ast } \mathcal{M} _{n} $. Nevertheless, and by bearing in mind that we concluded the previous Subsection by defining a new Hamiltonian function (\ref{h-nh-classic}), in which all the constraints identified until then were incorporated, it is important to continue applying the Dirac-Bergmann algorithm to evaluate whether other constraints are also present. And this new round of evaluation now requires us to calculate what follows from \cite{dirac}
			\begin{subequations} \label{Phi56-dot}
				\begin{align}
					\dot{\Phi } _{\left( 5 \right) } & = \left\{ \Phi _{\left( 5 \right) } , H_{\mathrm{total}} \right\} \approx 0 \quad \textnormal{and} \label{Phi5-dot} \\
					\dot{\Phi } _{\left( 6 \right) } & = \left\{ \Phi _{\left( 6 \right) } , H_{\mathrm{total}} \right\} \approx 0 \ . \label{Phi6-dot}
				\end{align}
			\end{subequations}
			
			In fact, as the expansion of the Poisson brackets, which appear in these last two consistency relations, shows us that
			\begin{subequations}
				\begin{align*}
					\left\{ \Phi _{\left( 5 \right) } , H_{\mathrm{total}} \right\} & = - \chi ^{\left( 1 \right) } \left( \sec ^{2} \phi \right) - \chi ^{\left( 2 \right) } \left( \sec \phi \right) \left( \tan \phi \right) - mg \quad \textnormal{and} \\
					\left\{ \Phi _{\left( 6 \right) } , H_{\mathrm{total}} \right\} & = - \chi ^{\left( 1 \right) } \left( \sec \phi \right) \left( \tan \phi \right) - \chi ^{\left( 2 \right) } \left[ \left( \sec ^{2} \phi \right) + \frac{a + 3}{2} \right] \ ,
				\end{align*}
			\end{subequations}
			it is immediately clear that
			\begin{subequations} \label{new-appoint-78}
				\begin{align}
					\dot{\Phi } _{\left( 5 \right) } & \approx 0 \ \Leftrightarrow \ \chi ^{\left( 1 \right) } \left( \sec ^{2} \phi \right) + \chi ^{\left( 2 \right) } \left( \sec \phi \right) \left( \tan \phi \right) + mg \approx 0 \quad \textnormal{and} \label{new-appoint-7} \\
					\dot{\Phi } _{\left( 6 \right) } & \approx 0 \ \Leftrightarrow \ \chi ^{\left( 1 \right) } \left( \sec \phi \right) \left( \tan \phi \right) + \chi ^{\left( 2 \right) } \left[ \left( \sec ^{2} \phi \right) + \frac{a + 3}{2} \right] \approx 0 \label{new-appoint-8} \ .
				\end{align}
			\end{subequations}
			But, despite being entirely valid to use these last two results to define new constraints (because, given our initial considerations, the Lagrange multipliers $ \chi ^{\left( 1 \right) } $ and $ \chi ^{\left( 2 \right) } $ were introduced as variables in our problem), these last two results also make it quite clear that, in fact, we are dealing with a linear system which, for example, already allows us to determine the values of $ \chi ^{\left( 1 \right) } $ and $ \chi ^{\left( 2 \right) } $ \cite{strang}. In these terms, and by noting that
			\begin{equation}
				\chi ^{\left( 1 \right) } \approx - \frac{mg \left[ \left( a + 3 \right) \left( \cos ^{2} \phi \right) + 2 \right] }{a + 5} \quad \textnormal{and} \quad \chi ^{\left( 2 \right) } \approx \frac{2mg \left( \sin \phi \right) }{a + 5} \label{multi-12}
			\end{equation}
			solve this linear system, it becomes clear that it is completely unnecessary to define new constraints using these Lagrange multipliers. After all, given that the values taken by $ \chi ^{\left( 1 \right) } $ and $ \chi ^{\left( 2 \right) } $ are two real constants, the application of the Dirac-Bergmann algorithm can be concluded here, since there are, in fact, no further constraints to be added to the Hamiltonian formulation of our problem. And it is thanks to this result that we can finally offer a fairly consistent explanation of why we constructed this Hamiltonian formulation with constraints based on the Lagrangian function (\ref{constrained-lagrangean-holo}). And why did we do this? Because, as (\ref{constrained-lagrangean-holo}) is the only Lagrangian function that consists of
			\begin{itemize}
				\item holonomic constraints and
				\item all two constraints necessary to implement the two conditions (i) and (ii) stated in the Introduction,
			\end {itemize}
			it is the only one that leads us to a Hamiltonian formulation whose constraints, being expressed by functions that depend only on the variables $ q = \left( x , y , \theta , \chi ^{\left( 1 \right) } , \chi ^{\left( 2 \right) } \right) $ and $ p = \left( P_{x} , P_{y} , P_{\theta } , \Pi _{\left( 1 \right) } , \Pi _{\left( 2 \right) } \right) $, take the form (\ref{dirac-parenteses}) which is necessary for dealing with Dirac brackets (\ref{dirac-parenteses}).
			
		\subsection{A new use of Poisson brackets}
		
			Incidentally, since the only constraints to be considered here are those defined by the expressions (\ref{Phi34}), (\ref{new-constraints}) and (\ref{Phi56}), one thing we can already do is to calculate the values of all the Poisson brackets
			\begin{equation}
				\left\{ \Phi _{\left( j \right) } , \Phi _{\left( \ell \right) } \right\} \ , \label{fifi}
			\end{equation}
			where $ j , \ell = 1 , 2 , \ldots , 6 $. And as tedious as it may be to continue calculating more and more Poisson brackets, this is still necessary, since this expression, which appears in (\ref{fifi}), is exactly the same as the one that defines the elements (\ref{terms}) of the matrix $ \Theta $ that we mentioned in the Introduction. However, as much as the fact that we are dealing with $ 6 $ constraints requires us to calculate the $ 36 $ elements of $ \Theta $ (i.e., $ 36 $ Poisson brackets), it is worth noting that expression (\ref{poisson}) makes this task considerably easier. After all, as this expression allows us to observe that
			\begin{equation}
                \left\{ \Phi _{\left( j \right) } , \Phi _{\left( \ell \right) } \right\} = - \left\{ \Phi _{\left( \ell \right) } , \Phi _{\left( j \right) } \right\} \ , \label{antissimetric}
			\end{equation}
            this antisymmetry, which is associated with this matrix $ \Theta $, already makes it quite clear, for example, that the elements of the main diagonal of $ \Theta $ are all zero.
			
			Another interesting point worth noting here is that, since the functions $ \Phi _{\left( 1 \right) } $ and $ \Phi _{\left( 2 \right) } $ depend solely on the position variables,
			\begin{equation*}
				\left\{ \Phi _{\left( 1 \right) } , \Phi _{\left( 2 \right) } \right\} = \left\{ \Phi _{\left( 2 \right) } , \Phi _{\left( 1 \right) } \right\} = 0 \ .
			\end{equation*}
			And this is a result that also extends to all Poisson brackets involving, exclusively, the functions $ \Phi _{\left( 3 \right) } $, $ \Phi _{\left( 4 \right) } $, $ \Phi _{\left( 5 \right) } $ and $ \Phi _{\left( 6 \right) } $: i.e., since these four functions depend only on momentum variables, this leads us to
			\begin{eqnarray*}
				\left\{ \Phi _{\left( 3 \right) } , \Phi _{\left( 4 \right) } \right\} & = & \left\{ \Phi _{\left( 4 \right) } , \Phi _{\left( 3 \right) } \right\} = \left\{ \Phi _{\left( 3 \right) } , \Phi _{\left( 5 \right) } \right\} = \left\{ \Phi _{\left( 5 \right) } , \Phi _{\left( 3 \right) } \right\} \\
				& = & \left\{ \Phi _{\left( 3 \right) } , \Phi _{\left( 6 \right) } \right\} = \left\{ \Phi _{\left( 6 \right) } , \Phi _{\left( 3 \right) } \right\} = \left\{ \Phi _{\left( 4 \right) } , \Phi _{\left( 5 \right) } \right\} \\
				& = & \left\{ \Phi _{\left( 5 \right) } , \Phi _{\left( 4 \right) } \right\} = \left\{ \Phi _{\left( 5 \right) } , \Phi _{\left( 6 \right) } \right\} = \left\{ \Phi _{\left( 6 \right) } , \Phi _{\left( 5 \right) } \right\} = 0 \ .
			\end{eqnarray*}
			
			Although the Poisson brackets that still need to be calculated involve
			\begin{itemize}
				\item one function that depends exclusively on the position variables and
				\item another function that depends only on the momentum variables,
			\end{itemize}
			it is also worth noting that, as $ \Phi _{\left( 3 \right) } $ and $ \Phi _{\left( 4 \right) } $ are the only functions that depend solely on $ \bigl( \chi ^{\left( 1 \right) } , \Pi _{\left( 1 \right) } \bigr) $ and $ \bigl( \chi ^{\left( 2 \right) } , \Pi _{\left( 2 \right) } \bigr) $, it also follows that
			\begin{eqnarray*}
				\left\{ \Phi _{\left( 1 \right) } , \Phi _{\left( 3 \right) } \right\} & = & \left\{ \Phi _{\left( 3 \right) } , \Phi _{\left( 1 \right) } \right\} = \left\{ \Phi _{\left( 1 \right) } , \Phi _{\left( 4 \right) } \right\} = \left\{ \Phi _{\left( 4 \right) } , \Phi _{\left( 1 \right) } \right\} \\
				& = & \left\{ \Phi _{\left( 2 \right) } , \Phi _{\left( 3 \right) } \right\} = \left\{ \Phi _{\left( 3 \right) } , \Phi _{\left( 2 \right) } \right\} = \left\{ \Phi _{\left( 2 \right) } , \Phi _{\left( 4 \right) } \right\} = \left\{ \Phi _{\left( 4 \right) } , \Phi _{\left( 2 \right) } \right\} = 0 \ .
			\end{eqnarray*}
			In other words, what all these results make clear is that $ \Theta $ is a matrix full of zero elements. And, due to the observation we made in the previous Subsection (namely, that the constraints in (\ref{sub-f-1}) form a cotangent subbundle of $ T^{\ast } \mathcal{M} _{n} $, whilst the pairs of constraints in (\ref{sub-f-2}) define another cotangent subbundle of this same $ T^{\ast } \mathcal{M} _{n} $), it becomes clear that the only four Poisson brackets we need to calculate here, explicitly, are those involving the functions that define these two pairs of constraints (\ref{sub-f-1}) and (\ref{sub-f-2}). And since the results of these four Poisson brackets are
			\begin{eqnarray*}
				\left\{ \Phi _{\left( 1 \right) } , \Phi _{\left( 5 \right) } \right\} & = & \left( \sec ^{2} \phi \right) \ , \\
				\left\{ \Phi _{\left( 2 \right) } , \Phi _{\left( 6 \right) } \right\} & = & \left( \sec ^{2} \phi \right) + \frac{a + 3}{2} \quad \textnormal{and} \\
				\left\{ \Phi _{\left( 1 \right) } , \Phi _{\left( 6 \right) } \right\} & = & \left\{ \Phi _{\left( 2 \right) } , \Phi _{\left( 5 \right) } \right\} = \left( \sec \phi \right) \left( \tan \phi \right) \ ,
			\end{eqnarray*}
			it becomes quite clear that
			\begin{equation}
				\Theta = \begin{pmatrix}
					\mathds{O} & \mathds{O} & \Theta _{C} \\
					\mathds{O} & \mathds{O} & \mathds{O} \\
					\Theta _{A} & \mathds{O} & \mathds{O}
				\end{pmatrix} \ , \label{n-i-matrix}
			\end{equation}
			where $ \mathds{O} $ is a zero matrix of order $ 2 $, and
			\begin{equation*}
				\Theta _{A} = \begin{pmatrix}
					\left\{ \Phi _{\left( 5 \right) } , \Phi _{\left( 1 \right) } \right\} & \left\{ \Phi _{\left( 5 \right) } , \Phi _{\left( 2 \right) } \right\} \\
					\left\{ \Phi _{\left( 6 \right) } , \Phi _{\left( 1 \right) } \right\} & \left\{ \Phi _{\left( 6 \right) } , \Phi _{\left( 2 \right) } \right\}
				\end{pmatrix} \quad \textnormal{and} \quad \Theta _{C} = \begin{pmatrix}
					\left\{ \Phi _{\left( 1 \right) } , \Phi _{\left( 5 \right) } \right\} & \left\{ \Phi _{\left( 2 \right) } , \Phi _{\left( 5 \right) } \right\} \\
					\left\{ \Phi _{\left( 1 \right) } , \Phi _{\left( 6 \right) } \right\} & \left\{ \Phi _{\left( 2 \right) } , \Phi _{\left( 6 \right) } \right\}
				\end{pmatrix} \ .
			\end{equation*}
			That is,
			\begin{equation*}
				\Theta _{A} = - \begin{pmatrix}
					\sec ^{2} \phi \ & \ \left( \sec \phi \right) \left( \tan \phi \right) \\
					\left( \sec \phi \right) \left( \tan \phi \right) \ & \ \sec ^{2} \phi + \left( a + 3 \right) / 2
				\end{pmatrix} = - \Theta _{C} \ .
			\end{equation*}
			
		\subsection{An important comment on these results}
		
			Note that, although these matrices $ \Theta _{A} $ and $ \Theta _{C} $ are clearly invertible, it is also crystal clear that $ \Theta $ does not have an inverse since, unlike $ \Theta _{A} $ and $ \Theta _{C} $, its determinant is zero. Indeed, it is worth noting that, if we construct a new matrix
			\begin{equation*}
				\tilde{\Theta } = \begin{pmatrix}
					\mathds{O} & \Theta _{C} \\
					\Theta _{A} & \mathds{O}
				\end{pmatrix} \ ,
			\end{equation*}
			simply by eliminating all the identically zero rows and columns from $ \Theta $, this new matrix will have
			\begin{equation}
				\tilde{\Theta } ^{-1} = \begin {pmatrix}
					\mathds{O} & \Theta ^{-1} _{A} \\
					\Theta ^{-1} _{C} & \mathds{O}
				\end{pmatrix} \label{general-inverse}
			\end{equation}
			as its inverse, where
			\begin{equation*}
				\Theta _{A} ^{-1} = \frac{1}{a + 5} \begin{pmatrix}
					\left( a + 3 \right) \left( \cos ^{2} \phi \right) + 2 \ & \ - 2 \left( \sin \phi \right) \\
					- 2 \left( \sin \phi \right) \ & \ 2
				\end{pmatrix} = - \Theta _{C} ^{-1} \ .
			\end{equation*}
			However, even though everything that we are saying may sound extremely technical, it is actually very welcome. Why? Because this is precisely in line with two important observations we made in the Introduction, the first of which is: the Dirac quantisation process requires the replacement of Poisson brackets with those bearing his name when the classical system to be quantised is described, in Hamiltonian terms, with the aid of a set of constraints (\ref{general-constraints}). The second observation, which is not independent of the first, is: although the Dirac brackets (\ref{dirac-parenteses}) are defined with the help of a matrix $ \Theta ^{-1} $, the matrix $ \Theta $ is not always invertible. In other words, as technical as all this may sound, it is actually very welcome because it is precisely all of this (mainly the matrix (\ref{n-i-matrix}) we have just derived) that makes it perfectly clear that the system we are analysing here leads precisely to a $ \Theta $ that is not invertible. Therefore, the question that needs to be answered is: what should be done in this situation?
			
			In order to understand the answer to this question, it is important to note that this same matrix $ \Theta $ can also be identified when, for example, we observe that the consistency equations (\ref{Phi34-dot}), (\ref{Phi12-dot}) and (\ref{Phi56-dot}) can also be written as
			\begin{equation}
				\dot{\Phi } _{\left( j \right) } = \left\{ \Phi _{\left( j \right) } , H_{\mathrm{total}} \right\} = \left\{ \Phi _{\left( j \right) } , H \right\} + \sum ^ {6} _{\ell = 1} \chi ^{\left( \ell \right) } \left\{ \Phi _{\left( j \right) } , \Phi _{\left( \ell \right) } \right\} \approx 0 \ . \label{hnd-time-evolution}
			\end{equation}
			After all, as
			\begin{itemize}
				\item the solution of the linear system (\ref{new-appoint-78}) made it clear that, although the Lagrange multipliers $ \chi ^{\left( 1 \right) } $ and $ \chi ^{\left( 2 \right) } $ were listed as variables, they take on constant values on the surface generated by all the constraints, and
				\item the remaining Lagrange multipliers (namely, $ \chi ^{\left( 3 \right) } $, $ \chi ^{\left( 4 \right) } $, $ \chi ^{\left( 5 \right) } $ and $ \chi ^{\left( 6 \right) } $) were not considered as variables in our system,
			\end{itemize}
			all these Lagrange multipliers can be interpreted as mere constants in the expansion of the term $ \left\{ \Phi _{\left( j \right) } , H_{\mathrm{total}} \right\} $ when all constraints (\ref{Phi34}), {\ref{new-phi-constraints}) and (\ref{Phi56}) are satisfied, which therefore justifies this result (\ref{hnd-time-evolution}). In this context, when we recognise, for instance, that (\ref{hnd-time-evolution}) is nothing more than the general expression of the rows of the matrix equation
			\begin{equation*}
				\begin{pmatrix}
					\Theta _{11} & \Theta _{12} & \ldots & \Theta _{16} \\
					\Theta _{21} & \Theta _{22} & \ldots & \Theta _{26} \\
					\vdots & \vdots & \vdots & \vdots \\
					\Theta _{61} & \Theta _{62} & \ldots & \Theta _{66} \\
				\end{pmatrix}
				\begin{pmatrix}
					\chi ^{\left( 1 \right) } \\
					\chi ^{\left( 2 \right) } \\
					\vdots \\
					\chi ^{\left( 6 \right) }
				\end{pmatrix} \approx - \begin{pmatrix}
					\left\{ \Phi _{\left( 1 \right) } , H \right\} \\
					\left\{ \Phi _{\left( 2 \right) } , H \right\} \\
					\vdots \\
					\left\{ \Phi _{\left( 6 \right) } , H \right\}
				\end{pmatrix} \ ,
			\end{equation*}
			it is immediately clear that the invertibility of $ \Theta $ is directly related to the possibility of solving all the Lagrange multipliers of our problem. However, it is worth noting that, due to the result (\ref{n-i-matrix}), we can also break this same matrix equation down into three others, which are independent of one another, namely
			\begin{subequations} \label{subeq}
				\begin{align}
					\Theta _{A} = \begin{pmatrix}
						\chi ^{\left( 1 \right) } \\
						\chi ^{\left( 2 \right) }
					\end{pmatrix} & \approx - \begin{pmatrix}
						\left\{ \Phi _{\left( 5 \right) } , H \right\} \\
						\left\{ \Phi _{\left( 6 \right) } , H \right\}
					\end{pmatrix} \ , \label{subeq-1} \\
					\Theta _{B} = \begin{pmatrix}
						\chi ^{\left( 3 \right) } \\
						\chi ^{\left( 4 \right) }
					\end{pmatrix} & \approx - \begin{pmatrix}
						\left\{ \Phi _{\left( 3 \right) } , H \right\} \\
						\left\{ \Phi _{\left( 4 \right) } , H \right\}
					\end{pmatrix} \ \ \textnormal{and} \label{subeq-2} \\
					\Theta _{C} = \begin{pmatrix}
						\chi ^{\left( 5 \right) } \\
						\chi ^{\left( 6 \right) }
					\end{pmatrix} & \approx - \begin{pmatrix}
						\left\{ \Phi _{\left( 1 \right) } , H \right\} \\
						\left\{ \Phi _{\left( 2 \right) } , H \right\}
					\end{pmatrix} \ , \label{subeq-3}
				\end{align}
			\end{subequations}
			where $ \Theta _{B} = \mathds{O} $. In this way, although the matrix (\ref{n-i-matrix}) is not actually invertible, the breakdown we have just carried out shows us that the invertibility of the matrices $ \Theta _{A} $ and $ \Theta _{C} $ allows us to write
			\begin{subequations} \label{multi-equation}
				\begin{align}
					\begin{pmatrix}
						\chi ^{\left( 1 \right) } \\
						\chi ^{\left( 2 \right) }
					\end{pmatrix} & \approx - \Theta ^{-1} _{A} \begin{pmatrix}
						\left\{ \Phi _{\left( 5 \right) } , H \right\} \\
						\left\{ \Phi _{\left( 6 \right) } , H \right\}
					\end{pmatrix} \ \ \textnormal{and} \label{multi-equation-1} \\
					\begin{pmatrix}
						\chi ^{\left( 5 \right) } \\
						\chi ^{\left( 6 \right) }
					\end{pmatrix} & \approx - \Theta ^{-1} _{C} \begin{pmatrix}
						\left\{ \Phi _{\left( 1 \right) } , H \right\} \\
						\left\{ \Phi _{\left( 2 \right) } , H \right\}
					\end{pmatrix} \ . \label{multi-equation-2}
				\end{align}
			\end{subequations}
			
			Note that, as the expansion of (\ref{multi-equation-1}) gives us
			\begin{equation}
				\begin{pmatrix}
					\chi ^{\left( 1 \right) } \\
					\chi ^{\left( 2 \right) }
				\end{pmatrix} \approx \frac{1}{a + 5} \begin{pmatrix}
					\left( a + 3 \right) \left( \cos ^{2} \phi \right) + 2 \ & \ - 2 \left( \sin \phi \right) \\
					- 2 \left( \sin \phi \right) \ & \ 2
				\end{pmatrix} \begin{pmatrix}
					- mg \\
					0
				\end{pmatrix} \label{second-multi-12}
			\end{equation}
			this is a result that comes as no surprise (and which, in fact, was already expected) because it is exactly the same result as that previously stated in (\ref{multi-12}). Now, in the case of the development of (\ref{multi-equation-2}), as it leads us to
			\begin{equation}
				\begin{pmatrix}
					\chi ^{\left( 5 \right) } \\
					\chi ^{\left( 6 \right) }
				\end{pmatrix} \approx \frac{1}{m \left( a + 5 \right) } \begin{pmatrix}
					\left( a + 3 \right) \left( \cos ^{2} \phi \right) + 2 \ & \ - 2 \left( \sin \phi \right) \\
					- 2 \left( \sin \phi \right) \ & \ 2
				\end{pmatrix} \begin{pmatrix}
					\Phi _{\left( 5 \right) } \\
					\Phi _{\left( 6 \right) }
				\end{pmatrix} \ , \label{multipliers-56-determination}
			\end{equation}
			it becomes quite clear that both $ \chi ^{\left( 5 \right) } $ and $ \chi ^{\left( 6 \right) } $ must be taken as zero. After all, since the description of our ball in terms of a Hamiltonian system with constraints requires us to recognise that its physics is restricted to the cotangent subbundle of $ T^{\ast } \mathcal{M} _{n} $, which is defined by
			\begin{equation}
				\Phi \left( q , p \right) = \left( \Phi _{\left( 1 \right) } \left( q , p \right) , \Phi _{\left( 2 \right) } \left( q , p \right) , \ldots , \Phi _{\left( 6 \right) } \left( q , p \right) \right) \approx \bigl( \underbrace{0 , 0 , \ldots , 0} _{6 \ \text{times}} \bigr) \ , \label{set-six-constraints}
			\end{equation}
			the fact that (\ref{multipliers-56-determination}) makes it clear that the Lagrange multipliers $ \chi ^{\left( 5 \right) } $ and $ \chi ^{\left( 6 \right) } $ are both linear superpositions of the functions (\ref{Phi56-functions}) (which are precisely the functions that determine part of the constraints defining (\ref{set-six-constraints})) implies that both multipliers are zero on such a subbundle.
			
		\subsection{The interpretation of the system in terms of a gauge theory}
		
			However, one very interesting point that these latest results are showing us is that, although the non-inversibility of $ \Theta $ leads us to an equality (\ref{subeq-2}) that does not allow us to determine $ \chi ^{\left( 3 \right) } $ and $ \chi ^{\left( 4 \right) } $ unambiguously, this same equality remains valid ``ad infinitum'' because, in addition to $ \Theta _{B} $ being a zero matrix, the Poisson brackets
			\begin{equation*}
				\left\{ \Phi _{\left( 3 \right) } , H \right\} \quad \textnormal{and} \quad \left\{ \Phi _{\left( 4 \right) } , H \right\}
			\end{equation*}
			are also zero. In plain English, what this equality (\ref{subeq-2}) is actually showing us is that it will always hold true, regardless of the values assigned to $ \chi ^{\left( 3 \right) } $ and $ \chi ^{\left( 4 \right) } $. And this is precisely what allows us to interpret this physical system as a gauge model, an interpretation that is strengthened by the fact that all the Poisson brackets
			\begin{equation*}
				\left\{ \Phi _{\left( 3 \right) } , \Phi _{\left( \ell \right) } \right\} \quad \textnormal{and} \quad \left\{ \Phi _{\left( 4 \right) } , \Phi _{\left( \ell \right) } \right\} \ ,
			\end{equation*}
			which can be calculated by considering $ \ell = 1 , 2 , \ldots , 6 $, are identically zero. After all, as the theory of Hamiltonian systems with constraints makes well clear, whenever a first-class constraint (i.e., whenever a constraint whose Poisson brackets, between itself and all those that define the theory (including itself), are identically zero) is present, this theory is a gauge theory \cite{gitman,mf-constraints,teitel}.
			
			Note that, from a physical point of view, recognising that (\ref{subeq-2}) remains valid regardless of the values of $ \chi ^{\left( 3 \right) } $ and $ \chi ^{\left( 4 \right) } $ is closely constrainted to the realisation that these Lagrange multipliers have no physical meaning. And a nice way to begin to understand why this is the case is, for example, by noting that the relations contained in (\ref{subeq}) already make it quite clear that this impossibility, of determining $ \chi ^{\left( 3 \right) } $ and $ \chi ^{\left( 4 \right) } $ unequivocally, does not affect the determination of the rest of the multipliers that appear in (\ref{h-nh-classic}). A second important point, which must be borne in mind in order to understand why $ \chi ^{\left( 3 \right) } $ and $ \chi ^{\left( 4 \right) } $ have no physical meaning is to remember that these same multipliers were only used to introduce, into the Hamiltonian function (\ref{h-st}) (and, consequently, into (\ref{h-nh-classic})), the unique constraints that depend on the momenta $ \Pi _{\left( 1 \right) } $ and $ \Pi _{\left( 2 \right) } $. And since these momenta $ \Pi _{\left( 1 \right) } $ and $ \Pi _{\left( 2 \right) } $ were only proposed under the assumption that they are conjugate to $ \chi ^{\left( 1 \right) } $ and $ \chi ^{\left( 2 \right) } $ respectively, it is precisely the unequivocal determination of these multipliers that makes it quite clear that such momenta also have no physical significance. After all, since both (\ref{multi-12}) and (\ref{second-multi-12}) show us that $ \chi ^{\left( 1 \right) } $ and $ \chi ^{\left( 2 \right) } $ take on univocal constant values in the subbundle determined by (\ref{set-six-constraints}), this constancy and univocality is consistent with the fact that $ \dot{\chi } ^{\left( 1 \right) } $ and $ \dot{\chi } ^{\left( 2 \right) } $ and, consequently, $ \Pi _{\left( 1 \right) } $ and $ \Pi _{\left( 2 \right) } $ are identically zero on this subbundle. That is, since $ \chi ^{\left( 1 \right) } $ and $ \chi ^{\left( 2 \right) } $ do not behave as variables in the subbundle determined by (\ref{set-six-constraints}) and, therefore, neither do $ \Pi _{\left( 1 \right) } $ and $ \Pi _{\left( 2 \right) } $ do not either, this is what allows us to affirm that neither of them (i.e., neither $ \left( \chi ^{\left( 1 \right) } , \Pi _{\left( 1 \right) } \right) $ nor $ \left( \chi ^{\left( 2 \right) } , \Pi _{\left( 2 \right) } \right) $) has any relevance to the parameterisation/description of our physical system. From a geometric point of view, this constancy of $ \left( \chi ^{\left( 1 \right) } , \Pi _{\left( 1 \right) } \right) $ and $ \left( \chi ^{\left( 2 \right) } , \Pi _{\left( 2 \right) } \right) $ is closely related to the fact that these pairs are, in fact, extrinsic parameters of the cotangent subbundle determined by (\ref{set-six-constraints}) \cite{gitman,mf-constraints,manfredo-gr}.
			
		\subsection{Obtaining the Dirac brackets}
		
			Given that the equality (\ref{subeq-2}) remains valid, for instance, under gauge fixation
			\begin{equation}
				\chi ^{\left( 3 \right) } = \chi ^{\left( 4 \right) } = 0 \ , \label{gauge-fixation}
			\end{equation}
			such a gauge fixing turns out to be very welcome. Why? Because it is this that allows us to replace the matrix $ \Theta $ with $ \tilde{\Theta } $ when, for example, we need to calculate the Dirac brackets between two functions $ F \left( q , p \right) $ and $ G \left( q , p \right) $ that depend only on the parameters defining the constraints in (\ref{new-phi-constraints}) and (\ref{Phi56}). After all, since the sum, which appears on the right-hand side of (\ref{dirac-parenteses}), can be identified as the result of a matrix multiplication, the same considerations that previously led us to (\ref{multi-equation}) also allow us to reduce such Dirac brackets to
			\begin{subequations} \label{dirac-reduction}
				\begin{align}
					\left\{ F , G \right\} _{D} = \left\{ F , G \right\} & - \begin{pmatrix}
						\left\{ F , \Phi _{\left( 1 \right) } \right\} & \left\{ F , \Phi _{\left( 2 \right) } \right\}
					\end{pmatrix} \Theta ^{-1} _{C} \begin{pmatrix}
						\left\{ \Phi _{\left( 1 \right) } , G \right\} & \left\{ \Phi _{\left( 2 \right) } , G \right\}
					\end{pmatrix} ^{T} \label{dirac-reduction-1} \\
					& - \begin{pmatrix}
						\left\{ F , \Phi _{\left( 5 \right) } \right\} & \left\{ F , \Phi _{\left( 6 \right) } \right\}
					\end{pmatrix} \Theta ^{-1} _{A} \begin{pmatrix}
						\left\{ \Phi _{\left( 5 \right) } , G \right\} & \left\{ \Phi _{\left( 6 \right) } , G \right\}
					\end{pmatrix} ^{T} \ . \label{dirac-reduction-2}
				\end{align}
			\end{subequations}
			
			As the expressions in (\ref{antissimetric}) and (\ref{general-inverse}) allow us to conclude that the antisymmetry of the Poisson brackets also extends to the Dirac brackets, it follows immediately that
			\begin{equation*}
				\begin{matrix}
					\left\{ x , x \right\} _{D} & \ , \ & \left\{ y , y \right\} _{D} & \ , \ & \left\{ \theta , \theta \right\} _{D} & \ , \\
					\left\{ P_{x} , P_{x} \right\} _{D} & , & \left\{ P_{y} , P_{y} \right\} _{D} & \textnormal{and} & \left\{ P_{\theta } , P_{\theta } \right\} _{D}
				\end{matrix}
			\end{equation*}
			are all zero. And, even though calculating the remaining Dirac brackets is a rather tedious task, all these calculations that still need to be done become a little less tedious when one realises that the Poisson brackets
			\begin{equation*}
				\begin{matrix}
					\left\{ x , \Phi _{\left( 1 \right) } \right\} & \ , \ & \left\{ x , \Phi _{\left( 2 \right) } \right\} & \ , \ & \left\{ y , \Phi _{\left( 1 \right) } \right\} & \ , \ & \left\{ y , \Phi _{\left( 2 \right) } \right\} \ , \\
					\left\{ y , \Phi _{\left( 6 \right) } \right\} & , & \left\{ \theta , \Phi _{\left( 1 \right) } \right\} & , & \left\{ \theta , \Phi _{\left( 2 \right) } \right\} & , & \left\{ \theta , \Phi _{\left( 5 \right) } \right\} \ , \\
					\left\{ P_{x} , \Phi _{\left( 5 \right) } \right\} & , & \left\{ P_{x} , \Phi _{\left( 6 \right) } \right\} & , & \left\{ P_{y} , \Phi _{\left( 2 \right) } \right\} & , & \left\{ P_{y} , \Phi _{\left( 5 \right) } \right\} \ , \\
					\left\{ P_{y} , \Phi _{\left( 6 \right) } \right\} & , & \left\{ P_{\theta } , \Phi _{\left( 1 \right) } \right\} & , & \left\{ P_{\theta } , \Phi _{\left( 5 \right) } \right\} & \textnormal{and} & \left\{ P_{\theta } , \Phi _{\left( 6 \right) } \right\}
				\end{matrix}
			\end{equation*}
			are all zero. After all, since the nullity of these results also extends to
			\begin{equation*}
				\begin{matrix}
					\left\{ \Phi _{\left( 1 \right) } , x \right\} & \ , \ & \left\{ \Phi _{\left( 2 \right) } , x \right\} & \ , \ & \left\{ \Phi _{\left( 1 \right) } , y \right\} & \ , \ & \left\{ \Phi _{\left( 2 \right) } , y \right\} \ , \\
					\left\{ \Phi _{\left( 6 \right) } , y \right\} & , & \left\{ \Phi _{\left( 1 \right) } , \theta \right\} & , & \left\{ \Phi _{\left( 2 \right) } , \theta \right\} & , & \left\{ \Phi _{\left( 5 \right) } , \theta \right\} \ , \\
					\left\{ \Phi _{\left( 5 \right) } , P_{x} \right\} & , & \left\{ \Phi _{\left( 6 \right) } , P_{x} \right\} & , & \left\{ \Phi _{\left( 2 \right) } , P_{y} \right\} & , & \left\{ \Phi _{\left( 5 \right) } , P_{y} \right\} \ , \\
					\left\{ \Phi _{\left( 6 \right) } , P_{y} \right\} & , & \left\{ \Phi _{\left( 1 \right) } , P_{\theta } \right\} & , & \left\{ \Phi _{\left( 5 \right) } , P_{\theta } \right\} & \textnormal{and} & \left\{ \Phi _{\left( 6 \right) } , P_{\theta } \right\}
				\end{matrix}
			\end{equation*}
			due to the antisymmetry of the Poisson brackets, the only non-zero matrices relevant to the derivation of (\ref{dirac-reduction}) are those defined by
			\begin{subequations}
				\begin{align*}
					\left\{ x , \Phi _{\left( 5 \right) } \right\} & = - \left\{ P_{x} , \Phi _{\left( 1 \right) } \right\} = \tan \phi \ , \ \ \left\{ x , \Phi _{\left( 6 \right) } \right\} = - \left\{ P_{x} , \Phi _{\left( 2 \right) } \right\} = \sec \phi \ , \\
					\left\{ y , \Phi _{\left( 5 \right) } \right\} & = - \left\{ P_{y} , \Phi _{\left( 1 \right) } \right\} = 1 \ , \ \ \left\{ \theta , \Phi _{\left( 6 \right) } \right\} = - \frac{a + 3}{2R} \ \ \textnormal{and} \ \ \left\{ P_{\theta } , \Phi _{\left( 2 \right) } \right\} = R \ .
				\end{align*}
			\end{subequations}
			Thus, since the only non-zero Poisson brackets between $ \left( x , y , \theta \right) $ and $ \left( P_{x} , P_{y} , P_{\theta } \right) $ are
			\begin{eqnarray*}
				\left\{ x , P_{x} \right\} & = & - \left\{ P_{x} , x \right\} = 1 \\
				\left\{ y , P_{y} \right\} & = & - \left\{ P_{y} , y \right\} = 1 \quad \textnormal{and} \\
				\left\{ \theta , P_{\theta } \right\} & = & - \left\{ P_{\theta } , \theta \right\} = 1 \ ,
			\end{eqnarray*}
			it is not difficult to see (although it remains rather tedious to carry out all the calculations that allow us to see this) that the only non-zero Dirac brackets between these same parameters are
			\begin{subequations}
				\begin{align*}
					\left\{ x , P_{x} \right\} _{D} & = \left( \frac{a + 3}{a + 5} \right) \cos ^{2} \phi \ , \ \ \left\{ y , P_{y} \right\} _{D} = \left( \frac{a + 3}{a + 5} \right) \sin ^{2} \phi \ , \\
					\left\{ x , P_{y} \right\} _{D} & = \left\{ y , P_{x} \right\} _{D} = - \left( \frac{a + 3}{a + 5} \right) \frac{\sin \left( 2 \phi \right) }{2} \ , \ \ \left\{ \theta , P_{\theta } \right\} _{D} = \frac{2}{a + 5} \\
					\left\{ x , P_{\theta } \right\} _{D} & = \left( \frac{2R}{a + 5} \right) \cos \phi \ , \ \ \left\{ y , P_{\theta } \right\} _{D} = - \left( \frac{2R}{a + 5} \right) \sin \phi \ , \\
					\left\{ \theta , P_{x} \right\} _{D} & = \left( \frac{a + 3}{a + 5} \right) \frac{\cos \phi }{R} \ \ \textnormal{and} \ \ \left\{ \theta , P_{y} \right\} _{D} = - \left( \frac{a + 3}{a + 5} \right) \frac{\sin \phi }{R} \ .
				\end{align*}
			\end{subequations}
			
		\subsection{Determining the equations of motion}	
		
			Based on everything we have just outlined, it must already be quite clear to you, the reader, that describing a physical system with constraints may be quite complicated when, instead of using these constraints for finding the intrinsic parameters that describe it, we retain all these constraints for a description in terms of non-independent parameters, do you agree? After all, although there are several examples in the literature, where dealing with non-intrinsic parameters greatly facilitates the formulation of a theory (as is the case, for instance, with Maxwell's electrodynamics and the Standard Model of elementary particles \cite{jackson,quigg,ait-hey}), there is no denying that the classical description of our system in intrinsic terms was far simpler than the one we are presenting in this Section, which we have not even finished yet.
			
			But since we have come this far, and now even have at our disposal the results of all the Dirac brackets from the previous Subsection, it is very important to recognise that this puts us in a position to finally find the equations of motion for our Hamiltonian system with constraints. And how can this be done? By recognising that the time evolution of any physical observable $ F \left( q , p \right) $, which can be associated with the description of this Hamiltonian system, is expressed by
			\begin{equation}
				\dot{F} = \left\{ F , H_{\mathrm{total}} \right\} = \left\{ F , H \right\} _{D} \ . \label{dirac-movement}
			\end{equation}
			And, as the application of (\ref{dirac-movement}) to the parameters $ \left( x , y , \theta \right) $ and $ \left( P_{x} , P_{y} , P_{\theta } \right) $ makes it clear that these equations of motion can be expressed, in matrix form, as
			\begin{subequations} \label{moviment-constrained}
				\begin{align}
					& \begin{pmatrix}
						\dot{x} \\
						\dot{y} \\
						\dot{\theta }
					\end{pmatrix} = \frac{1}{m} \left( \frac{a+3}{a+5} \right) \begin{pmatrix}
						\cos ^{2} \phi & - \left[ \sin \left( 2 \phi \right) \right] / 2 & \left( \cos \phi \right) / R \\
						- \left[ \sin \left( 2 \phi \right) \right] / 2 & \sin ^{2} \phi & - \left( \sin \phi \right) / R \\
						\left( \cos \phi \right) / R & - \left( \sin \phi \right) / R & 1 / R^{2}
					\end{pmatrix} \begin{pmatrix}
						P_{x} \\
						P_{y} \\
						P_{\theta }
					\end{pmatrix} \ , \ \ \textnormal{and} \label{x-movement-constrained} \\
					& \begin{pmatrix}
						\dot{P} _{x} \\
						\dot{P} _{y} \\
						\dot{P} _{\theta }
					\end{pmatrix} = \left( \frac{mg}{a+5} \right) \begin{pmatrix}
						\left( a + 3 \right) \left[ \sin \left( 2 \phi \right) \right] / 2 \\
						- \left( a + 3 \right) \sin ^{2} \phi \\
						2R \left( \sin \phi \right)
					\end{pmatrix} \ , \label{p-movement-constrained}
				\end{align}
			\end{subequations}
			is precisely a further development of these same equations of motion that already allows us to compare them with those obtained in the previous Sections. To this end, it is important to note that, from the matrix equation (\ref{x-movement-constrained}), it follows that
			\begin{equation}
				\begin{pmatrix}
					\ddot{x} \\
					\ddot{y} \\
					\ddot{\theta }
				\end{pmatrix} = \frac{1}{m} \left( \frac{a+3}{a+5} \right) \begin{pmatrix}
					\cos ^{2} \phi & - \left[ \sin \left( 2 \phi \right) \right] / 2 & \left( \cos \phi \right) / R \\
					- \left[ \sin \left( 2 \phi \right) \right] / 2 & \sin ^{2} \phi & - \left( \sin \phi \right) / R \\
					\left( \cos \phi \right) / R & - \left( \sin \phi \right) / R & 1 / R^{2}
				\end{pmatrix} \begin{pmatrix}
					\dot{P} _{x} \\
					\dot{P} _{y} \\
					\dot{P} _{\theta }
				\end{pmatrix} \ . \label{x-movement-derivative-constrained}
			\end{equation}
			And why is it important to note this? Because, as the column matrix, which appears on the left-hand side of (\ref{x-movement-derivative-constrained}), is the same as the one that already appeared on the right-hand side of (\ref{p-movement-constrained}), it is precisely the combination of these two results that shows us that
			\begin{subequations} \label{final-equations}
				\begin{align}
					\ddot{x} & = g \left( \frac{a+3}{a+5} \right) \frac{\sin \left( 2 \phi \right) }{2} \ , \label{x-final-equations} \\
					\ddot{y} & = - g \left( \frac{a+3}{a+5} \right) \sin ^{2} \phi \ \ \textnormal{and} \label{y-final-equations} \\
					\ddot{z} & = g \left( \frac{a+3}{a+5} \right) \frac{\sin \phi }{R} \ . \label{z-final-equations}
				\end{align}
			\end{subequations}
			In other words, we are faced with a result that is entirely consistent with everything we have previously set out because, in addition to (\ref{x-final-equations}) being identified as the same equation of motion that we have already found in (\ref{mov-eq}), (\ref{mov-equation-l2}) and (\ref{h-intrinsic-mov-equation}), the ratios
			\begin{equation*}
				\frac{\ddot{x}}{\ddot{y}} = - \cot \phi \quad \textnormal{and} \quad \frac{\ddot{x}}{\ddot{\theta }} = R \left( \cos \phi \right) \ ,
			\end{equation*}
			which can be obtained using all the results contained in (\ref{final-equations}), also follow from the constraints $ \Phi _{\left( 1 \right) } \left( x , y \right) \approx 0 $ and $ \Phi _{\left( 2 \right) } \left( x , \theta \right) \approx 0 $ respectively.
			
	\section{\label{q-constraints-description}Quantum description of the Hamiltonian in terms of constraints}
	
		Another point that the Dirac brackets in the penultimate Subsection also make clear is that, since they do not identify as Poisson brackets due to the fact that we are dealing with a Hamiltonian system with constraints, the quantisation $ \mathcal{Q} $ of this system must be guided by the correspondence (\ref{correspondence-dirac}). And because all the physical observables obtained above are described by real-valued functions, this is precisely what reinforces the fact that all the observables, which take the form of $ F \left( q , p \right) $ and $ G \left( q , p \right) $ in (\ref{correspondence-dirac}), must correspond to the respective self-adjoint operators $ \hat{F} = \mathcal{Q} \left( F \right) $ and $ \hat{G} = \mathcal{Q} \left( G \right) $. Anyhow, given that
        \begin{itemize}
            \item all the Dirac brackets obtained above turn out to be mere constants, whether they are zero or not, and
			\item the real functions, which describe all physical observables, do not contain any products of conjugate parameters,
        \end{itemize}
        it is crucial to note that we are dealing with a scenario that admits a quantisation that, once again, is univocal: more specifically, we are dealing with a situation that admits a quantisation that, under the assumption that each of the components of
		\begin{equation}
			\bigl( \hat{x} , \hat{y} , \hat{\theta } \bigr) = \left( \mathcal{Q} \left( x \right) , \mathcal{Q} \left( y \right) , \mathcal{Q} \left( \theta \right) \right) \quad \textnormal{and} \quad \bigl( \hat {P} _{x} , \hat{P} _{y} , \hat{P} _{\theta } \bigr) = \left( \mathcal{Q} \left( P_{x} \right) , \mathcal{Q} \left( P_{y} \right) , \mathcal{Q} \left( P_{\theta } \right) \right) \label{constrained-components}
		\end{equation}
		are identified as self-adjoint operators, leads us to a Schrödinger equation that is univocally dominated by the Hamiltonian operator
		\begin{eqnarray}
			\hat{H} & = & H \bigl( \left( \mathcal{Q} \left( x \right) , \mathcal{Q} \left( y \right) , \mathcal{Q} \left( \theta \right) \right) , \left( \mathcal{Q} \left( P_{x} \right) , \mathcal{Q} \left( P_{y} \right) , \mathcal{Q} \left( P_{\theta } \right) \right) \bigr) \label{h-quantum-constrained} \\
			& & \hspace*{6.0cm} = \frac{\hat{P} _{x} ^{2}}{2m} + \frac{\hat{P} _{y} ^{2}}{2m} + \frac{\left( a + 3 \right) \hat{P} _{\theta } ^{2}}{4mR^{2}} + mg \hat{y} \ .
		\end{eqnarray}
		
		\subsection{\label{subsub-phy}Defining a physical subspace}
		
			Moreover, since the quantisation of this constrained Hamiltonian system must be guided by (\ref{correspondence-dirac}), it is important to note that all the components appearing in (\ref{constrained-components}) are necessarily such that
			\begin{subequations} \label{dirac-constrained-comutation}
				\begin{align}
					\bigl[ \hat{x} , \hat{P} _{x} \bigr] & = i \hbar \left( \frac{a + 3}{a + 5} \right) \left( \cos ^{2} \phi \right) \hat{I} \ , \ \ \bigl[ \hat{y} , \hat{P} _{y} \bigr] = i \hbar \left( \frac{a + 3}{a + 5} \right) \left( \sin ^{2} \phi \right) \hat{I} \ , \\
					\bigl[ \hat{x} , \hat{P} _{y} \bigr] & = \bigl[ \hat{y} , \hat{P} _{x} \bigr] = - i \hbar \left( \frac{a + 3}{a + 5} \right) \left( \frac{\sin \left( 2 \phi \right) }{2} \right) \hat{I} \ , \ \ \bigl[ \hat{\theta } , \hat{P} _{\theta } \bigr] = \left( \frac{2i \hbar }{a + 5} \right) \hat{I} \\
					\bigl[ \hat{x} , \hat{P} _{\theta } \bigr] & = i \hbar \left( \frac{2R}{a + 5} \right) \left( \cos \phi \right) \hat{I} \ , \ \ \bigl[ \hat{y} , \hat{P} _{\theta } \bigr] = - i \hbar \left( \frac{2R}{a + 5} \right) \left( \sin \phi \right) \hat{I} \ , \\
					\bigl[ \hat{\theta } , \hat{P} _{x} \bigr] & = i \hbar \left( \frac{a + 3}{a + 5} \right) \left( \frac{\cos \phi }{R} \right) \hat{I} \ \ \textnormal{and} \ \ \bigl[ \hat{\theta } , \hat{P} _{y} \bigr] = - i \hbar \left( \frac{a + 3}{a + 5} \right) \left( \frac{\sin \phi }{R} \right) \hat{I} \ .
				\end{align}
			\end{subequations}
			And why is it important to note this? Because this result only reinforces that all components that define (\ref{constrained-components}) have some kind of dependency among themselves. Scilicet, it is a dependency inherited from the classical description since, just as quantisation $ \mathcal{Q} $ raises the variables $ \bigl( x , y , \theta \bigr) $ and $\bigl( P_{x} , P_{y} , P_{\theta } \bigr) $ to operators (\ref{constrained-components}) that act on a Hilbert space, it also does the same with the functions present in (\ref{reexpress}) and (\ref{Phi56-functions}). That is, they raise such functions to the operators
			\begin{subequations} \label{quantum-operators}
				\begin{align}
					\hat{\Phi } _{\left( 1 \right) } & = \Phi _{\left( 1 \right) } \bigl( \left( \mathcal{Q} \left( x \right) , \mathcal{Q} \left( y \right) \right) \bigl) = \left( \tan \phi \right) \hat{x} + \hat{y} \ , \label{quantum-operators-1} \\
					\hat{\Phi } _{\left( 2 \right) } & = \Phi _{\left( 2 \right) } \bigl( \left( \mathcal{Q} \left( x \right) , \mathcal{Q} \left( \theta \right) \right) \bigl) = \left( \sec \phi \right) \hat{x} - R \hat{\theta } \ , \label{quantum-operators-2} \\
					\hat{\Phi } _{\left( 5 \right) } & = \Phi _{\left( 5 \right) } \bigl( \left( \mathcal{Q} \left( P_{x} \right) , \mathcal{Q} \left( P_{y} \right) \right) \bigl) = \left( \tan \phi \right) \hat{P} _{x} + \hat{P} _{y} \ \ \textnormal{and} \label{quantum-operators-5} \\
					\hat{\Phi } _{\left( 6 \right) } & = \Phi _{\left( 6 \right) } \bigl( \left( \mathcal{Q} \left( P_{x} \right) , \mathcal{Q} \left( P_{\theta } \right) \right) \bigl) = \left( \sec \phi \right) \hat{P} _{x} - \frac{\left( a + 3 \right) \hat{P} _{\theta }}{2R} \ . \label{quantum-operators-6}
				\end{align}
			\end{subequations}
			Thus, since the constraints given in (\ref{new-phi-constraints}) and (\ref{Phi56}) define, together, the cotangent subbundle of $ T^{\ast } \mathcal{M} _{n} $ to where the dynamics of our Hamiltonian system are restricted, this correspondence $ \mathcal {Q} $ also requires that the set, comprising all eigenstates, with zero eigenvalues, of all the operators listed in (\ref{quantum-operators}), defines the Hilbert subspace $ \mathfrak{H} _{\mathrm{ph}} $ in which the physical state of our system lies \cite{dirac,gitman}.
			
			In order to understand a little better what this $ \mathfrak{H} _{\mathrm{ph}} $ is, it is important to note two things. The first is that, since
            \begin{equation*}
                \left\{ \Phi _{\left( j \right) } , \Phi _{\left( \ell \right) } \right\} _{D} = 0
			\end{equation*}
            is an equality that is valid for all values of $ j, \ell = 1 , 2 , 5 , 6 $, the correspondence $ \mathcal{Q} $ replicates this validity in terms of commuting
			\begin{equation}
                \bigl[ \hat{\Phi } _{\left( j \right) } , \hat{\Phi } _{\left( \ell \right) } \bigr] = 0 \ . \label{replication-1}
            \end{equation}
            The second thing to note that, as the consistency equations of the Dirac-Bergmann algorithm can also be rewritten as
			\begin{equation}
                \dot{\Phi } _{\left( \ell \right) } = \left\{ \Phi _{\left( \ell \right) } , H \right\} _{D} \approx 0 \label{dirac-bergmann-alt}
            \end{equation}
			using Dirac brackets, the same applies here. In plain English, since the Dirac-Bergmann algorithm requires that (\ref{dirac-bergmann-alt}) be a valid relation for all values of $ \ell = 1 , 2 , 5 , 6 $, this correspondence $ \mathcal{Q} $ also replicates the validity of this relation in terms of commuting
			\begin{equation}
				\dot{\hat{\Phi }} _{\left( \ell \right) } = \bigl[ \hat{\Phi } _{\left( \ell \right) } , \hat{H} \bigr] \approx 0 \ . \label{replication-2}
			\end{equation}
			And why is it important to note these two things? Because, as the spectral theorem guarantees that, whenever two operators commute in Quantum Mechanics, they share a common basis of eigenstates \cite{hilbert-halmos,spectral-halmos,moretti}, it is precisely (\ref{replication-1}) and (\ref{replication-2}) that ultimately ensure, for example, that all eigenstates of (\ref{h-quantum-constrained}) belong to $ \mathfrak {H} _{\mathrm{ph}} $ \cite{dirac,gitman}. Therefore, since the fact that the Hamiltonian operator (\ref{h-quantum-constrained}) does not depend explicitly on time allows us, as before, to recognise that the solution to the Schrödinger equation in which it appears can be expressed as
			\begin{equation}
				\psi \left( \left( x , y , \theta \right) , t \right) = \Psi \left( x , y , \theta \right) e^{-i E \left( t - t_{0} \right) / \hbar } \ , \label{general-solution-constrained}
			\end{equation}
			where $ E $ and $ \Psi \left( x , y , \theta \right) $ are the respective eigenenergies and eigenstates of the equation
			\begin{equation}
				\hat{H} \hspace*{0.04cm} \Psi \left( x , y , \theta \right) = E \hspace*{0.04cm} \Psi \left( x , y , \theta \right) \ , \label{position-solution-constrained}
			\end{equation}
			this is what allows us to assert that all stationary eigenstates $ \Psi _{\mathrm{ph}} \left( x , y , \theta \right) $ that are physical and, consequently, that all physical solutions $ \psi _{\mathrm{ph}} \left( \left( x , y , \theta \right) , t \right) $ of the time-dependent Schrödinger equation will always be contained in $ \mathfrak{H} _{\mathrm{ph}} $. Note that it is precisely this ``physical'' predicate that allows us to explain, somewhat more clearly, what we said at the end of the previous paragraph. After all, given the commutativity, which is present on the right-hand side of (\ref{replication-2}), that guarantees that the physical stationary eigenstates satisfy the relation
			\begin{equation}
				\hat{\Phi } _{\left( \ell \right) } \hspace*{0.04cm} \Psi _{\mathrm{ph}} \left( x , y , \theta \right) \approx 0 \label{ph-subspace}
			\end{equation}
			for all values $ \ell = 1 , 2 , 5 , 6 $, it is precisely these physical stationary eigenstates that illustrate this statement, which we made at the end of the previous paragraph, since they are eigenstates with zero eigenvalues of all the operators listed in (\ref{quantum-operators}).
			
		\subsection{A further comment on the physical subspace}
		
			Of course, due to the way we chose to refer to the solutions (\ref{general-solution-constrained}) and (\ref{position-solution-constrained}) of the respective time-dependent and time-independent Schrödinger equations, a more attentive reader may be wondering whether it is actually possible to find a representation for the Hamiltonian operator (\ref{h-quantum-constrained}) in position space. However, given the commutation relations announced at the beginning of Subsubsection \ref{subsub-phy}, it is not difficult to see that the self-adjoint operators that define them can be represented as
			\begin{subequations} \label{representation-constrained}
				\begin{align}
					& \hat{x} = x \quad , \quad \hat{P} _{x} = - i \hbar \left( \frac{a + 3}{a + 5} \right) \left[ \left( \cos ^{2} \phi \right) \hspace*{0.04cm} \frac{\partial }{\partial x} - \frac{\sin \left( 2 \phi \right) }{2} \hspace*{0.04cm} \frac{\partial }{\partial y} + \frac{\cos \phi }{R} \hspace*{0.04cm} \frac{\partial }{\partial \theta } \right] \ , \label{representation-constrained-1} \\
					& \hat{y} = y \quad , \quad \hat{P} _{y} = - i \hbar \left( \frac{a + 3}{a + 5} \right) \left[ - \frac{\sin \left( 2 \phi \right) }{2} \hspace*{0.04cm} \frac{\partial }{\partial x} - \left( \sin ^{2} \phi \right) \hspace*{0.04cm} \frac{\partial }{\partial y} - \frac{\sin \phi }{R} \hspace*{0.04cm} \frac{\partial }{\partial \theta } \right] \ , \label{representation-constrained-2} \\
					& \hat{\theta } = \theta \quad \textnormal{and} \quad \hat{P} _{\theta } = - i \hbar \left[ \left( \frac{2R}{a + 5} \right) \cos \phi \hspace*{0.04cm} \frac{\partial }{\partial x} - \left( \frac{2R}{a + 5} \right) \sin \phi \hspace*{0.04cm} \frac{\partial }{\partial y} + \left( \frac{2}{a + 5} \right) \hspace*{0.04cm} \frac{\partial }{\partial \theta } \right] \label{representation-constrained-3}
				\end{align}
			\end{subequations}
			in position space. In other words, although it is entirely unnecessary for us to refer to a particular choice for the representation of the operators when discussing the properties of the Schrödinger equation, the expressions above make it quite clear that the Hamiltonian operator (\ref{h-quantum-constrained}) can, in fact, be represented in position space, and this is what ensures the consistency of everything we have said above.
			
			Nonetheless, and still paying attention to the reasons why a state can be described as ``physical'', it is essential to note that, just as is the situation with the Hamiltonian function (\ref{h-nh-classic}) in the classical context, the Hamiltonian operator (\ref{h-quantum-constrained}), alone, is not capable of describing our physical system quantum mechanically. After all, just as the Hamiltonian function (\ref{h-nh-classic}) (because it is expressed by a greater number of parameters than are physically necessary due to the constraints (\ref{new-phi-constraints}) and (\ref{Phi56})) can only describe the physics of our system when the functions (\ref{reexpress}) and (\ref{Phi56-functions}) take on zero values, the Hamiltonian operator (\ref{h-quantum-constrained}) (by also being expressed by a greater number of operators than are physically necessary) can only describe the physics of our system quantum mechanically when the action of the operators, on the various eigenstates of (\ref{h-quantum-constrained}), results in zero eigenvalues. In other words, since the operators listed in (\ref{quantum-operators}), being also expressed by the same operators as in (\ref{h-quantum-constrained}), act on a Hilbert space $ \mathfrak{H} $, which is higher-dimensional than $ \mathfrak{H} _{\mathrm{ph}} $ in which the physical states of our Hamiltonian system are contained, the quantum equivalent of the constraints must be strictly defined by (\ref{ph-subspace}) and not in terms of the equalities
			\begin{equation}
				\hat{\Phi } _{\left( \ell \right) } \approx \hat{0} \ , \label{quantum-constraints-relations}
			\end{equation}
			since, by definition, these equalities are only valid in $ \mathfrak{H} _{\mathrm{ph}} \subset \mathfrak{H} $.
			
		\subsection{\label{equivalence-ic}On the equivalence of intrinsic and non-intrinsic formulations}
		
			However, as we have just pointed out that all these equalities (\ref{quantum-constraints-relations}) hold in $ \mathfrak{H} _{\mathrm{ph}} $, it is also interesting to note that the same equalities already stated in (\ref{representation-constrained}), which allow us to represent the operators $ \hat{P} _{x} $, $ \hat{P} _{y} $ and $ \hat{P} _{\theta } $ in position space, can be rewritten in matrix form as
			\begin{equation}
				\begin{pmatrix}
					\hat{P} _{x} \\
					\hat{P} _{y} \\
					\hat{P} _{\theta }
				\end{pmatrix} = - i \hbar \left( \frac{a + 3}{a + 5} \right) \begin{pmatrix}
					\cos ^{2} \phi & - \left[ \sin \left( 2 \phi \right) \right] / 2 & \left( \cos \phi \right) / R \\
					- \left[ \sin \left( 2 \phi \right) \right] / 2 & \sin ^{2} \phi & - \left( \sin \phi \right) / R \\
					2 R \left( \cos \phi \right) / \left( a + 3 \right) & - 2 R \left( \sin \phi \right) / \left( a + 3 \right) & 2 / \left( a + 3 \right)
				\end{pmatrix} \begin{pmatrix}
					\partial / \partial x \\
					\partial / \partial y \\
					\partial / \partial \theta
				\end{pmatrix} \ . \label{matrix-ppp}
			\end{equation}
			And why is it also interesting to draw attention to this fact? Because, as the determinant of the square matrix appearing on the right-hand side of (\ref{matrix-ppp}) is zero, we cannot use this matrix equation to, for instance, express the differential operators $ \partial / \partial x  $, $ \partial / \partial y $ and $ \partial / \partial \theta  $ in terms of $ \hat{P} _ {x} $, $ \hat{P} _{y} $ and $ \hat{P} _{\theta } $. But, as boring as it may be to calculate this determinant explicitly to show that it is zero, observe that a neat way to confirm this result is showing that this matrix contains rows that are proportional to one another \cite{strang}. And since, when multiplying the entire first row of this matrix by
			\begin{itemize}
				\item $ - \tan \phi $, this leads us to realise that $ \hat{P} _{y} = - \left( \tan \phi \right) \hat{P} _{x} $, and
				\item $ \left( a + 3 \right) / \left[ 2 R \left( \cos \phi \right) \right] $, this leads us to $ \hat{P} _{\theta } = \left[ 2 R \left( \cos \phi \right) / \left( a + 3 \right) \right] \hat{P} _{x} $,
			\end{itemize}
			this is precisely what reinforces that the expressions
			\begin{subequations} \label{relation-56-reason}
				\begin{align}
					\hat{\Phi } _{\left( 5 \right) } \approx \hat{0} & \ \Rightarrow \ \left( \tan \phi \right) \hat{P} _{x} + \hat{P} _{y} \approx \hat{0} \ \ \textnormal{and} \label{relation-5-reason} \\
					\hat{\Phi } _{\left( 6 \right) } \approx \hat{0} & \ \Rightarrow \ \left( \sec \phi \right) \hat{P} _{x} - \frac{\left( a + 3 \right) \hat{P} _{\theta }}{2R} \approx \hat{0} \label{relation-6-reason}
				\end{align}
			\end{subequations}
			need only hold in $ \mathfrak{H} _{\mathrm{ph}} $.
			
			Note that this observation we have just made was, in fact, to be expected. After all, given that the representation described in (\ref{representation-constrained}) ensures that the operators $ \hat{P} _{x} $, $ \hat{P} _{y} $ and $ \hat{P} _{\theta } $ satisfy all the commutation relations (\ref{dirac-constrained-comutation}) between themselves and the operators $ \hat{x} $, $ \hat{y} $ and $ \hat{\theta } $, the fact that these commutators are mere fruits of the quantisation of the Dirac brackets is what ensures that this requirement, of the (\ref{relation-56-reason}) relations being valid, emerges from the matrix manipulation we have just performed above. And since all this allows us, for instance, to recognise that
			\begin{equation*}
				\hat{P} ^{2} _{y} = \left( \tan ^{2} \phi \right) \hat{P} ^{2} _{x} \quad \textnormal{and} \quad \hat{P} ^{2} _{\theta } = \frac{4R^{2} \left( \sec ^{2} \phi \right) \hat{P} ^{2} _{x}}{\left( a + 3 \right) ^{2}} \ ,
			\end{equation*}
			it is already quite significant to note that this leads us to
			\begin{equation}
				\frac{\hat{P} ^{2} _{x}}{2m} + \frac{\hat{P} ^{2} _{y}}{2m} + \frac{\left( a + 3 \right) \hat{P} ^{2} _{\theta }}{4mR^{2}} = \frac{\left( \sec ^{2} \phi \right) \hat{P} ^{2} _{x}}{2m} \left( \frac{a + 5}{a + 3} \right) = \frac{\hat{P} ^{2} _{x}}{2m \left( \sec ^{2} \phi \right) } \left( \frac{a + 3}{a + 5} \right) \label{kinetic-constrained}
			\end{equation}
			because this result is nothing more than the expression that the kinetic part of the Hamiltonian operator (\ref{h-quantum-constrained}) takes on in the physical subspace $ \mathfrak{H} _{\mathrm{ph}} $. Thus, if we also note that, in this same subspace, the relation
			\begin{equation}
				\hat{y} = - \left( \tan \phi \right) \hat{x} \label{potential-constrained}
			\end{equation}
			is also valid, this is precisely what allows us to conclude that the quantisation of our Hamiltonian system with constraints describes the same physics as the Hamiltonian operator (\ref{h-operator}) which was expressed in intrinsic terms. After all, despite (\ref{kinetic-constrained}) and (\ref{potential-constrained}) allowing us to reduce the expression of the Hamiltonian operator (\ref{h-quantum-constrained}) of our constrained Hamiltonian system to
			\begin{equation}
				\hat{H} = \frac{\left( \sec ^{2} \phi \right) \hat{P} ^{2} _{x}}{2m} \left( \frac{a + 5}{a + 3} \right) - mg \left( \tan \phi \right) \hat{x} \label{reexpress-h-constrained}
			\end{equation}
			in $ \mathfrak{H} _{\mathrm{ph}} $, it is not difficult to see that (\ref{h-quantum-constrained}) takes the same form as (\ref{h-operator}) under the consideration that
			\begin{equation}
				\hat{P} = \left( \frac{a + 5}{a + 3} \right) \left( \sec ^{2} \phi \right) \hat{P} _{x} \ . \label{ppx}
			\end{equation}
			In other words, the Hamiltonian operators (\ref{h-operator}) and (\ref{reexpress-h-constrained}) actually describe the same quantum mechanics, a fact further reinforced by the observation that it is precisely this relation (\ref{ppx}) that allows us to observe that
			\begin{equation*}
				\bigl[ \hat{x} , \hat{P} _{x} \bigr] = i \hbar \left( \frac{a + 3} {a + 5} \right) \left( \cos ^{2} \phi \right) \hat{I} \ \Leftrightarrow \ \bigl[ \hat{x} , \hat{P} \bigr] = i \hbar \hspace*{0.04cm} \hat{I} \ .
			\end{equation*}
			
	\section{\label{final-comments}Final remarks}
	
		Even though Section \ref{q-constraints-description} is not as extensive as Section \ref{constraints-h-description}, its Subsection \ref{equivalence-ic} makes it clear that the quantisation we have performed, on the classical Hamiltonian formulation of our system with constraints, leads to the same physical result as Section \ref{intrinsic-q-description} has already shown. In other words, although we could present further developments here, regarding the results that follow from this quantisation with constraints, they will all be physically equivalent to those given in Section \ref{intrinsic-q-description}. Nevertheless, it is important to emphasize that this manuscript should be interpreted, amongst other things, as a comprehensive guide to the significance and general characteristics of the Dirac quantisation process. After all, no matter how simple the physical system we have been considering here may be, it has enabled us to discuss a number of issues of paramount importance relating to both its classical and quantum formulations.
		
		Indeed, while the Hamiltonian description of our massive ball (which moves in accordance with points (i) and (ii) of the Introduction) has led us to unambiguous quantisations, it is not wrong to claim that everything we have developed so far has raised a number of questions that are currently highly topical. And in order to understand why this is so, it suffices to notice, for example, that right at the beginning of Section \ref{sec:vak}, we have already drawn your attention, the reader, to the fact that the first Lagrangian function we encountered pointed to the need to deal with Vakononic Mechanics \cite{kozlov-1,kozlov-2,kozlov-3,cardin}: i.e., the Lagrangian function (\ref{constrained-lagrangean-vak}) pointed to the need to deal with an alternative variational formulation to standard Classical Mechanics that, despite not being the ``apple of physicists' eye'', plays a central role in various current issues involving, for example, the theory of geometric control \cite{martinez,leon-2,cortes,enomoto}. Of course, it would not be too difficult to proceed with this Lagrangian function, by analysing the resulting outcomes in the light of the principles of Vakonomic Mechanics. But, since going into all the details would make this manuscript even longer than it already is, we chose not to go into such an analysis, leaving it for a future paper.		
		
		Another point worth highlighting, which also underscores how this manuscript addresses current issues, concerns the various studies that have been (and continue to be) carried out using quantum models with linear potentials. After all, remember that, in addition to gravitational force, the force due to the action of an electric field, on an electrically charged particle, also leads us to a scenario that requires us to deal with linear potentials \cite{jackson}. And in the case of this research, which deals with linear electrostatic potentials in quantum contexts in some way, it is worth noting, for example, that it is of paramount importance for a proper understanding of
        \begin{itemize}
            \item the tunnelling phenomena that occur at semiconductor junctions \cite{liu,hashemi} and
			\item the Fowler-Nordheim effect, which is characterised by the transfer of electrons, from the surface of a material (which may be a metal or a superconductor) to some insulating medium (such as, for example, a vacuum), due to the action of an extremely intense electric field \cite{fowler,nordhein,lepetit,karaoulanis}.
        \end{itemize}
		Nonetheless, something that particularly catches our attention stems from certain results involving the use of linear gravitational potentials in quantum contexts, especially those derived from the bouncing ball model \cite{haar,sakurai,langhoff,gibbs,banacloche}. After all, as this theoretical model deals with the quantum mechanics of a ball of mass $ m $, which moves in a single direction $ z $ (perpendicular to the ground) due to the Earth’s gravitational potential
		\begin{equation*}
            V \left( z \right) = \begin{cases}
                \ mgz \ , \ \ \text{when} \ \ z > 0 \ , \ \ \textnormal{and} \\
                \hspace*{0.30cm} \infty \hspace*{0.24cm} \ , \ \ \text{otherwise} \ ,
            \end{cases}
		\end{equation*}
        it is impossible not to recognise that this model is analogous to the one we presented, as an example, in Subsection \ref{boucing}. And recognising this analogy only serves to reinforce, even further, that everything we have set out in this manuscript points to highly current research questions because, given the results of this bouncing ball model, subsequent investigations have made it crystal clear that the neutron (being a subatomic particle with mass and no electric charge) was the perfect particle for testing the experimental validity of this model. And thanks to the successful results of the experiments, which were carried out at the Laue-Langevin Institute using ultracold neutrons \cite{nesvizhevsky-1}, subsequent investigations were undertaken (both theoretically and experimentally), some of which involved the search for new physics. After all, given that the eigenenergies (\ref{eigenenergies-p-infinity}) depend (in magnitude) on the gravitational acceleration, the same applies to the eigenenergies of this bouncing ball model, and, therefore, any deviation, no matter how small, from the values predicted by this model, can be interpreted, for instance, as indicative of the existence of dark matter and/or dark energy, and/or of deviations from Newton's inverse square law \cite{jenke,abele,voronin}.
		
		Of course, as much as we have only focused on the analogy between the model in Subsection \ref{boucing} and that of a bouncing ball in the previous paragraph, it is well worth noting that other, quite recent studies also explore different confinement geometries, which include some that bear a certain resemblance to the model in Subsection \ref{symm-pot} \cite{nesvizhevsky-2,nesvizhevsky-3,allam,altarawneh}. And since we have just mentioned the exploration of these other geometries, it seems worth noting here that, towards the end of the 20th century, considerable efforts were made to assess whether the quantisation of a particle in a curved space would become unambiguous if, for example, such a curved space could be regarded as a surface of another Euclidean space \cite{art-fujii,art-chep,art-hom,art-shi,art-saa}. After all, as the univocity of the quantisations discussed in this manuscript results from the fact that the classical formulations presented here do not contain products of conjugate variables, note that this is not the reality, for example, for a particle situated in a curved space. That is, because the metric tensor of a curved space is parameterised by the position variables \cite{manfredo-gr}, the kinetic term of the Hamiltonian function, which describes any particle intrinsically within such a curved space, is necessary determined by \cite{syn}
        \begin{equation*}
            g^{\mu \nu } \left( q \right) p_{\mu } p_{\nu } \ .
		\end{equation*}
		And to conclude this manuscript by drawing your attention, dear reader, to this fact is noteworthy, given that, as we have just observed, other geometries are being explored for the experimental realisation of quantum models with linear potentials, the application of the results found in Refs. \cite{art-fujii,art-chep,art-hom,art-shi,art-saa} to some specific geometry, which can be explored in the laboratory, may help us to understand something more visceral, related to the non-univocality of quantisation processes in curved spaces. In other words, this is another relevant evaluation, whose discussion we shall leave for future publications.
	
	\section{Acknowledgments}
	
		We thank P. A. S. Autreto for providing some of the logistical support during the drafting of this manuscript. We also thank the artificial intelligence behind ChatGPT for helping us create the images shown in Figures \ref{plano-inclinado}, \ref{reta}, \ref{rolando}, \ref{barreira} and \ref{cunha}, which we later edited using TikZ/LaTex. In fact, as Brazilians, we are very grateful to this artificial intelligence for creating such beautiful images featuring footballs.

\bigskip{\small \smallskip\noindent Updated: \today.}


\end{document}